\documentclass[superscriptaddress,10pt,nature,twocolumn,longbibliography, notitlepage]{revtex4-1}
\usepackage{float}
\floatstyle{boxed}
\usepackage{array}
\usepackage{graphicx}
\usepackage{amsbsy,stmaryrd}
\usepackage[utf8x]{inputenc}
\usepackage{epstopdf}
\usepackage{amsmath,amssymb,amsfonts,amsthm}
\usepackage{indentfirst}
\usepackage{dsfont}
\usepackage{soul}
\usepackage[T1]{fontenc}
\usepackage[dvipsnames]{xcolor}
\usepackage{url}
\usepackage[colorlinks]{hyperref}
\usepackage{lipsum}  
\usepackage{siunitx} 
\usepackage{gensymb} 
\usepackage{footmisc}
\usepackage{cleveref}
\usepackage{easyReview}

\usepackage{titlesec}
\titleformat{\section} 
	{\normalfont\large\bfseries}{\makebox[20pt][l]{\thesection}}{0pt}{} 
\titleformat{\subsection} 
	{\normalfont\normalsize\bfseries}{\makebox[20pt][l]{\thesubsection}}{0pt}{}
\titleformat{\subsubsection} 
    {\normalfont\itshape\bfseries}{\makebox[20pt][l]{\thesubsubsection}}{0pt}{}
 
\titlespacing*{\section}{0pt}{20pt}{2pt} 
\titlespacing*{\subsection}{0pt}{10pt}{2pt}
\titlespacing*{\subsubsection}{0pt}{10pt}{2pt} 

\hypersetup{
	plainpages=true,
	breaklinks=true,
	hypertexnames=false,
	pageanchor=true,
	colorlinks=true,
	linkcolor={blue},
	citecolor={red},
	urlcolor={blue},
	anchorcolor={black}
}

\graphicspath{{figs/}} 

\newcommand{\bra}[1]{\langle #1|}
\newcommand{\ket}[1]{|#1\rangle}

\newcommand{\ketbra}[2]{\left| #1 \rangle \langle #2 \right|}
\newcommand{\brakket}[3]{\left\langle #1\left| #2 \right| #3\right\rangle}
\newcommand{\expec}[1]{\left\langle #1 \right\rangle}

\newcommand{\figref}[1]{\mbox{Fig.~\ref{#1}}}
\newcommand{\figsref}[1]{\mbox{Figs.~\ref{#1}}}
\newcommand{\tabref}[1]{\mbox{Table~\ref{#1}}}

\renewcommand{\eqref}[1]{\mbox{Eq.~(\ref{#1})}}

\newcommand{\be}{\begin{equation}}
\newcommand{\ee}{\end{equation}}
\newcommand{\bea}{\begin{eqnarray}}
\newcommand{\eea}{\end{eqnarray}}

\newcommand{\LL}{\mathcal{L}}
\newcommand{\DD}{\mathcal{D}}
\newcommand{\rhot}{\hat{\rho}}
\newcommand{\sss}{\hat{\rho}_{\rm ss}}

\newcommand{\de}{{\rm d}}

\usepackage{dsfont}

\newcommand{\HH}{\hat{H}}
\newcommand{\trhot}{\hat{\tilde\rho}(t)}
\newcommand{\Bessel}[2]{J_{#1}\left( #2 \right)}
\newcommand{\SParam}{{S}_{21}}

\begin{document}

\title{Landau-Zener without a Qubit: Multiphoton Sidebands Interaction and Signatures of Dissipative Quantum Chaos} 

\author{Léo Peyruchat}
\affiliation{Hybrid Quantum Circuits Laboratory (HQC), Institute of Physics, \'{E}cole Polytechnique F\'{e}d\'{e}rale de Lausanne (EPFL), 1015 Lausanne, Switzerland}
\affiliation{Center for Quantum Science and Engineering, \\ \'{E}cole Polytechnique F\'{e}d\'{e}rale de Lausanne (EPFL), CH-1015 Lausanne, Switzerland}
\author{Fabrizio Minganti}
\affiliation{Center for Quantum Science and Engineering, \\ \'{E}cole Polytechnique F\'{e}d\'{e}rale de Lausanne (EPFL), CH-1015 Lausanne, Switzerland}
\affiliation{Laboratory of Theoretical Physics of Nanosystems (LTPN), Institute of Physics, \'{E}cole Polytechnique F\'{e}d\'{e}rale de Lausanne (EPFL), 1015 Lausanne, Switzerland}
\author{Marco Scigliuzzo}
\affiliation{Center for Quantum Science and Engineering, \\ \'{E}cole Polytechnique F\'{e}d\'{e}rale de Lausanne (EPFL), CH-1015 Lausanne, Switzerland}
\affiliation{%
Laboratory of Photonics and Quantum Measurements (LPQM),  Institute of Physics, EPFL, CH-1015 Lausanne, Switzerland
}%
\author{Filippo Ferrari}
\affiliation{Center for Quantum Science and Engineering, \\ \'{E}cole Polytechnique F\'{e}d\'{e}rale de Lausanne (EPFL), CH-1015 Lausanne, Switzerland}
\affiliation{Laboratory of Theoretical Physics of Nanosystems (LTPN), Institute of Physics, \'{E}cole Polytechnique F\'{e}d\'{e}rale de Lausanne (EPFL), 1015 Lausanne, Switzerland}
\author{Vincent Jouanny}
\affiliation{Hybrid Quantum Circuits Laboratory (HQC), Institute of Physics, \'{E}cole Polytechnique F\'{e}d\'{e}rale de Lausanne (EPFL), 1015 Lausanne, Switzerland}
\affiliation{Center for Quantum Science and Engineering, \\ \'{E}cole Polytechnique F\'{e}d\'{e}rale de Lausanne (EPFL), CH-1015 Lausanne, Switzerland}
\author{Franco Nori}
\affiliation{Theoretical Quantum Physics
Laboratory,  Cluster for Pioneering Research, RIKEN,  Wako-shi,
Saitama 351-0198, Japan} \affiliation{Quantum Information Physics
Theory Research Team, Quantum Computing Center, RIKEN, Wakoshi,
Saitama, 351-0198, Japan} \affiliation{Physics Department, The
University of Michigan, Ann Arbor, Michigan 48109-1040, USA}
\author{Vincenzo Savona}
\affiliation{Center for Quantum Science and Engineering, \\ \'{E}cole Polytechnique F\'{e}d\'{e}rale de Lausanne (EPFL), CH-1015 Lausanne, Switzerland}
\affiliation{Laboratory of Theoretical Physics of Nanosystems (LTPN), Institute of Physics, \'{E}cole Polytechnique F\'{e}d\'{e}rale de Lausanne (EPFL), 1015 Lausanne, Switzerland}
\author{Pasquale Scarlino}
\email[E-mail: ]{pasquale.scarlino@epfl.ch}
\affiliation{Hybrid Quantum Circuits Laboratory (HQC), Institute of Physics, \'{E}cole Polytechnique F\'{e}d\'{e}rale de Lausanne (EPFL), 1015 Lausanne, Switzerland}
\affiliation{Center for Quantum Science and Engineering, \\ \'{E}cole Polytechnique F\'{e}d\'{e}rale de Lausanne (EPFL), CH-1015 Lausanne, Switzerland}

\begin{abstract}
Landau-Zener-Stückelberg-Majorana (LZSM) interference occurs when qubit parameters are periodically modulated across avoided level crossings. 
We explore this phenomenon in nonlinear multilevel bosonic systems, where interference is influenced by multiple energy levels. 
We fabricate two superconducting resonators with flux-tunable Josephson junction arrays. 
The first device, exhibiting weak nonlinearity, behaves like a linear resonator under weak driving but shows LZSM interference akin to two-level systems. 
With stronger driving, nonlinear effects alter the interference pattern. 
We theoretically demonstrate that merging LZSM peaks can lead to dissipative quantum chaos. 
In the second device, where nonlinearity exceeds photon-loss rates, we observe additional LZSM peaks from Kerr multiphoton resonances. 
Under Floquet theory, these resonances represent synthetic modes of coupled nonlinear cavities, revealing effective coupling as modulation parameters vary. 
Our findings advance the understanding of LZSM physics and emphasize the control of nonlinear Floquet states and the emergence of chaos in engineered systems, with significant implications for novel applications in quantum dynamics and quantum control.
\end{abstract}

\date{\today}

\maketitle

\setreviewsoff

\begin{figure}
    \includegraphics[width=.5\textwidth]{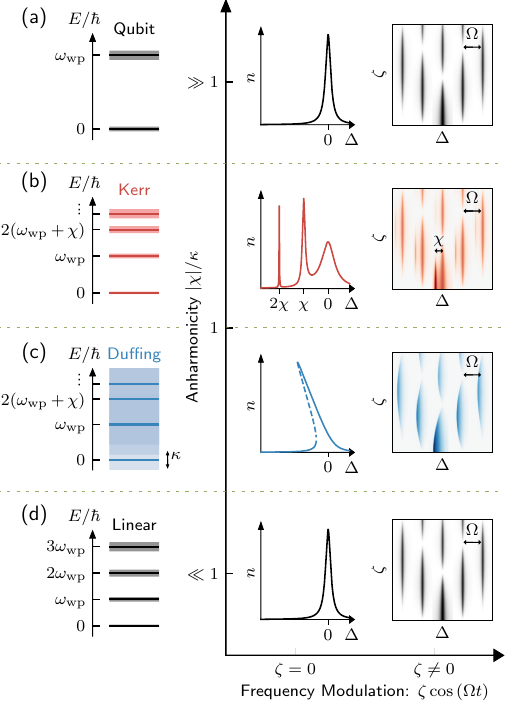} 
    \caption{\label{fig:concept}
    Landau-Zener-St\"uckelberg-Majorana (LZSM) interference mechanisms in bosonic systems.
    For cases studied in this article, we show the level structure of the undriven system (left); the photon number $n$ of the driven, but not-modulated, system as a function of the pump-to-cavity detuning $\Delta$ (center); and the LZSM pattern emerging when the cavity eigenfrequency is periodically modulated with strength $\zeta$ and frequency $\Omega$ (right).    
    (a) In the qubit regime of infinite nonlinearity, the system consists only of the ground and excited states. 
    This level structure gives rise to a single excitation peak ($\ket{0} \to \ket{1}$) at detuning $\Delta = 0$. Thus, the standard LZSM interference pattern emerges. 
    (b) In the Kerr regime, where the anharmonicity is larger than the loss ($|\chi|>\kappa$), the system consists of many uneven-spaced states with different numbers of excitations. 
    When $\Delta =  n \chi$, multiphoton transitions $\ket{0} \to \ket{n}$ occur for large-enough drive.
    This multi-photonic transition structure is periodically repeated around each standard LZSM peak.
    (c) In the Duffing regime, where the anharmonicity is smaller than the loss ($|\chi|<\kappa$), the uneven-spaced states are broadened by dissipation and cannot be distinguished. 
    The drive excites multiple levels, resulting in a deviation from a Lorentzian shape, and the Kerr nonlinearity competes with detuning, giving rise to bistability.
    Such a deviation and the presence of bistability are imprinted in each LZSM peak.
    (d)
    In the linear regime ($\chi=0$), all levels are equispaced. 
    When driven, only a Lorentzian peak appears at $\Delta =0$, similar to the qubit regime.
    Upon modulation of the resonator frequency, the LZSM interference is also indistinguishable from that in the qubit regime.
    }
\end{figure}

Qubits -- two-level systems -- are the building blocks of digital quantum computers and simulators, as well as an essential paradigm for describing many quantum systems~\cite{BlaisRMP21,AltmanPRXQ21}.
Understanding and controlling their dynamics is thus pivotal to the progress of quantum technologies.
When the qubit's energy-level splitting is varied in such a way that the two levels become almost degenerate, the Landau-Zener-St\"uckelberg-Majorana (LZSM)~\cite{landau1932theorie,zener1932non,stuckelberg1932theorie,majorana1932atomi} transition probability dictates the likelihood of non-adiabatic transitions between the ground and excited states. 
When the variation of the splitting is periodic in time, a rich \textit{LZSM interference pattern} arises, as schematically shown in \figref{fig:concept}~(a) (see Ref.~\cite{IvakhnenkoPHYSREP23} for a recent overview of the field). 
At each oscillation period, the transition paths can interfere constructively or destructively to determine the final probability of the qubit to reach the excited state, as observed in, e.g., superconducting qubit architectures~\cite{OliverSCIENCE05, sillanpaaContinuousTime2006}, semiconductor quantum dots~\cite{stehlikLandauZenerSt2012, forsterCharacterization2014}, and nitrogen-vacancy center in diamond~\cite{childressMultifrequency2010}.

Historically, the understanding of LZSM transitions was a foundational step in the development of non-relativistic quantum mechanics~\cite{landau1932theorie,zener1932non,stuckelberg1932theorie,majorana1932atomi}.
Recently, LZSM interference gained also considerable attention, as a versatile tool for the study of quantum systems. 
Examples include the characterization of the frequency noise of superconducting resonators~\cite{niepce_stability_2021} and the decoherence properties of charge states from steady-state measurements~\cite{dupont-ferrierCoherent2013, forsterCharacterization2014, heQuantifying2024}.
LZSM interferometry was also employed for the fast coherent control of charge~\cite{caoUltrafast2013, chatterjeesiliconbased2018} \add{and spin~\cite{boganLandauZenerSt2018, khomitskySinglespin2022, khomitskyControllable2023} qubits}, and to mediate and control the coupling \add{of two flux qubits~\cite{munyaevControl2021} or} of a single qubit to multiple mechanical modes~\cite{kervinenLandauZenerSt2019}.
Finally, LZSM interferometry has also been proposed as a tool for efficient quantum parameter estimation~\cite{yangQuantum2017} and for the preparation of exotic quantum states, such as two-level systems with tunable absorption properties~\cite{wenLandauZenerStuckelbergMajorana2020}, correlated photons~\cite{changCircuit2022} and Schr\"odinger-cat states~\cite{lidalGeneration2020, wangSchrodingerCat2021}.
The physics of LZSM interference beyond the two-level approximation has been marginally investigated, and often focuses on isolated avoided level crossing within a larger multilevel structure~\cite{IvakhnenkoPHYSREP23}.
Furthermore, coupled classical oscillators have been proposed~\cite{ivakhnenkoSimulating2018} and studied~\cite{zhouDynamic2019, lorenzClassical2023} as classical systems displaying LZSM interference.
Indeed, in the presence of tailored modulation, these \emph{multimode classical systems} display the same equation of motion of a qubit~\cite{bernazzaniFluctuating2024}, and thus exhibit LZSM interference due to the presence of an effective two-level system.

In bosonic systems, the qubit limit can be reached by the introduction of a Kerr nonlinearity (anharmonicity) $\chi$, permitting, in principle, to address only the ground and first excited states~\cite{BlaisRMP21}.
This description applies to several platforms, including superconducting circuits~\cite{CarusottoNATPHYS20}, polaritonic microcavities~\cite{CarusottoRMP13}, mechanical resonators~\cite{HuberPRX20}, and the vibration of trapped ions~\cite{DingPRL17}. 
A realistic description of these systems must include the effects of dissipative processes, which blur the distinction between energy levels and thus hinder the possibility of addressing them singularly.
Depending on the magnitude of the total loss rate $\kappa$, one can thus determine two distinct regimes that we dub the Kerr ($|\chi| > \kappa$) and Duffing ($|\chi| < \kappa$) regimes~\cite{andersenQuantum2020, YamajiPRA22}.
In the Kerr regime, depicted in \figref{fig:concept}~(b), the energy quantization of the bosonic mode is still accessible despite the presence of dissipation~\cite{WinkelPRX20}. 
The system can absorb $n$ photons from a drive and transition to the $n$th excited level, in a process known as \textit{Kerr multiphoton resonance} (or multiphoton transition) \footnote{Note that here multiphoton resonance refers to the fact that absorbing $n$ photons leads to the $n$th excited state of the resonator. This is not the multiphoton Rabi resonance, where $n$ driving photons are absorbed to populate the excited level of the qubit.}.
In the Duffing regime [\figref{fig:concept}(c)], instead, dissipation blurs these multiphoton resonances, giving rise to a single spectral feature, where the energy quantization of the underlying bosonic mode can't be resolved.
The effect of the nonlinearity, in this case, is to shift this resonance, leading to phenomena such as bistability and hysteresis~\cite{dykmanFluctuating2012, ChenNATCOM23, BeaulieuARXIV23}.
Multimodal Duffing oscillators, where multiple bistabilities are present, display emergent phenomena, such as the formation of domain walls and dissipative phase transitions~\cite{foss-feigEmergent2017, VicentiniPRA18,LiPRL22}, as well as dissipative quantum chaos~\cite{FerrariPRR25}.
The latter is triggered by the combined presence of classical and quantum fluctuations and the competition of unitary dynamics and dissipative processes~\cite{DahanNPJ22,CohenPRXQ23}.

More in general, the study of periodically modulated systems through all regimes of nonlinearity is a topic attracting growing attention in the community of superconducting circuits~\cite{nguyenProgrammable2024}.
These studies sit within the broader context of Floquet physics, which has proven crucial in describing periodically-modulated quantum systems, finding diverse applications such as quantum control~\cite{gandonEngineering2022, nguyenProgrammable2024} and band topology engineering~\cite{rudnerAnomalous2013, maczewskyObservation2017, okaFloquet2019, weitenbergTailoring2021}. 
This approach enables the construction of synthetic lattices~\cite{ozawaTopological2019, meierExploring2019, arnalChaosassisted2020}, allowing, for example, the implementation of controllable LZSM transitions between Floquet states~\cite{ikedaFloquetLandauZener2022, wangPhotonic2023}. 
Recent research focused on expanding Floquet theory to encompass open quantum systems~\cite{satoFloquet2020, moriFloquet2023}, and investigating its effects in nonlinear systems~\cite{shanGiant2021, mukherjeeObservation2020, luFloquet2021, goldmanFloquetEngineered2023a}. 

In this article, we extend the paradigm of LZSM physics, through the study of two nonlinear superconducting resonators, one in the Kerr and one in the Duffing regime, investigating strongly driven and dissipative nonlinear Floquet systems.
Given the high degree of tunability of the drive, the modulation, and the other system parameters,
we determine the whole LZSM interference diagram for nonlinear bosonic systems.
We present a simple unified model that captures the relevant features of the system under consideration.

The main results of this work can be summarized as follows.

First, we experimentally demonstrate and theoretically clarify that, at low driving amplitude, the LZSM interference pattern is independent of the nonlinearity of the system [c.f. the rightmost panels of Figs.~\ref{fig:concept}(a) and (d)].
Namely, there is no distinction in the LZSM interference pattern between a completely linear resonator and a qubit.

Second, we show novel effects due to the competition between the modulation and the nonlinearity at larger pumping power, demonstrating the role of dissipation [Figs.~\ref{fig:concept} (b) and (c)]. 
In particular (i) In the Kerr regime, we observe how Kerr multiphoton resonances add structure to the LZSM interference. 
These resonances and the associated quasi-energy (Floquet) states can be interpreted as the modes of a multimode synthetic cavity array, with effective interference between these multiple transitions resulting in avoided level crossings.
(ii) In the Duffing regime, we show how bistability and hysteresis come into play in determining the state of the system, suggesting the emergence of dissipative quantum chaos in a Floquet regime, i.e., Floquet-dissipative quantum chaos.

\begin{figure*}
    \includegraphics[width=1\textwidth]{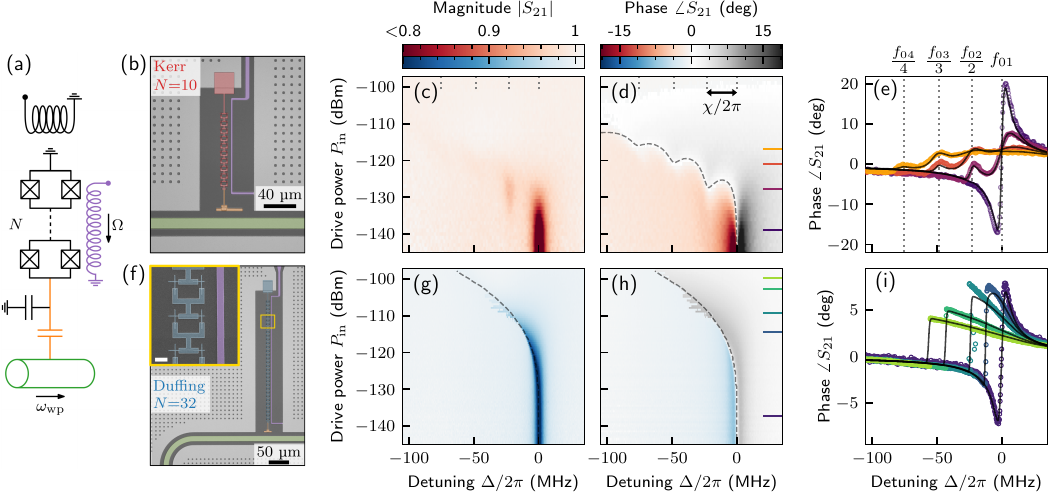} 
    \caption{\label{system}
    SQUID array characterization.
    (a) Equivalent lumped electrical circuit of the device composed of $N$ SQUIDs in series. 
    The static flux is controlled via an external coil and a microwave signal can be applied to the flux line (purple) for fast frequency modulation. 
    The resonator is coupled to a feedline in a notch configuration.
    (b, f) Optical micrograph of the two SQUID array resonators studied in this work with $N=10$ (red) and $N=32$ (blue) SQUIDs, with details on the Josephson junction shown in the inset where the white scale bare represents $\SI{5}{\um}$.
    (c-d) For the Kerr case ($N=10$), the magnitude and phase of the transmission coefficient through the feedline as a function of drive power at the sample. 
    The black dashed curve indicates where the phase of $\SParam$ is zero according to numerical simulations, highlighting the position of the multiphoton resonances. 
    (e) Selected traces of the data reported in panel (d).
    Experimental data are shown with circle markers whose colors correspond to the ticks in panel (d).
    Numerical fits to a full quantum model are shown in black solid lines.
    The grey dotted vertical lines indicate the position of the multiphoton resonances obtained from numerical simulations. 
    (g-i) Same measurement as in (c-e), but for the Duffing case. 
    The dashed curves in (g,h) indicate the minima of $|S_{21}|$ obtained from a full quantum simulation.
    }
\end{figure*}

Our work establishes a comprehensive framework for understanding LZSM and Floquet physics, clarifying the role of nonlinearity and dissipation in determining the interference patterns. It paves the way to their control, with perspectives for \textit{synthetic dimension engineering} in Floquet configurations. This platform can be used as a quantum simulator to investigate quantum chaos and critical phenomena in highly controllable superconducting systems.

The article begins with a presentation of the experimental system and the model used to describe it. It then explores LZSM interference in the qubit and linear regimes, followed by an analysis of photon-resolved effects in the Kerr regime, where multiphoton resonances influence the emergent LZSM interference. The discussion continues with an examination of the Duffing regime, characterized by weak nonlinearity, highlighting the emergence of Floquet dissipative quantum chaos. The methods section provides details on device fabrication, measurement setup, and parameter extraction. Supplementary material includes theoretical insights and additional experimental data that enhance the understanding of our findings.

\begin{table*}[]
\caption{
Summary of the relevant SQUID array parameters at flux operating point $\Phi_{\rm wp}$. \add{The characterization of the two devices is detailed in the methods section.}
}
\begin{tabular}{|c|c|c|c|}
\hline
 Parameters & $N=10$ device & $N=32$ device & Description  \\
 & (Kerr/Qubit regime) & (Duffing/Linear regime) & \\
 \hline
$|\chi|/\kappa$ & 5 & 0.05 & Photon-number distinguishability \\  \hline  \hline
$\omega_c/2\pi$ &  $\approx 13$ GHz         &    $\approx 6.4$ GHz       & Zero-flux frequency \\ \hline
$\omega_{\rm wp}/2\pi$ & 4.502 GHz & 4.306 GHz & Frequency at $\Phi_{\rm wp}$ \\ \hline
$\Phi_{\rm wp}/\Phi_0$ & 0.45 & 0.32 & Flux operating point \\ \hline \hline 
$\chi/2\pi$ & \SI{-23.5}{\MHz} & $\SI{-0.35}{\MHz}$ & Kerr nonlinearity \\ \hline \hline
$\kappa_\mathrm{in}/2\pi$ & 1.1 MHz & 4.92 MHz & Internal loss rate \\ \hline
$\kappa_\mathrm{ext}/2\pi$ & 3.75 MHz & 1.49 MHz & External loss rate\\ \hline 
$\kappa/2\pi$ & 4.85 MHz & 6.41 MHz & Total loss rate\\ \hline 
$\kappa_\phi/2\pi$ & 0.75 MHz & 0.4 MHz & Dephasing rate \\ \hline

\end{tabular}
\label{tab:param_main}
\end{table*}
    
\section*{Results}
\subsection*{Experimental system and model}

We aim to investigate all regimes of nonlinearity and dissipation, namely, qubit, Kerr, Duffing, and linear, as shown in \figsref{fig:concept}~(a-d), respectively.
To this extent, we design and fabricate two frequency-tunable nonlinear resonators that can operate in these different regimes according to the driving amplitude.
These are superconducting SQUID arrays~\cite{maslukMicrowave2012a, weisslKerr2015, krupkoKerr2018}, galvanically connected to ground on one side, and capacitively shunted to the ground on the other side, as shown in Figs.~\ref{system}~(a),~(b), and (f). 
A detailed summary of their parameters is reported in \tabref{tab:param_main}.

The frequency of the resonators can be tuned by a dedicated flux line that uniformly threads the magnetic fields in each SQUID loop (in purple) and an external superconducting coil.
The two resonators differ in the number of SQUIDs in each array, as highlighted by red and blue false colors, determining the two orders of magnitude difference in their Kerr nonlinearity $\chi$. 
Increasing the number of SQUIDs in an array leads to a decrease in nonlinearity due to the reduced Josephson inductance required to maintain a fixed frequency and the diminished phase fluctuations across each junction~\cite{sivakJosephson2020}.
Hereafter, the \textcolor{red}{red} and \textcolor{blue}{blue} color schemes will always indicate measurements of the devices in the \textcolor{red}{Kerr/qubit} and \textcolor{blue}{Duffing/linear} regimes, respectively.
Each resonator is also capacitively coupled to a feedline (in green) in a notch configuration, resulting in an external coupling $\kappa_{\rm ext}$ close to the internal dissipation rate  $\kappa_{\rm int}$ (critically coupled regime).
We define the total dissipation rate as $\kappa = \kappa_{\rm ext} + \kappa_{\rm int}$.

Each device is thermally and mechanically anchored at the mixing chamber plate of a dilution refrigerator, reaching an average base temperature of $\SI{15}{\milli \kelvin}$.
The devices are probed by a coherent drive with amplitude $F$ at the sample, injected in the feedline through highly attenuated coaxial lines. 
The drive amplitude is related to the input power $P_{\rm in}$ by $F=\sqrt{P_{\rm in} \kappa_{\rm ext} / \hbar\omega_{\rm d}}$, where $\omega_{\rm d}$ is the drive frequency.
Although the frequency of the untuned cavity is $\omega_c$, through the paper, the frequency working point of the resonators, $\omega_{\rm wp}$, is set by a static flux generated by direct current through an external superconducting coil.  
The flux operating point is $\Phi_{\rm wp} = 0.45 \Phi_0$ for the $N = 10$ device, while $\Phi_{\rm wp} = 0.32 \Phi_0$ for the $N=32$ one, where $\Phi_0=h/2e$ is the magnetic flux quantum.
Finally, driving the fluxline at a frequency $\Omega$ periodically modulates the frequency of the resonator, approximately between $\omega_{\rm wp}\pm \zeta$, with $\zeta$ representing the strength of the modulation.
The single-tone spectroscopy of the resonators at low-driving power as a function of the external magnetic flux is reported in the Supplementary Information. 

As explained in the Supplementary Information, both devices can be modeled as a bosonic mode with the following time-dependent Hamiltonian:
\begin{equation}\label{Eq:Hamiltonian}
\HH / \hbar = - \Delta \hat{a}^\dagger \hat{a} + \chi \hat{a}^\dagger \hat{a}^\dagger \hat{a} \hat{a} + F \, (\hat{a} +\hat{a}^\dagger ) + \zeta \cos(\Omega \, t) \hat{a}^\dagger \hat{a} \, ,
\end{equation}
where $\Delta =  \omega_{\rm d} - \omega_{\rm wp}$ is the detuning between the working point of the devices ($\omega_{\rm wp}$) and the drive frequency ($\omega_{\rm d}$).
Beyond the total dissipation rate $\kappa$, the system is subject to dephasing with rate $\kappa_{\phi}$. We include them in the time evolution of the density matrix $\rhot$ using the Lindblad master equation
\begin{equation}\label{Eq:Liouvillian}
 \hbar \, \partial_t \rhot =- i \left[\HH, \rhot \right] + \kappa \DD[\hat{a}] \rhot + \kappa_{\phi} \DD[\hat{a}^\dagger \hat{a}] \rhot \,.
\end{equation}
Here, $\DD[\hat{L}] \rhot \equiv \hat{L} \rhot \hat{L}^\dagger - \{\hat{L}^\dagger \hat{L}, \rhot \}/2$ is the Lindblad dissipator~\cite{LidarARXIV19}.

In \figref{system} we characterize the coherent response of the resonators in the absence of modulation (i.e., $\zeta=0$) and use it to determine the parameters of the two devices. 
In \figsref{system} (c) and (d) we report the magnitude and phase of the transmission coefficient $\SParam$ (see Methods) as a function of the driving power $P_{\rm in} \propto F^2$ in the Kerr regime. 
At low power, only a single dip around $\Delta=0$ is visible, representing the transition to the first excited state (noted as $\ket{0} \to \ket{1}$, where $\ket{n}$ is the photon number state of the resonator).
At larger values of the drive, several dips appear, representing the so-called Kerr multiphoton transitions between the ground state and the higher-excited levels ($\ket{0} \to \ket{n}$), highlighted in the single traces shown in \figref{system} (e).
According to \eqref{Eq:Hamiltonian}, all the dips should appear at $\Delta \simeq (n-1) \chi$.
Small deviations from this prediction are due to higher-order nonlinearities
\footnote{
The resonator exhibits deviation from the pure Kerr nonlinearity prediction due to non-negligible effects of higher nonlinearities, e.g., terms of the form $\chi^{(5)} (\hat{a}^\dagger)^3 \hat{a}^3$. We estimate $\chi^{(5)}/2\pi \approx -1.1\,$MHz$\simeq 5\% \chi/2\pi$.
As such, although the $\chi^{(5)}$ term produces small changes compared to the model in \eqref{Eq:Hamiltonian}, all the relevant physical features of the LZSM interference are captured by the Kerr model.
We also note additional nonlinear effects due to the large values of flux modulation used to obtain the wanted $\zeta$.}.
For even larger powers, the nonlinearity suppresses the intracavity photon number with respect to the input power, resulting in an almost unitary transmission $\SParam$.

\begin{figure*}[ht!]
    \includegraphics[width=1.0\textwidth]{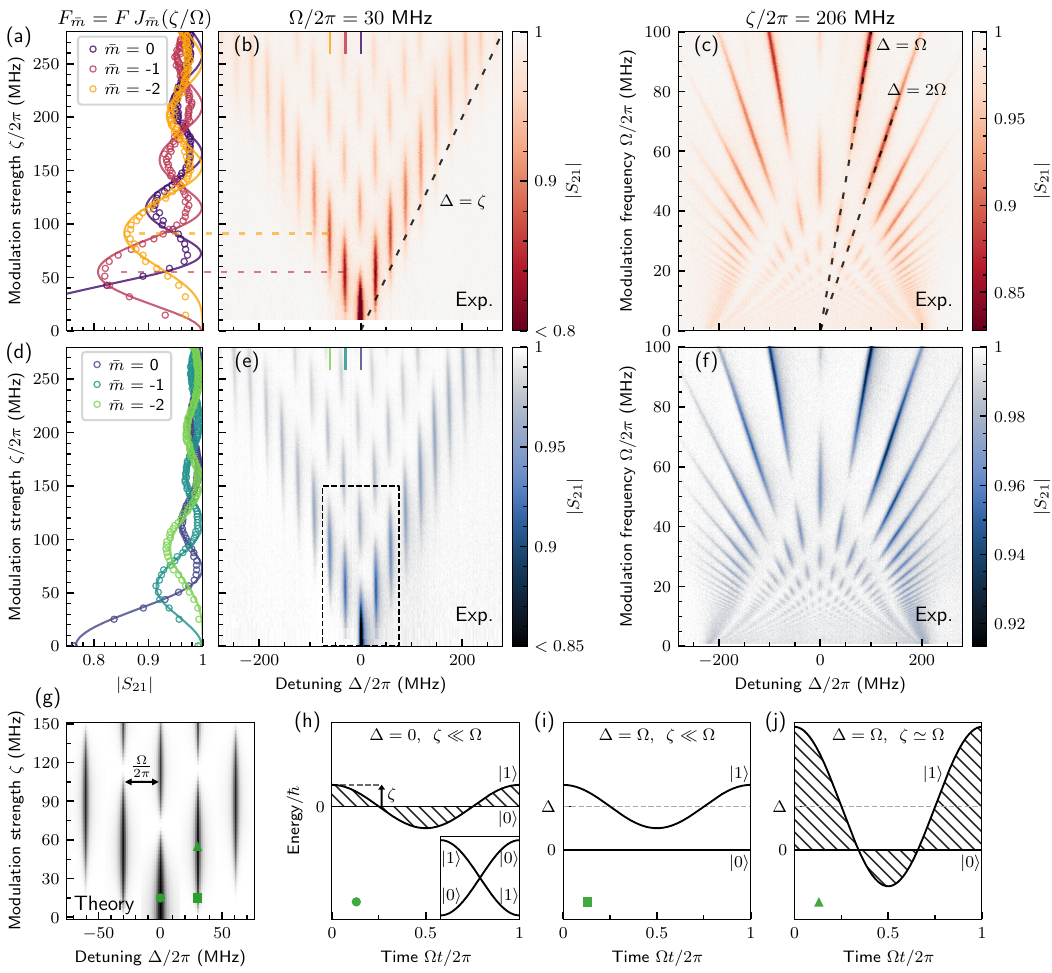}
    \caption{\label{Fig:LZSM_qubit_linear}
    LZSM interference patterns.
    (a-c, red) Analysis of the $N=10$ device, with input power $P_{\rm in}=-138.8$ dBm, ensuring that we are in the \textit{qubit regime}.
    (b) The transmission coefficient $|\SParam|$  as a function of the detuning $\Delta$ and the modulation strength $\zeta$, for fixed modulation frequency $\Omega/2\pi = 30$ MHz [see \eqref{Eq:Hamiltonian}]. 
    (a) Comparison of the experimental and theoretical data for $|\SParam|$.
    Solid lines represent the results of the numerical simulations of the full quantum model obtained at $\Delta = 0$, $\Delta = - \Omega$, and $\Delta = - 2 \Omega$ (see Supplementary Information). 
    The circles are the experimental data obtained from panel (b), in which $\Delta$ is slightly re-scaled to account for the nonlinear flux-dependency of the resonator frequency (see Supplementary Information). 
    (c) $|\SParam|$ as a function of $\Delta$ and $\Omega$.
    (d-f, blue) As in (a-c), but for the $N=32$ device, with $P_{\rm in}=-133.3$ dBm to ensure that the system is in the \textit{linear regime}.
    From these plots, the two regimes appear almost indistinguishable.
    (g) The photon number versus $\Delta$ and $\zeta$ is obtained from a simulation using the effective model of~\eqref{Eq:effective_hamiltonian} that reproduces the interference pattern in panels (b) and (e).
    (h-j) Depiction of the time evolution of the energy level $\ket{1}$, in the frame rotating at the drive frequency $\omega_{\rm d}$, if $F=0$.
    A finite drive $F$ opens gaps at each crossing between $\ket{0}$ and $\ket{1}$, allowing a non-adiabatic passage between the two.
    The parameters $\Delta$ and $\zeta$ are indicated by green markers in (g). 
    (h) At $\Delta=0$, the level $\ket{1}$ becomes resonant with $\ket{0}$ (they form a level crossing, see the inset). The values of $\zeta$, $F$, and $\kappa$ then determine the probability of transitioning out of the vacuum.
    (i) For non-zero detuning (e.g. $|\Delta| = \Omega$) and small modulation ($\zeta\ll \Omega$), the level $\ket{1}$ is never resonant with $\ket{0}$ and it cannot be populated.
    (j) For strong enough modulation $\zeta>|\Delta|$, the level $\ket{1}$ can form again an avoided level crossing, and constructive interference is possible again.
   }
\end{figure*}

We report the same measurements for the resonator in the Duffing regime in \figsref{system} (g-i). 
In this case, dissipation smears the multiphoton resonances, resulting in an indistinguishable level structure. 
Increasing the drive, the single dip of $\SParam$ originally at $\Delta =0$ moves to negative detunings, indicating that the drive is exciting higher levels.
Scanning the detuning from negative to positive values, as done in \figref{system}~(i), reveals the presence of a sharp jump, where the resonator passes from a highly- to a lowly-populated phase.
This behavior is associated with optical bistability, i.e., the presence of two metastable states that require a long time to decay to the steady state~\cite{foss-feigEmergent2017, VicentiniPRA18}. 
This phenomenon gives rise to hysteresis~\cite{ChenNATCOM23,BeaulieuARXIV23} and makes it difficult to properly resolve the exact detuning where the transition occurs.

\subsection*{Linear and qubit LZSM interference}

We can investigate the linear and qubit regimes using the $N=10$ and $N=32$ resonators described above.
Indeed, for the $N=10$ resonator, the second-excited level is not significantly populated if $F^2 \ll |\chi|\kappa$ \footnote{
One shows that, assuming at most two-photon in the system, the maximum of the two-photon population occurs at multiphoton resonance $\Delta =  \chi$, where
$$
\brakket{2}{\sss}{2} =\frac{2 F^4}{9 F^4+2 \kappa ^2 \left[2 \left(\kappa ^2+\chi
   ^2\right)-5 F^2\right]} \simeq \frac{ F^4}{2 \kappa ^2
   \chi ^2}.
$$
It follows that $F^2 \ll |\chi|\kappa$ ensures the validity of the qubit approximation
}.
For the $N=10$ device parameters and the drive $F/2\pi \simeq \SI{1.6}{\MHz}$ considered here, the third level is predicted to be populated less than $0.03\%$.
For the values used in this first part of the experiment, the system effectively behaves like an ideal qubit subject to dissipation and dephasing.
We report the experimental data in \figsref{Fig:LZSM_qubit_linear} (b) and (c).
In \figref{Fig:LZSM_qubit_linear} (b) we show the norm of the scattering coefficient $\SParam$ sweeping the detuning $\Delta$, for a fixed modulation frequency $\Omega$, and varying the modulation strength $\zeta$.
One observes the LZSM pattern emerging, with populated regions at $\Delta = m \Omega$, for integer $m$.
Fixing $\zeta$ and scanning $\Omega$, in \figref{Fig:LZSM_qubit_linear}~(c) we observe again the interference pattern at $\Delta = m \Omega$.
We thus confirm the presence of LZSM interference and the control over the modulation of the resonator in the Kerr regime.

We now consider the $N=32$ device.
In this case, the oscillator approximately behaves as a purely linear resonator if $F < \sqrt{\kappa^3 / |\chi|}$ \footnote{
To justify this approximation, consider the semiclassical (coherent state approximation) $\sss = \ketbra{\alpha}{\alpha}$. One finds that the photon number $n=|\alpha|^2$ satisfies
\begin{equation*}
\begin{split}
&\left[ \frac{\kappa ^2}{4}  +(\Delta +n \chi)^2 \right] n  -F^2 
\\
&\quad \simeq 2 \Delta  n^2 \chi +n \left(\Delta ^2+\frac{\kappa
   ^2}{4}\right)-F^2=0.
\end{split}
\end{equation*}
The solution to this equation can be expanded in powers of $\chi$ as
$$
n = n_0 \left(1 - n_0 \frac{8 \Delta \chi}{4 \Delta^2 + \kappa^2} \right), \, \text{ with } \, n_0 = \frac{F^2}{\Delta^2 + \kappa^2/4}.
$$
The deviation from the linear regime $\delta n = 1- n / n_0 $ is then maximal for $\Delta = \kappa / (2\sqrt{3})$, and leading to 
$$
\delta n = \frac{3 \sqrt{3} F^2 \chi }{\kappa ^3}.
$$
As we are interested in $\delta n \ll 1$, we find back the formula in the main text.
}.
For the $N=32$ device parameters and the drive $F/2\pi \simeq \SI{3}{\MHz}$ considered here, we estimate a relative photon-nu
mber deviation from a completely linear resonator of less than $3\%$.
Within this regime, we repeat the previous measurements and report them in \figsref{Fig:LZSM_qubit_linear} (e) and (f).
Surprisingly, we observe the same interference pattern emerging, with \textit{no distinguishable differences between the qubit and the completely linear case}.
This feature indicates that only the energy difference between $\ket{0}$ and $\ket{1}$ determines the interference pattern in both the qubit and the linear regimes  [c.f. \figsref{Fig:LZSM_qubit_linear} (g-j)].
This similarity may be expected from linear response theory; indeed, weakly driven nonlinear oscillators should behave similarly even if they have widely different anharmonicities.
The presence of LZSM interference seems even more general, as it should be observable even in a purely linear cavity and for arbitrarily large number of photons. 
We remark that the frequency modulation of almost linear oscillators has been used to perform mode-conversion~\cite{leePropagation2020}, parametric amplification~\cite{lecocqNonreciprocal2017} and squeezing~\cite{zagoskinControlled2008}. 
However, to the best of our knowledge, LZSM interferences were not studied in linear oscillators, and this discussion is missing from recent reviews on the topic~\cite{silveriQuantum2017, IvakhnenkoPHYSREP23}.

To provide a more quantitative reasoning, we choose $\bar{m}$ minimizing $\Delta - \bar{m}{\Omega}$ and, following the procedure derived in the Supplementary Information, and passing in the frame rotating at the frequency $\bar{m}{\Omega}$, we have
\begin{equation}
\label{Eq:effective_hamiltonian}
\begin{split}
\HH_{\bar m}/\hbar \simeq -\Delta_{\bar{m}} \hat{a}^\dagger \hat{a} + \chi \hat{a}^\dagger \hat{a}^\dagger \hat{a} \hat{a} +
F_{\bar{m}} \left(\hat{a} + \hat{a}^\dagger \right) \, ,
\end{split}
\end{equation}
where the renormalized detuning $\Delta_{\bar{m}}$ and renormalized drive $F_{\bar{m}}$ are
\begin{equation}
\Delta_{\bar{m}} = (\Delta -  \bar{m} {\Omega} )\, , \qquad F_{\bar{m}} = F  \Bessel{\bar{m}}{\frac{\zeta}{\Omega}},
\end{equation}
with $\Bessel{\bar{m}}{\zeta/\Omega}$ indicating the Bessel function of the first kind.
All dissipative terms maintain their form as in \eqref{Eq:Liouvillian}.
In other words, when we can single out a single relevant frequency $\Delta_{\bar{m}} $ for each of the LZSM interference dips, the devices behave as a collection of independent nonlinear resonators, whose driving amplitudes $F_{\bar{m}}$ are modulated via Bessel functions. 
For the parameters we consider here, and if we also assume a weak enough drive to be in the linear and qubit regime \cite{Note3, Note4}, we obtain 
\begin{equation}
    \expec{\hat{a}^\dagger \hat{a}} \simeq \frac{4 F_{\bar{m}}^2}{\kappa}\frac{\kappa +\beta \kappa_{\phi} }{4 \Delta_{\bar{m}} ^2+(\kappa +\beta \kappa_{\phi}
   )^2} \, ,
\end{equation}
with $\beta=1$ for a linear resonator regime and $\beta=4$ in the weakly driven qubit limit.
Namely, the \textit{two regimes have identical interference patterns}, only slightly modulated by the dephasing rate $\kappa_{\phi}$.
To further demonstrate the validity of these results, additional LZSM interference patterns are reported in the Supplementary Information, highlighting the precise control of the number and frequency spacing of modes over a broad range of modulation strengths and frequencies. 

The approximation of the effective model correctly captures the value of the photon number, but not that of the field $\hat{a}$ (and thus cannot be used to quantitatively study $S_{21}$). 
As is discussed in the Supplementary Information, to correctly capture this feature, one has to resort to a full quantum simulation of the Floquet model. 
This is shown in \figsref{Fig:LZSM_qubit_linear}~(a) and (d), where we plot $|S_{21}|$ of the first three LZSM lobes, comparing the experimental data with the theoretical predictions both for the qubit and linear regimes.
In both cases, we find an \textit{excellent agreement between theory and experiments}. 
We note that the maxima and the minima of $|S_{21}|$ of the $\bar{m}$th LZSM mode coincides with the extremes of the associated Bessel function $J_{\bar m}$, showing the qualitative validity of \eqref{Eq:effective_hamiltonian} in describing also $S_{21}$.

We remark here that the Hamiltonian in \eqref{Eq:effective_hamiltonian} could be obtained approximating the response of an array of nonlinear resonators, each at a frequency $\Delta_{\bar m}$. Therefore, we can interpret each of the LZSM dips as the response of a different \textit{Floquet synthetic mode}.
As we show below, by increasing the drive, these initially non-interacting modes will begin to interact.

\subsection*{LZSM beyond the qubit approximation: Kerr regime}

\begin{figure*}[]
    \includegraphics[width=1.0\textwidth]{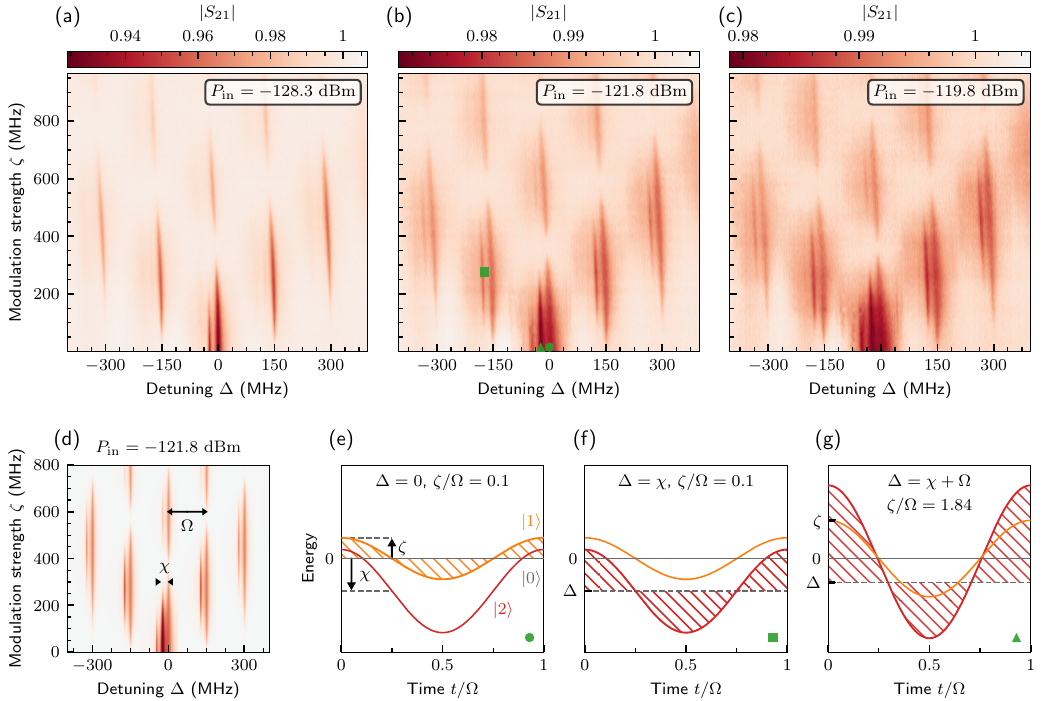} 
    \caption{\label{fig:multiphotonic_Kerr}
    LZSM interferometry for the $N=10$ device, in the Kerr regime and strongly-modulated case $\Omega\gg|\chi|$.
    (a-c) The magnitude of $\SParam$ is measured versus $\Delta$ and $\zeta$ for fixed modulation frequency $\Omega/2\pi=150$ MHz.
    As the drive power $P_{\rm in}$ is increased, Kerr multiphoton resonances from $\ket 0$ to $\ket n$ appear detuned by $(n-1)\chi$ on the left of bare LZSM resonances.
    For large $\zeta$, notice the shift of the pattern to negative detuning, due to the nonlinear dependence of the SQUID array frequency on the flux, as explained in the Supplementary Information. 
    (d) Photon-number simulation using the effective model of \eqref{Eq:effective_hamiltonian} for the same parameters as in panel (b), recovering the same interference pattern. 
    (e-g) In the drive frame, energy versus time for different values of $\zeta$ and $\Delta$ including the first three levels of an undriven Kerr resonator ($F=0$). 
    Green markers indicate the corresponding value of $\Delta$ and $\zeta$ in (d).
    (e) For $\Delta=0$, although multiple levels cross with $\ket 0$, only the level $\ket 1$ form a constructive interference. 
    (f) For $\Delta=\chi$, the second level $\ket 2$ crosses $\ket 0$, and an appropriate choice of parameters leads to constructive interference.  
    (g) For $\Delta=\chi+\Omega$, similar LZSM interference can be constructive again and the level $\ket 2$ can be populated.
    We verified that both the data and full numerical simulations recover that the interference patterns are fully constructive at $\Delta = \Omega$ and $\zeta\approx 1.84 \Omega$, where the Bessel function $J_1(\zeta/\Omega)$ is at a maximum,
    confirming the prediction of the effective model in \eqref{Eq:effective_hamiltonian}.
    }
\end{figure*}

We now focus on those phenomena emerging due to the simultaneous presence of the multilevel structure of nonlinear resonators and the modulation of their eigenenergies, studying the devices beyond their qubit and linear regimes.

In the Kerr regime and for strong enough drives to probe the multiphoton transitions~\cite{Note3}, we investigate how the frequency and amplitude of the modulation modifies the multiphoton resonances.

The system's behavior around the multiphoton resonance $\ket{0} \to \ket{n}$ occurring for $\Delta \simeq \chi (n - 1) $ can be described by a $2 \times 2$ matrix. For instance, the $\ket{0} \to \ket{2}$ multiphoton transition can be described as
\begin{equation}\label{Eq:multiphoton_Hamiltonian}
\begin{split}
    \hat{H}^{(2)}/\hbar = &  2 [- \Delta + \chi  + \zeta \cos(\Omega \, t) ]  \ketbra{2}{2}  \\ &+ G^{(2)} (\ketbra{0}{2} + {\rm h.c.}) ,
\end{split}
\end{equation}
where $G^{(2)}$ represents the effective drive between the vacuum and the state $\ket{2}$. 
For $\zeta=0$, one has $G^{(2)} = F^2 /\chi$ for $\Delta = \chi$.
This formula can be generalized to obtain $G^{(n)}$ for an arbitrary $\ket{0} \to \ket{n}$ transitions \cite{leboiteTHESIS15}.
The dissipation maintains its form, instead.
\add{As we show below, the fundamental parameter to describe these phenomena is the ratio between the modulation frequency $\Omega$, determining the position of the LZSM sidebands, and the nonlinearity $\chi$, determining the position of the multiphoton resonance.}

\hspace{6pt}

\emph{Strong modulation case} \textbf{--} We first choose $\Omega \gg |\chi|$ (strongly modulated case).
In \figsref{fig:multiphotonic_Kerr}~(a-c) we report the scattering coefficient $|\SParam|$ as a function of the detuning $\Delta$ and the strength $\zeta$ of the modulation.
As the drive amplitude $F$ is increased, several additional dips appear, signaling the transitions between the photon number states $\ket{0}$ and $ \ket{n}$ of the resonator.
These dips occur at a frequency lower than each main LZSM dip associated with the transition $\ket{0} \to \ket{1}$. 
Each new additional dip is detuned by the same frequency as the unmodulated multiphoton resonances shown in \figsref{system}~(c-e). 
Within a first approximation, this effect is due to the interplay between the modulation in \eqref{Eq:effective_hamiltonian} and the nonlinearity of the system, as shown in \figref{fig:multiphotonic_Kerr}~ (d) reporting the result of a numerical simulation.

To explain this behavior, we can assume that, around each of the LZSM dips, we again have a drive of the same form as \eqref{Eq:effective_hamiltonian}.
When we then match the condition for a multiphoton resonance, it is this effective drive that leads to the excitation of the state $\ket{2}$.
One then obtains
\begin{equation}\label{Eq:effective_hamiltonian_multiphotonic}
\begin{split}  \hat{H}_{\bar{m}}^{(2)}/\hbar = & 2 [- \Delta_{\bar{m}}^{(2)} + \chi  ] \ketbra{2}{2}  + G^{(2)}_{\bar{m}} (\ketbra{0}{2} + {\rm h.c.}),
\end{split}
\end{equation}
where~\cite{leboiteTHESIS15}
\begin{equation}\label{Eq:multiphoton_param_strong}
    \Delta_{\bar{m}}^{(2)} = \Delta -  \bar{m} {\Omega}, \quad G^{(2)}_{\bar{m}} \simeq \frac{F_{\bar{m}}^2 }{\chi} = \frac{F^2}{\chi}  \left[\Bessel{\bar{m}}{\frac{\zeta}{\Omega}} \right]^2.
\end{equation}
This formula can be generalized to arbitrary $n$-photon resonances with $\Delta_{\bar{m}}^{(n)} = \Delta -  \bar{m} {\Omega}$ and $G_{\bar{m}}^{(n)} \propto (\Bessel{\bar{m}}{\zeta/\Omega})^n $. 
We conclude that when the rescaled detuning matches the condition for the $n$th multiphoton resonance, and if the rescaled drive $F_{\bar m}$ is strong enough, an additional dip appears.
Therefore, we can treat each of the multiphoton resonances for each LZSM dip as a yet separate phenomenon.

As sketched in \figsref{fig:multiphotonic_Kerr} (e-g), at the multiphoton resonance, i.e., at $\Delta=m \Omega + (n-1)\chi$, the states $\ket{0}$ and $\ket{2}$ can satisfy the conditions for the development of constructive interference.
In other words, around each of the main LZSM dips, and for large enough drive amplitude, several multiphoton resonances emerge with the same characteristics as those shown in \figsref{system}~(c-e).

\begin{figure*}[]
    \includegraphics[width=1\textwidth]{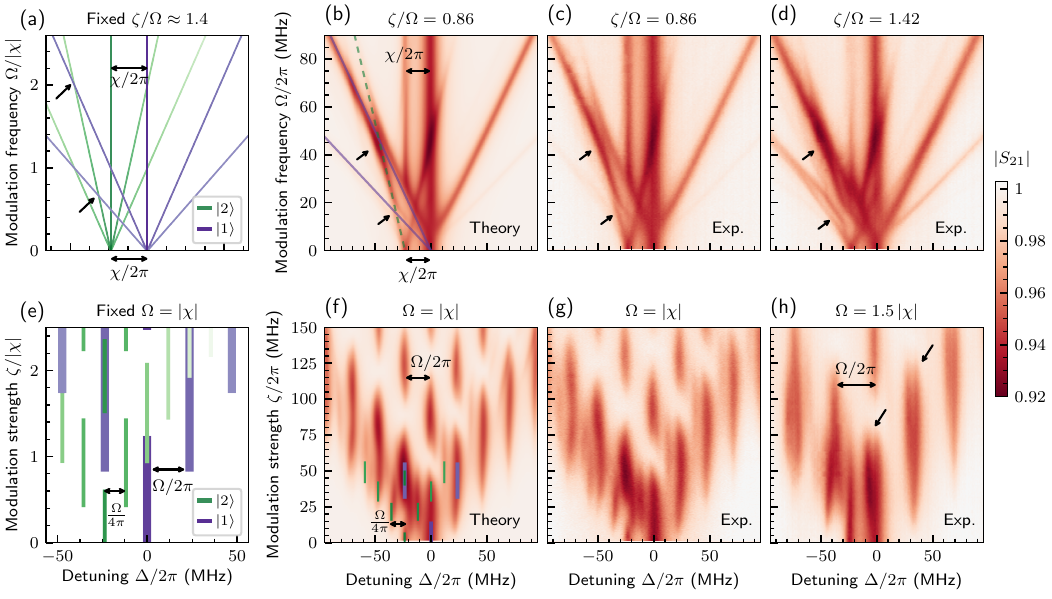} 
    \caption{\label{fig:non_perturbative}
    Controllable Floquet states with the $N=10$ device in the Kerr regime. Through the figure, we set the drive input power to $P_{\rm in}=-128$ dBm.
    (a-d) Both modulation strength $\zeta$ and frequency $\Omega$ are swept together to maintain a constant ratio $\zeta/\Omega$. 
    This choice ensures that the effective drives in Eqs.~(\ref{Eq:effective_hamiltonian}) and (\ref{Eq:effective_hamiltonian_multiphotonic}) are kept constant. 
    This allows enhancing the visibility of the transition  between the strongly- and weakly-modulated cases.
    (a) Sketch of the results of \eqref{Eq:effective_hamiltonian_multiphotonic} for the transition $\ket{0}\to\ket{1}$ (purple, labeled $\ket{1}$) and $\ket{0}\to\ket{2}$ (green, labeled $\ket{2}$).
    For $\ket{1}$, the pattern radiates from $\Delta=0$ with frequency modulation $\Omega$.
    For the multiphoton transition to $\ket{2}$, the LZSM interference pattern is centered at $\Delta=\chi$ and scales with $\Omega/2$.
    (b) Simulation of $|\SParam|$ as a function of $\Delta$ and $\Omega$ with the full quantum model in \eqref{Eq:Liouvillian} (see Supplementary Information for details on the simulation method), having fixed $\zeta/\Omega=0.86$.  
    The corresponding measurement is shown in panel (c) and perfectly overlaps with the results of the numerical simulation.
    The black arrows in (a-d) mark the position of two avoided crossings, where the ``bare levels'' in (a) interact and hybridize in (b-d).
    The crossings are further highlighted by the solid (associated with $\ket 1$) and dashed ($\ket 2$) lines in (b).
    The amplitude of the different avoided crossings can be controlled by modulating the Bessel functions $J_{\bar m}(n\zeta/\Omega)$ as shown in (d), where a larger ratio $\zeta/\Omega$ is chosen.
    (e) As in panel (a), the sketch of the results of \eqref{Eq:effective_hamiltonian_multiphotonic} for $\Omega=|\chi|$ and as a function of $\Delta$ and $\zeta$.
    In this ``bare picture'', the two independent LZSM interference patterns scale with $\Omega$ and $\Omega/2$ for $\ket 1$ and $\ket 2$, respectively.
    (f) Full quantum simulation and (g) corresponding measurement of $|\SParam|$ for $\Omega=|\chi|$.
    The position of some LZSM resonances in the bare picture is superimposed in (f) as a guideline for the eye.
    (h) Repeating the measurement for $\Omega=1.5|\chi|$, we observe line splittings, indicating a modulation of the coupling between different Floquet states.
    }
\end{figure*}

\hspace{6pt}

\emph{Weak modulation case} \textbf{--} When $\Omega \ll |\chi|$ (weakly modulated case), instead, one can capture the system's behavior around the second multiphoton resonance via the Hamiltonian in \eqref{Eq:multiphoton_Hamiltonian}, with $G^{(2)} = F^2/ \chi$ representing the effective drive between the vacuum and the state $\ket{2}$ if $\zeta =0$.
Removing the modulation using the same approximation as in \eqref{Eq:effective_hamiltonian} leads to an equation identical to \eqref{Eq:effective_hamiltonian_multiphotonic}, where now 
\begin{equation}\label{Eq:multiphoton_param_weak}
    \Delta_{\bar{m}}^{(2)} = (\Delta -  \bar{m} {\Omega}/2 )\, , \quad G^{(2)}_{\bar m} = \frac{F^{2}}{\chi} \Bessel{\bar{m}}{\frac{2 \zeta}{\Omega}}.
\end{equation}
This formula can be generalized to arbitrary $n$-photon resonances, with $\Delta_{\bar{m}}^{(n)} = \Delta -  \bar{m} {\Omega}/n$ and $G_{\bar{m}}^{(n)} \propto \Bessel{\bar{m}}{n \zeta/\Omega} $. 
Thus, for detunings close to the $n$th multiphoton transition, a new LZSM interference pattern should emerge, characterized by an effective modulation frequency $\Omega/n$.
It is this scaling that differentiates the weakly and strongly modulated cases, c.f.~\figsref{fig:non_perturbative}~(a,e) and \figref{fig:multiphotonic_Kerr}~(d). 
While previously, for $\Omega \gg |\chi|$, the standard LZSM sidebands were dressed by Kerr multiphoton resonances, we now find that, for $\Omega \ll |\chi|$, each Kerr $n$-th multiphoton resonance is dressed by LZSM sidebands with effective modulation frequency $\Omega/n$.

For the device under consideration, accessing the weakly modulated case would require $\kappa \ll \Omega/n $ to distinguish between the different LZSM dips.
To better resolve this feature, we propose the following driving scheme.
We fix the ratio $\zeta/\Omega$ to have a constant effective drive,  according to both effective theories in Eqs.~(\ref{Eq:effective_hamiltonian}) and (\ref{Eq:effective_hamiltonian_multiphotonic}).
We then increase $\Omega$ and $\zeta$, to cross from the weakly modulated $|\chi|> \Omega$ to the strongly modulated case  $|\chi|<\Omega$.
This is shown in \figsref{fig:non_perturbative}~(b-d) where, for small $\Omega$, we distinctly see the expected LZSM dips associated with the second multiphoton resonance $\ket{0} \to \ket{2}$ and with a slope $\Omega/2$.
\add{We note that such two-photon LZSM transitions were recently reported in a linearly-modulated three-level system~\cite{bjorkmanObservation2024}, with a similar factor two in the LZSM velocity compared to regular single-photon LZSM transitions.}

\hspace{6pt}

\emph{Non-perturbative regime} \textbf{--} The weak- and strong-modulation regimes have very different scaling from each other [c.f. the effective models in Eqs.~(\ref{Eq:multiphoton_param_strong}) and (\ref{Eq:multiphoton_param_weak})]. 
We thus expect that there is a  \textit{non-perturbative passage} from weak- to strong-modulation through some effective interaction, and the transition between these two regimes cannot be explained using any of the two effective theories alone.

Particularly interesting are the values of $\Omega \simeq n |\chi|$, where the system passes from the weak- to the strong-modulated case for a specific state $\ket{n}$.
At these values, it is possible for a $n$-photon resonance to exactly match the LZSM dips of a different $m$-photon resonance. 
We observe the signatures of avoided level crossings between resonances in Figs.~\ref{fig:non_perturbative}(b-c), indicating that the $\ket{0} \to \ket{1}$ and $\ket{0} \to \ket{2}$ resonances interact through the action of an effective emergent coupling.
In this sense, these different resonances constitute a \textit{controllable synthetic Floquet space}, where changing $\Omega$ and $\zeta$ allows selecting an effective interaction between these multiphoton resonances.
This is also evident in Fig.~\ref{fig:non_perturbative} (d), where the ratio $\Omega/\zeta$ is changed, leading both to different interference patterns and different splittings between the Floquet states.

To further highlight an example of these non-perturbative effects, in \figsref{fig:non_perturbative}~(f-g) we fix $\Omega = |\chi|$.
First, we numerically simulate the interplay of these effects in \figref{fig:non_perturbative}~(f). We predict a partial overlap between the second multiphoton transition $\ket{0} \to \ket{2}$ with the first LZSM dip associated with the $\ket{0} \to \ket{1}$ transition at $\Delta = -\Omega$.
For increasing modulation strength $\zeta$, the LZSM structure predicted by \eqref{Eq:effective_hamiltonian_multiphotonic} is observed, although strongly deformed compared to the prediction of the effective model due to the presence of the LZSM lobe associated with the $\ket{0} \to \ket{1}$ resonance.
These theoretical predictions are completely recovered in the data in \figref{fig:non_perturbative}~(g).
Finally, in \figref{fig:non_perturbative}~(h) we fix $\Omega =  1.5 |\chi|$, and we observe a line splitting of several resonances, indicating again the merging and interaction between $\ket{0} \to \ket{2}$ and $\ket{0} \to \ket{1}$ transitions.
\remove{For larger drive amplitudes (not shown), the system shows an extremely rich structure that cannot be simply assigned to any of these original phenomena.}
Note also the asymmetric nature of the interference pattern, determined by the negative sign of the Kerr nonlinearity.

\begin{figure*}[ht]
    \includegraphics[width=1.0\textwidth]{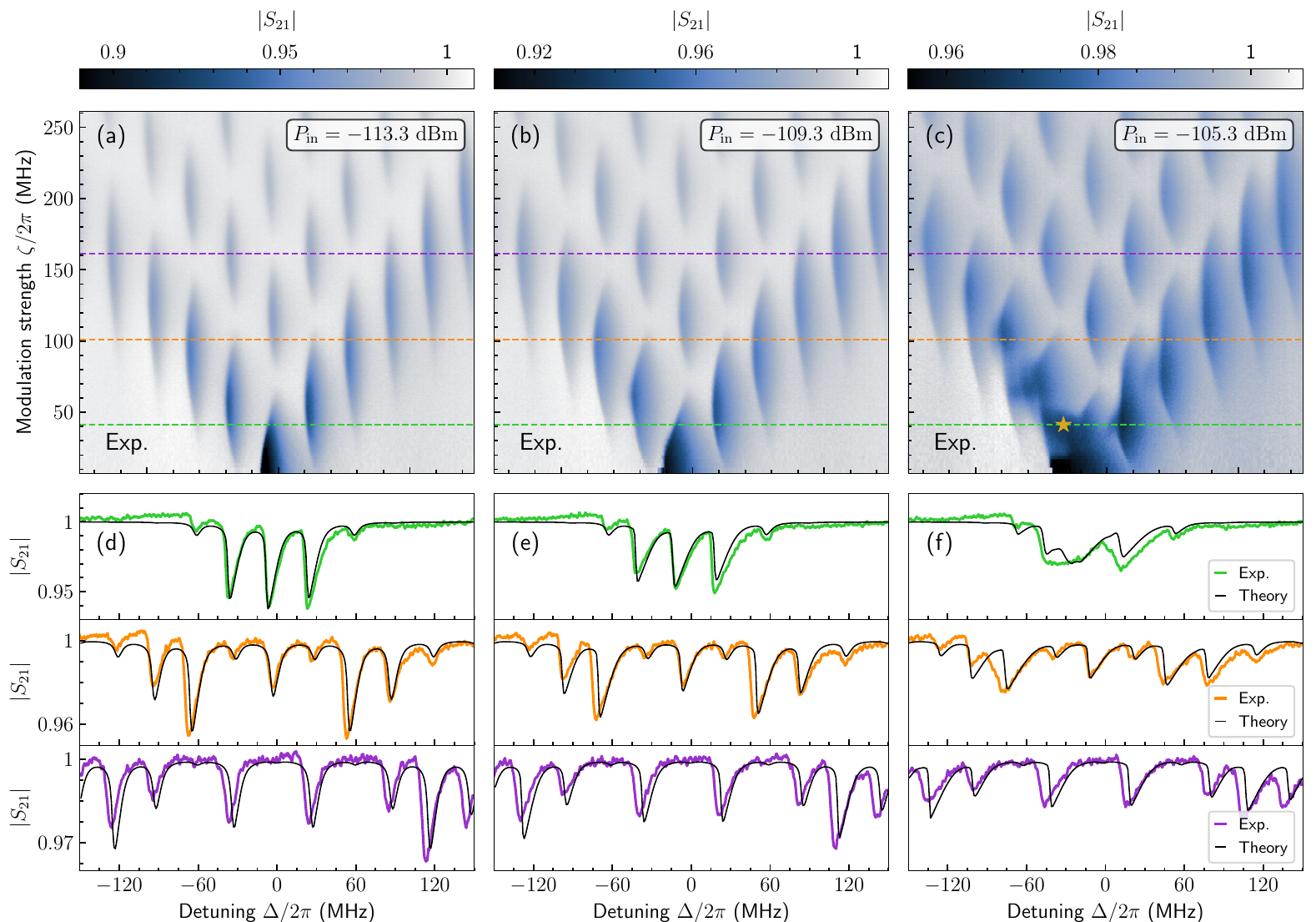}
    \caption{\label{fig:Duffing}
    LZSM interferometry with the $N=32$ device in the Duffing regime.
    (a-c) Measured magnitude of $\SParam$ versus $\Delta$ and $\zeta$ for increasing drive power $P_{\rm in}$.
    The dashed color lines refer to the values of $\zeta$ chosen for panels (d-f).
    The star in (c) indicates the value where the analysis of chaos is performed in \figref{fig:Duffing_chaos}(d).
    (d-f) Measured $\SParam$ as a function of detuning and for three specific values of $\zeta/2\pi$: $\SI{41.3}{\MHz}$ (green curve), $\SI{101.1}{\MHz}$ (orange curve), and $\SI{161.2}{\MHz}$ (purple curve).
    The black superimposed curves are the results of the numerical simulation of the full quantum model for the parameters in Table \ref{tab:param_main} (detailed in Methods).
    The modulation frequency is set to $\Omega/2\pi=\SI{30}{\MHz}$.
    The systematic discrepancy in the position of the dips between theory and experiments is due to the nonlinear dependence of the modulation of the flux amplitude discussed in the Supplementary Information. 
}
\end{figure*}

\subsection*{LZSM beyond the qubit approximation: Duffing regime}

Finally, we investigate the Duffing regime $\kappa > |\chi|$ for a drive amplitude sufficiently large to deviate from the linear regime~\cite{Note4}.
For the intermediate drive amplitudes shown in \figsref{fig:Duffing}~(a) and (b), the various dips are well separated despite showing an asymmetric bending of $|\SParam|$.
When compared with \figsref{system}~(g)~and~(h), we observe a similar deformation of the transmission dips.
Therefore, we assign this feature to the emergence of \textit{bistability triggered by the competition between detuning and Kerr nonlinearity}.
For these parameters, we find that the formula in \eqref{Eq:effective_hamiltonian} captures the deformation of the dips, as discussed more in detail in the Supplementary Information. 
Thus, the system behaves as a collection of independent Duffing oscillators and the overall effect of the modulation is to rescale the drive amplitude $F$ of each sideband. 

When the driving power is further increased in \figref{fig:Duffing}~(c), several of the neighboring LZSM dips eventually overlap. 
This case cannot be simply captured as separated LZSM interferences, and it is qualitatively different from all the previously studied cases.
The simplified picture of \eqref{Eq:effective_hamiltonian} thus breaks down, and the system becomes multimodal and behaves as a set of interacting nonlinear cavities.
Nonetheless, the full simulation of the quantum Floquet model matches the data in all regimes, as shown in \figsref{fig:Duffing}~(d-f).

As detailed in the methods section, the merging of several modes and the qualitative change in the system's behavior can be associated with the emergence of dissipative quantum chaos. 
At weak pump power, the LZSM dips correspond to distinct Fourier modes, each characterized by its own frequency. 
As the pump power increases, these modes begin to interact and merge, analogous to phenomena observed in strongly driven resonators~\cite{DahanNPJ22, FerrariPRR25}, thereby suggesting the onset of chaos in the Floquet system. 
Classical chaos is characterized by a system's sensitive dependence on initial conditions, often quantified by a positive Lyapunov exponent~\cite{strogatz_nonlinear_2018}.
On the other hand, the characterization of quantum chaos often relies on the spectral properties predicted by random matrix theory~\cite{bohigas_characterization_1984, haake_quantum_2001, dalessio_quantum_2016, GrobePRL88}.
In open quantum systems, quantum chaos can be extended through the analysis of the Liouvillian superoperator, which governs the dynamics of the density matrix and provides insights into integrability and chaos. 
The complex spacing ratio is an efficient criterion for distinguishing between integrable and chaotic regimes, as it assesses the distribution of spacings between eigenvalues~\cite{SaPRX20}.
However, as shown in the Methods section, applying the usual complex spacing ratio criterion to Floquet systems fails to capture the nuances of dissipative quantum chaos.
Instead, we generalize the spectral statistics of quantum trajectories (SSQT) criteria introduced in Ref.~\cite{FerrariPRR25} to Floquet systems, demonstrating its relevance and correctness in identifying chaotic behavior~(see \figref{fig:floquet_chaos}).
This refined approach allows for a precise analysis of the system's dynamics, accurately pinpointing the transition to chaotic phases.

\begin{figure*}[]
    \includegraphics[width=0.98\textwidth]{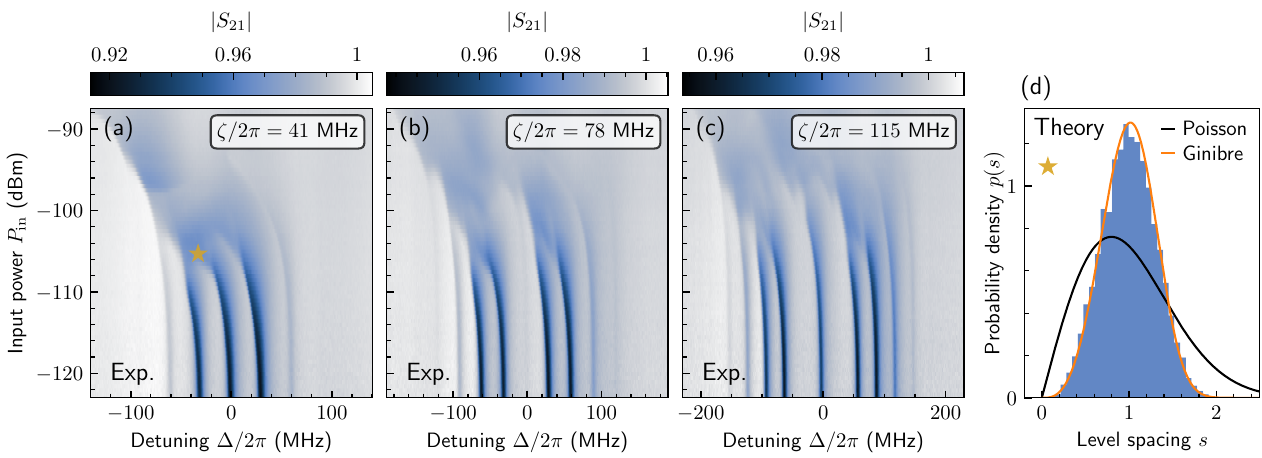} 
    \caption{\label{fig:Duffing_chaos}
    Multimodal-like behavior in the Duffing regime.
    (a-c) Measurement of $|\SParam|$ for increasing drive power, fixed frequency modulation $\Omega/2\pi=\SI{30}{\MHz}$ and increasing ratios $\zeta/\Omega\approx 1.4$ (a), $\zeta/\Omega\approx 2.6$ (b) and $\zeta/\Omega\approx 3.8$ (c).
    For low input powers the system behaves as a collection of noninteracting nonlinear modes, each one well separated from the others.
    For larger values of $P_{\rm in}$, the system enters a phase characterized by a single broad response where the notion of isolated mode is lost. Such a response can be observed in multimode nonlinear systems and has been associated with a transition from integrability to dissipative quantum chaos~\cite{FerrariPRR25}.
    To show that this is indeed a dissipative quantum chaotic phase we plot in (d) the histogram of the probability density $p(s)$ of the level spacings $s$ obtained by diagonalizing the Floquet Liouvillian in the broad-response region indicated by the star in panel (a).
    Parameters are set to $\zeta/2\pi=\SI{41.3}{\MHz}$, $\Omega/2\pi=\SI{30}{\MHz}$, $\Delta=-1.1\Omega$ and $F/2\pi=\SI{49.5}{\MHz}$ ($P_{\rm in}\approx-105$ dBm).
    The cutoff in the Hilbert space is set to $90$.
    The solid black (orange) curve represents the ideal Poisson (Ginibre) distribution given by \eqref{eqs:poisson} [\eqref{eqs:ginibre}] associated with integrability (chaos).
}
\end{figure*}

\hspace{6pt}

\emph{Merging of Sidebands: Signature of Dissipative Quantum Chaos} \textbf{--} Utilizing the novel criterion developed in the methods for Floquet dissipative quantum systems, we demonstrate that the parameters predicting chaos in the model correspond precisely with those where the sidebands merge.

We investigate LZSM interference for three values of modulation strength $\zeta$ as a function of the driving power $P_\mathrm{in}$, as shown in \figsref{fig:Duffing_chaos}~(a-c).
As in all the other experiments presented in this work, measurements are done in the steady-state and are thus independent on the initial conditions of the system.
At low driving power, we find a linear regime where $m$ dips appear separated by the frequency $\Omega/2\pi=\SI{30}{\MHz}$ and with visibility given by Bessel functions $J_{\bar m}(\zeta/\Omega)$.
This regime is remarkably similar to that of several nonlinear modes separated by the same frequency $\Omega$.
As the driving power increases, each of these dips initially follows the typical Duffing behavior of a single resonator, as already mentioned.
For high enough input power, however, these individual dips disappear and merge, leading to a very broad response of the system.
At this point, one completely loses the notion of individual synthetic modes and their bistability.

The merging and broadening of the dips of the scattering coefficient $|S_{21}|$ in Fig.~\ref{fig:Duffing_chaos}~(a) occur for an input power $P_\mathrm{in}\approx -108$ dBm, which coincides with the point where dissipative quantum chaos emerges according to our theory, as reported in Fig.~\ref{fig:floquet_chaos} of the Methods section.
In Fig.~\ref{fig:Duffing_chaos}~(d), we plot the probability density of the level spacings obtained by diagonalizing the Floquet Liouvillian for the parameters indicated by a star in Fig.~\ref{fig:Duffing_chaos}~(a) where LZSM dips have merged.
It has been shown that integrable systems exhibit Poisson-distributed level spacings, indicating no level repulsion, while chaotic systems follow Ginibre statistics, characterized by level repulsion and non-Hermitian random matrix behavior~\cite{GrobePRL88, FerrariPRR25}.
We find that, upon the merging of the Floquet modes, the Floquet Liouvillian level statistics conform to the Ginibre distribution [see \eqref{eqs:ginibre}], indicating a clear transition to the dissipative quantum chaos regime.

The onset of the chaotic phase can also be controlled by tuning the spacing between LZSM resonances through the modulation frequency $\Omega$, as shown in the Supplementary Information. 
For instance, the separated bistable regions of \figref{fig:Duffing}~(b) would start overlapping by decreasing $\Omega$, potentially resulting in a chaotic state.

\section*{Discussion}
\label{Sec:Conclusions}

This article investigates the physics of Landau-Zener-Stückelberg-Majorana (LZSM) interference beyond the conventional two-level approximation. 
By employing two nonlinear superconducting resonators---one in the Kerr (nonlinearity larger than dissipation rate) and the other in the Duffing (nonlinearity smaller than dissipation rate) regime---we have established a general paradigm for studying LZSM interference in bosonic systems. 
We have developed a unified model that accurately describes the observed phenomena across all parameter regimes before the onset of many-body-like effects. 

At low driving powers, we have shown that interference patterns remain independent of the system nonlinearity, preventing the distinction between linear and nonlinear resonators. 
However, at higher driving powers, we have uncovered novel effects arising from the interplay between modulation and nonlinearity, with the dissipation rate playing a crucial role in shaping the emergent features.
For large enough modulation frequency $\Omega$ with respect to the nonlinearity $|\chi|$, the sidebands remain well separated and the standard LZSM picture can be extended to account for nonlinear effects.
The nonlinearity of the resonator dresses each LZSM interference lobe by the nonlinear features observed in~\figref{system}.
For $|\chi| < \kappa$, we observe continuous bending of the LZSM interference pattern.
On the other hand, for $|\chi| > \kappa$, we observe how multiphoton resonances are reproduced all through the LZSM interference pattern.
For $\Omega \ll |\chi|$, we observed a different regime of multiphoton LZSM interferences.
For instance, when $|\chi| > \kappa$, each Kerr multiphoton resonance is dressed by LZSM sidebands.
The resulting pattern is determined by an effective $n$-photon absorption equivalent to a $n$-photon drive that is dressed by the modulation.
This results in a characteristic $\Omega/n$ modulation of the $n$-photon LZSM intereference pattern, similar to those recently observed in~\cite{bjorkmanObservation2024} for a single passage through the avoided crossing.

All of these diverse phenomena occurring across a wide range of device parameters are effectively captured by our extension of the standard LZSM transition paradigm, see~\eqref{Eq:effective_hamiltonian}.
This demonstrates the efficiency of the LZSM formalism in predicting nonlinear resonator dynamics in Floquet regimes.

Beyond this paradigm, we investigated regimes where different sidebands begin to overlap and interact. 
All observed features in this regime are quantitatively reproduced through full quantum simulations of the Floquet model, which is detailed in Sec.~I. of the Supplementary Information.
In the Kerr regime, we demonstrated that when the modulation frequency is commensurate with the nonlinearity, avoided level crossings form between LZSM sidebands. 
Moreover, the interaction between these Floquet states can be tuned via drive and modulation parameters.
Conversely, in the Duffing regime, we theoretically predicted and experimentally observed the overlap and merging of different sidebands, and how it coincides with quantum chaotic behavior.
The significance of this finding is twofold.
Theoretically, it contributes to recent efforts to provide an operational definition of chaos tied to measurable quantities. 
Our extension of dissipative quantum chaos to Floquet systems is general and can be applied to other periodically modulated quantum systems.
Experimentally, our work is relevant to superconducting quantum circuits, a leading platform for quantum computing and error correction. 
Recent studies predict that quantum chaos can impair quantum information storage and manipulation~\cite{berke_transmon_2022, CohenPRXQ23, FerrariPRR25, dumasMeasurementInduced2024}. 
Our work provides one of the first indirect observations of DQC in a fundamental component of superconducting quantum hardware.

From a fundamental point of view, the time features of the system remain to be investigated.
Indeed, in the absence of frequency modulation, switching dynamics of conventional Duffing oscillators in the bistable regime have been thoroughly studied, including phenomena such as the quantum-to-classical transition~\cite{andersenQuantum2020, ChenNATCOM23} and the two-photon driven case~\cite{wangQuantum2019, berdouOne2023, BeaulieuARXIV23}.
Applying frequency modulation is expected to modify the critical phase diagram, potentially offering new ways to control and shape dissipative phase transitions with possible applications in quantum sensing~\cite{VicentiniPRA18, dicandiaCritical2023,  montenegroReview2024}.
Additionally, in the regime where sidebands merge and lead to chaos, the system dynamics may become significantly richer and depart from standard bistable behaviors.
Furthermore, while our current analysis primarily utilizes spectral statistics to investigate chaos, other tools such as out-of-time-order correlations (OTOCs) could be developed within an open and dissipative formalism to investigate even more general features, such as quantifying quantum information scrambling and sensitivity to initial conditions~\cite{liMeasuring2017, braumullerProbing2022, xuScrambling2024}.
While the open system formulation of OTOCs to dissipative Kerr resonators has been used \cite{FerrariPRR25,DahanNPJ22}, extending these techniques to Floquet systems requires a robust definition of time reversion in the presence of periodic modulations.

Overall, our work significantly advances the current understanding of LZSM and Floquet physics, shedding light on the intricate interplay between interference and nonlinear effects. 
While many studies have demonstrated the applications of LZSM phenomena in two-level systems~\cite{IvakhnenkoPHYSREP23}, we anticipate our work enabling similar benefits for multilevel nonlinear bosonic systems. 
Our findings offer exciting perspectives for controlling and engineering Floquet states and synthetic dimensions~\cite{leePropagation2020, hungQuantum2021}, with potential extensions to systems involving multiple cavities~\cite{gomez-leonFloquetBloch2013, leon-montielObservation2018a, yamajiCorrelated2023, heugelrole2023a} and higher-dimensional synthetic spaces~\cite{ozawaTopological2019, duttsingle2020}. 
Moreoever, our platform is well-suited to investigate the rich interplay between Floquet physics and topology~\cite{rudnerAnomalous2013, maczewskyObservation2017}, with potential extensions to nonlinear topology~\cite{mukherjeeObservation2020, coenNonlinear2024, smirnovaNonlinear2020}. 
The merging of multiple interference peaks, both in the Kerr and Duffing regimes, offers several potential applications. 
In the Kerr regime, we show the presence of an ``effective interaction'' between Floquet states~\cite{clarkInteracting2019}, that can be either enabled or suppressed by tuning the modulation parameters.
These could be used to, e.g., engineer transition and interaction between states with different decay rates, and provide opportunities to simulate non-Markovian baths~\cite{lemmertrappedion2018}.
Conversely, in the Duffing regime, this Floquet approach to dissipative chaos has reduced susceptibility to disorder and fabrication mismatches when compared to alternative implementations in extended systems~\cite{underwoodLowdisorder2012, fedorovPhoton2021, jouannyBand2024}.
This opens possibilities to use this LZSM interference to simulate emergent chaotic features in engineered dissipative and time-dependent configurations, such as ultrastrongly coupled light-matter systems~\cite{friskkockumUltrastrong2019, bonifacioLandauZenerSt2020}, devices in the noisy intermediate-scale quantum (NISQ) era \cite{CohenPRXQ23,DahanNPJ22}, and two-photon driven systems~\cite{Lescanne2020, BeaulieuARXIV23, bjorkmanObservation2024}.
Finally, LZSM protocols have been used as quantum simulators of Kibble-Zurek mechanisms~\cite{BogdanPRL05, higuera-quinteroExperimental2022b}.
The extension of a similar protocol to multilevel phenomena is still lacking. 

\section*{Methods}

\subsection*{Device Fabrication}

The devices are fabricated on a \SI{525}{\micro\meter} thick high-resistivity intrinsic 4 inch silicon wafer. 
The substrate is cleaned using piranha solution, followed by the removal of native oxide via a \SI{1}{\percent} hydrofluoric acid treatment. 
Immediately after, a \SI{150}{\nano\meter} thick layer of aluminum is deposited by e-beam evaporation at a rate of \SI{0.2}{\nano\meter\per\second}.
Alignment markers are defined through photolithography, e-beam evaporation of a \SI{5}{\nano\meter} thick Ti layer and a \SI{55}{\nano\meter} thick Pt layer, and subsequent lift-off. 
The waveguide and control lines are patterned via photolithography and wet etching for \SI{2}{\minute} \SI{30}{\second} in TechniEtch Alu80 etchant. 
E-beam lithography is employed to define the Josephson junctions of the SQUID array. 
The wafer is coated with a bilayer resist stack consisting of \SI{500}{\nano\meter} of MMA EL9 and \SI{450}{\nano\meter} of PMMA 495K A8.
The mask is then patterned using e-beam lithography (Raith EBPG5000+ at \SI{100}{\kilo\electronvolt}) and developed in a 1:3 MIBK:IPA solution for \SI{2}{\minute}. 
The Josephson junctions have a square shape with a width of approximately $\SI{350}{\nm}$.
The Josephson junctions are formed by double-angle evaporation in an ultra-high vacuum Plassys MEB550SL3 system using the Manhattan technique~\cite{kreikebaumImproving2020}. 
This involves the deposition of \SI{50}{\nano\meter} of aluminum at \SI{0.5}{\nano\meter\per\second} at $+45\degree$ tilt angle, followed by an oxidation step of \SI{10}{\minute} in 0.15 Torr of pure dioxygen, a second aluminum deposition of \SI{120}{\nano\meter} at \SI{0.5}{\nano\meter\per\second} and $-45\degree$ tilt angle, and a capping oxidation layer formed during \SI{10}{\minute} in 4 Torr of pure O\textsubscript{2}. 
Lift-off is performed in \SI{80}{\degreeCelsius} 1165 remover for \SIrange{4}{8}{\hour}. 
A final patching step is carried out to close the loops of the isolated Josephson junctions formed with the Manhattan technique and to connect one side of the SQUID array to the ground plane. 
The same bilayer resist stack is used, and e-beam lithography is employed to expose the patch areas. 
The native oxide of the bottom aluminum layer is removed in the Plassys system by argon ion plasma milling, and a \SI{200}{\nano\meter} thick aluminum layer is deposited directly after at a rate of \SI{0.5}{\nano\meter\per\second}. 
Finally, the wafer is diced into 4x\SI{7}{\milli\meter\squared} chips using a nickel-bonded diamond blade.

\subsection*{Measurement setup}

\begin{figure}[]
    \includegraphics[width=.5\textwidth]{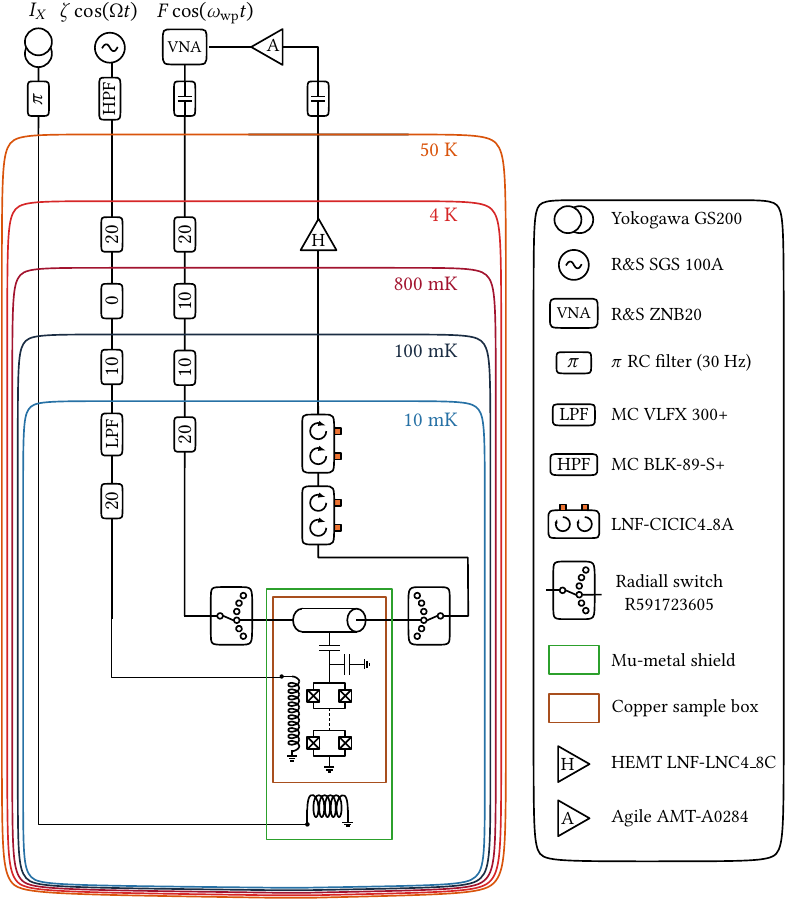} 
    \caption{\label{fig:full_wiring}
    Schematics of the full wiring of the cryostat and room-temperature electronics. 
    }
\end{figure}

A schematic of the measurement setup is shown in~\figref{fig:full_wiring}.
The 4x7 mm$^2$ die is wire bonded with aluminum wire on custom-printed circuit board. 
The die is then glued directly on a high-purity copper sample holder that is thermally anchored at the mixing chamber stage of a LD Bluefors cryostat with a typical base temperature of 15 mK.
The sample holder is protected against external magnetic fields using two mu-metal shields.
The SQUID array is coupled to a 50 Ohm coplanar waveguide in a notch configuration.
The input signal is generated by a vector network analyzer (VNA) R$\&$S ZNB20 and transmitted via a heavily attenuated coaxial line to the device feedline.
The output signal passes through two double-circulators before being amplified at 4K by a LNF-LNC4$\_$8C HEMT amplifier and at room temperature by an Agile AMT-A0284 low-noise amplifier.
The signal is collected and demodulated in the VNA.
Six-ports Radiall R591723605 coaxial switches are placed on the mixing chamber plate on both sides of the feedline to allow switching between different devices.
Both $N=10$ and $N=32$ devices presented in this work were connected between the same switches and thus shared the same input and output lines.
The static flux of the SQUID array is controlled by applying a direct current to a NbTi external coil mounted underneath the sample holder.
The direct current is applied via twisted NbTi pairs using a Yokogawa GS200 source.
The frequency modulation of the SQUID array is performed by applying a signal generated by a R$\&$S SGS 100A signal generator to the local flux line of the device.
The DC noise is attenuated using a high-pass filter with a cutoff frequency of 100 kHz at room temperature.
The line is further attenuated and filtered at the mixing chamber stage with a Minicircuits  VLFX-300+ low-pass filter (LPF).
We found that without this LPF, the internal loss rate of the SQUID arrays was increased by up to a factor of ten.
We also included an additional 20 dB of attenuation between the LPF and the flux line to eliminate spurious standing wave modes between the LPF and the on-chip ground termination of the flux line.
Devices $N=10$ and $N=32$ were housed in different sample holders in separate shields and thus did not share the same external coil and flux lines.

\subsection*{Estimation of Device Parameters}

We then carefully characterize the two devices at their chosen flux operating points.
The parameters of the devices, reported in \tabref{tab:param_main}, are obtained by fitting $\SParam$ without modulating the frequency.
The SQUID array is modeled as a Kerr resonator according to the Hamiltonian of \eqref{Eq:Hamiltonian} with $\zeta=0$.
First we fit the transmission at low enough power to ensure an average occupation of less than one photon.
This allows us to neglect the Kerr nonlinearity and the dephasing.
The expression of the linear transmission coefficient $\SParam$ is obtained from standard notch configuration input-output theory~\cite{ chenScattering2022a,BeaulieuARXIV23} as
\begin{equation}
\SParam  = 1 - \frac{\kappa_\text{ext}}{\kappa_\text{ext} + \kappa_\text{int} + 2i\Delta }  \times \frac{e^{i\phi}}{\cos \phi}.
\end{equation}
Following the diameter correction method~\cite{khalilAnalysisMethodAsymmetric2012}, we add the last term to compensate for impedance mismatch.
To fit the measured transmissions to this expression, we first normalize the data by a background transmission measured with the SQUID threaded by a different flux, such that its frequency lies outside of the measurement range.
All experimental data reported in this work are normalized this way.
We then extract the precise operating frequency as well as the internal and external loss rates of each device.

To determine the Kerr nonlinearity $\chi$ and the dephasing rate $\kappa_\phi$, we need to fit the power dependence of the transmission which is reported in \figref{system}.
A simple analytical model could be used for weakly anharmonic devices satisfying $|\chi| \ll \kappa$~\cite{eichlerControlling2014, BeaulieuARXIV23}.
Instead we directly solve the Lindblad master equation [\eqref{Eq:Liouvillian}] to find the intra-cavity field $\alpha$, again setting $\zeta=0$ in the Hamiltonian.
This model is valid for both devices studied in this work and accounts for dephasing.
Using input-output theory, we convert $\alpha$ to the transmission scattering parameter using the following relation
\begin{equation}
    S_{21} = 1 - i \frac{\kappa_{\rm ext} \, \alpha}{2 F}.
\end{equation}
The drive amplitude $F$ is related to the input power $P_{\rm in}$ as
\begin{equation}
    F = \sqrt{\frac{P_{\rm in} \kappa_{\rm ext}}{\hbar \omega_d}}.
\end{equation}

We start by fitting the device $N=10$ in the Kerr regime.
We perform a global simultaneous fit of approximately ten frequency sweeps at different driving powers.
We use the parameters obtained from the low-power fit and keep three independent fitting parameters: $\kappa_\phi$, $\chi$, and the attenuation of the input drive line.
Because single multiphoton transitions are well-resolved with the Kerr device, we can obtain all three parameters without prior calibration of the input attenuation.
The Kerr multiphoton resonances reported in \figref{system}~(c-e) are not equispaced by $\chi$, instead the spacing increases for larger $\ket n$.
We attribute this effect to non-negligible higher-order nonlinearities from the expansion of the Josephson cosine potential.
To accurately reproduce the experimental data of the $N=10$ device, we also include a term of the form $\chi^{(5)} (\hat{a}^\dagger)^3 \hat{a}^3$ in the model, and find a value of $\chi^{(5)}\approx 5\% \chi$~\cite{Note2}.

Finding the Kerr nonlinearity of the Duffing device, however, requires knowing the input attenuation.
But the feedline of the Kerr and Duffing devices are connected on the same microwave switch, as depicted in \figref{fig:full_wiring}.
Therefore we assume that the input attenuation obtained from the fit of the Kerr device is also valid for the Duffing device.
We perform a similar global fit of the power dependence of the transmission of the Duffing device, this time with only two free fitting parameters: $\chi$ and $\kappa_\phi$.
Simulations of the Kerr shift of both devices are shown in \figref{system}.

\subsection*{Linear and qubit regime approximations}

Under sufficiently weak drive, the two devices can be approximated as respectively qubit and linear resonators.
It is in this regime that we observed standard LZSM interferences as shown in Fig.~\ref{Fig:LZSM_qubit_linear}.
We now give explicit conditions for the linear approximation to hold.

For the strongly nonlinear $N=10$ resonator, one can show that, assuming at most two photons in the system, the maximum of the two-photon population occurs at the multiphoton resonance $\Delta =  \chi$, where
\begin{equation}
\begin{split}
\brakket{2}{\sss}{2} &=\frac{2 F^4}{9 F^4+2 \kappa ^2 \left[2 \left(\kappa ^2+\chi
   ^2\right)-5 F^2\right]} \\
   & \simeq \frac{ F^4}{2 \kappa ^2 \chi ^2}.
\end{split}
\label{eq:footnote72}
\end{equation}
It follows that $F^2 \ll |\chi|\kappa$ ensures the validity of the qubit approximation

For the weakly nonlinear $N=32$ resonator, we consider the semiclassical (coherent state approximation) $\sss = \ketbra{\alpha}{\alpha}$. 
One finds that the photon number $n=|\alpha|^2$ satisfies
\begin{equation}
\begin{split}
&\left[ \frac{\kappa ^2}{4}  +(\Delta +n \chi)^2 \right] n  -F^2 
\\
&\quad \simeq 2 \Delta  n^2 \chi +n \left(\Delta ^2+\frac{\kappa
   ^2}{4}\right)-F^2=0.
\end{split}
\end{equation}
The solution to this equation can be expanded in powers of $\chi$ as
\begin{equation}
n = n_0 \left(1 - n_0 \frac{8 \Delta \chi}{4 \Delta^2 + \kappa^2} \right), \, \text{ with } \, n_0 = \frac{F^2}{\Delta^2 + \kappa^2/4}.
\end{equation}
The deviation from the linear regime, defined as $\delta n = 1- n / n_0 $ is then maximal for $\Delta = \kappa / (2\sqrt{3})$, which leads to 
\begin{equation}
\delta n = \frac{3 \sqrt{3} F^2 \chi }{\kappa ^3}.
\label{eq:footnote73}
\end{equation}
Since we are interested in the regime $\delta n \ll 1$, we recover the condition $F < \sqrt{ \kappa ^3 / |\chi|}$ for the linear approximation, as given in the main text.

\subsection*{Analysis of Dissipative Quantum Chaos}
\label{methods:chaos}

Classical chaos is defined by the sensitivity of a system's dynamics to initial conditions, often characterized by a positive Lyapunov exponent \cite{strogatz_nonlinear_2018}. Quantum chaos, in both isolated and open systems, is typically described through the \textit{quantum chaos conjecture} \cite{bohigas_characterization_1984, haake_quantum_2001, dalessio_quantum_2016, GrobePRL88}, i.e., assuming the system has a meaningful classical limit that exhibits chaos, one can conjecture that the spectral properties of the time evolution generator align with the universal predictions of random matrix theory. For models without a classical limit, random matrix theory predictions are still employed to define quantum chaos \cite{kos_many-body_2018, AkemannPRL19, SaPRX20} due to their success in forecasting the properties of quantum systems without a classical counterpart \cite{serbyn_spectral_2016, bordia_periodically_2017, abanin_colloquium_2019}.

\vspace{6pt}

\emph{A criterion for dissipative quantum chaos in Floquet systems} \textbf{--} For time-independent Liouvillian systems, integrability and dissipative quantum chaos in the open quantum system are often characterized via the spectral properties of the Liouvillian.
For a time-independent system, the equation of motion reads $\partial \hat{\rho}/\partial t = \LL \hat{\rho}$
where $\LL$ is the non-Hermitian Liouvillian superoperator.
As $\hat\rho(t) = \exp(\LL t) \hat\rho(0)$, 
the eigendecomposition of $\LL$ fully characterizes the dynamics of the density matrix.
The right eigenoperators $\hat{\eta}_j$ and left eigenoperators $\hat{\sigma}_j$ of $\LL$ \cite{breuer_theory_2007} are defined by
\begin{equation}\label{Eq:Liouvillian_eigenvalues}
\mathcal{L}\hat{\eta}_j = \lambda_j\hat{\eta}_j,\,\,\,\,\,\,\,\,\,\mathcal{L}^{\dagger}\hat{\sigma}_j = \lambda_j^*\hat{\sigma}_j,
\end{equation}
where $\lambda_j$ are complex, ${\rm Re} (\lambda_j)\leq0$, and  $\operatorname{Tr}(\hat{\sigma}_j^{\dagger}\hat{\eta}_k) = \delta_{jk}$.

Chaos is then characterized through the statistical distribution of the spacings of the complex Liouvillian eigenvalues $\{\lambda_j\}$~\cite{GrobePRL88}. 
In particular, one studies the distribution of nearest-neighbor eigenvalue spacings
\begin{equation}\label{eqs:distribution}
     p(s) = \sum_{j} \delta(s_{j} -s),
\end{equation}
where $s_j=|\lambda_j - \lambda_j^{\rm NN}|$, with $\lambda_j^{\rm NN}$ the eigenvalue closest to $\lambda_j$ in the complex plane. 
In integrable dissipative systems, $s$ follows a 2D Poisson distribution
\begin{equation}\label{eqs:poisson}
     p_{2\textrm{D}}(s) = \frac{\pi}{2}se^{-\frac{\pi}{4}s^2},
\end{equation}
while for chaotic dissipative systems, the level spacing distribution follows the Ginibre distribution of Gaussian non-Hermitian random matrices ensembles
\begin{equation}\label{eqs:ginibre}
   p_{\textrm{GinUE}}(s) = \left(\prod_{k=1}^{+\infty}\frac{\Gamma(1+k, s^2)}{k!}\right)\sum_{j=1}^{+\infty}\frac{2s^{2j+1}e^{-s^2}}{\Gamma(1+j, s^2)}.
\end{equation}
An unfolding procedure, in which the uncorrelated part is removed from $p(s)$ in \eqref{eqs:distribution}, is required to evaluate the level statistics from the spectrum and for the proper characterization of chaos~\cite{MarkumPRL99}.
We adopt that described in Ref.~\cite{AkemannPRL19}.

An alternative, efficient way to perform this analysis is the complex spacing ratio~\cite{SaPRX20}
\begin{equation}
    z_j  = \frac{\lambda_j^{\rm NN} - \lambda_j}{\lambda_j^{\rm NNN} - \lambda_j}= r_j e^{i \theta_j},
\end{equation} 
with $\lambda_j^{\rm NN}$ the eigenvalue closest to $\lambda_j$ in the complex plane, and $\lambda_j^{\rm NNN}$   the second-nearest neighbor to $\lambda_j$.
The average values $\langle r \rangle$ of $r_j$  and $\langle \cos\theta \rangle$ of $\cos\theta_j$, can be used as indicators of dissipative quantum chaos.
For a 2D Poisson distribution, associated with an integrable system,  $\langle r \rangle=0.66$, and  $-\langle \cos\theta \rangle=0$.
For the Ginibre distribution, i.e., chaos, $\langle r \rangle=0.74$, and  $-\langle \cos\theta \rangle=0.24$.

The spectral definition of DQC presented above can be extended to Floquet systems through the introduction of the Floquet Liouvillian superoperator $\LL_{\rm F} $. Also $\LL_{\rm F} $ can be diagonalized obtaining its right (left) eigenvectors $\hat{\eta}_j$ ($\hat{\sigma}_j$) and the Liouvillian spectrum $\{\lambda_j\}$
\begin{equation}\label{eqs:eigendecomposition}
\LL_{\rm F}\hat{\eta}_j = \lambda_j\hat{\eta}_j,\,\,\,\,\,\,\,\,\, \LL_{\rm F}^{\dagger}\hat{\sigma}_j = \lambda_j^*\hat{\sigma}_j,
\end{equation}
which satisfy the bi-orthonormality condition $\operatorname{Tr}\{\hat{\sigma}_j^{\dagger}\hat{\eta}_l\} = \delta_{jl}$.
The very same spectral criteria can then be applied to the Floquet eigenvalues.

\begin{figure}[t!]
    \includegraphics[width=0.45\textwidth]{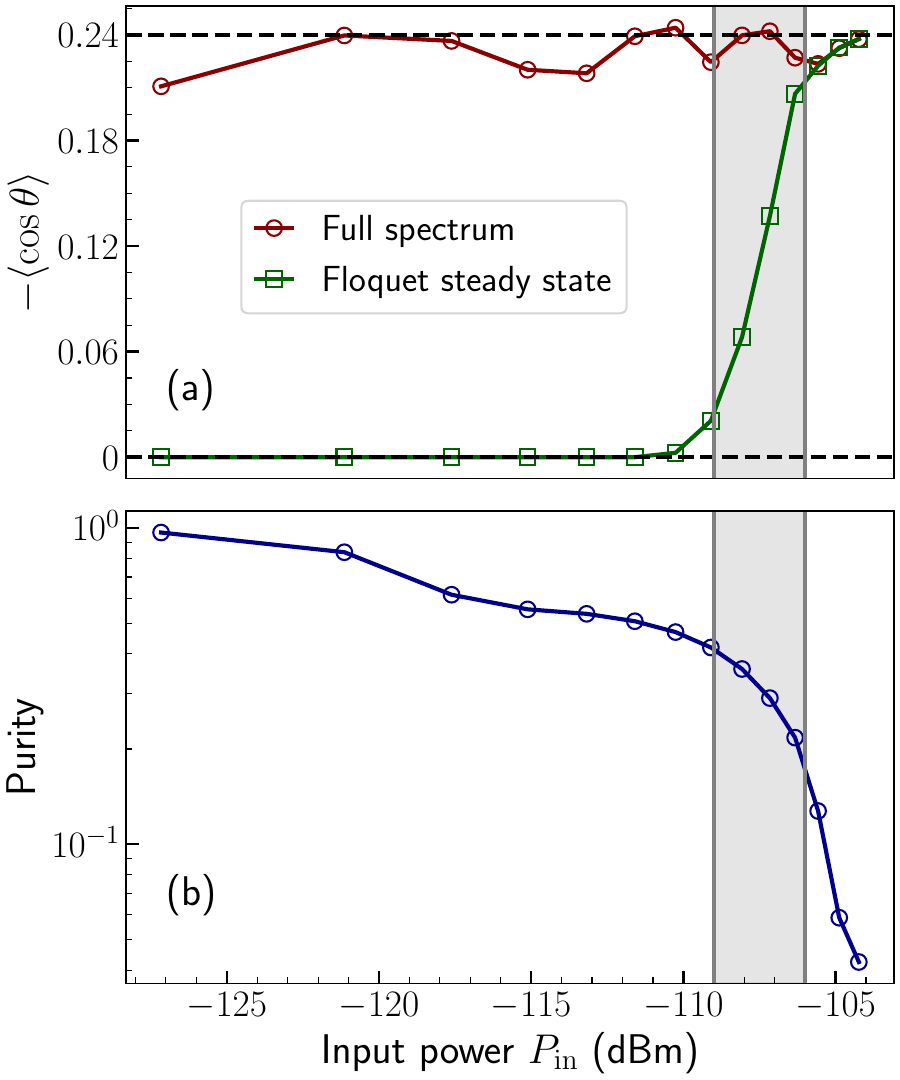} 
    \caption{\label{fig:floquet_chaos}
    Analysis of dissipative quantum chaos using the SSQT criterion detailed in Ref.~\cite{FerrariPRR25} and generalized to Floquet states here.
    (a) Theoretical indicator of chaos $\langle\cos\theta\rangle$ introduced in Ref.~\cite{SaPRX20} computed on the full Floquet-Liouvillian spectrum (red line and circles) and on the eigenvalues selected by the SSQT criterion (green line and squares).
    While the spectral analysis on the full Liouvillian indicates the presence of chaos independently of the drive amplitude for the parameters considered in the plot, the SSQT criterion identifies the broadening of the Duffing peaks in Fig.~\ref{fig:Duffing_chaos} (a-c) (grey rectangle) with a dissipative quantum chaotic phase for the Floquet steady state $\sss^{F}$. 
    When the number of selected eigenvalues is smaller than $100$ a statistically significant analysis can not be carried out, and we set $\langle\cos\theta\rangle=0$.
    (b) Purity $\operatorname{Tr}([\hat{\rho}^{\textrm{F}}_{\textrm{ss}}]^2)$ of the Floquet steady state $\hat{\rho}^{\textrm{F}}_{\textrm{ss}}$. 
    The onset of steady-state quantum chaos in panel (a) coincides with the drop of the purity of the steady state below $0.1$. 
    We use the parameters of Fig.~\ref{fig:Duffing_chaos} (a), the cutoff in the Hilbert space is fixed to $90$, and $c_{\rm min}$ is selected according to~\cite{Note5}.
    } 
\end{figure}

\hspace{6pt}

\emph{Extending the spectral statistics of quantum trajectories for Floquet systems} \textbf{--} The above spectral signatures alone, however, do not correctly capture the emergence of dissipative quantum chaos in the system considered in this work. 
For instance, in Fig.~\ref{fig:floquet_chaos} (a) we plot the indicator $\langle\cos\theta\rangle$ as a function of the input power $P_{\rm in}$. 
For these input powers, the prediction of the spectral analysis of the Floquet Liouvillian is that the system is \textit{always} in a chaotic phase, despite it being almost a pure state for weak $P_{\rm in}$ [c.f. Fig.~\ref{fig:floquet_chaos} (b)].
We conclude that this straightforward analysis of chaos cannot capture the relevant features of the model under consideration.

Given the lack of predictive results, here we generalize the theoretical framework of the spectral statistics of quantum trajectories (SSQT) introduced in Ref.~\cite{FerrariPRR25}.
First, one remarks that the Lindblad master equation admits also a stochastic unraveling in terms of quantum trajectories $\ket{\psi(t)}$, combining the Hamiltonian dynamics with a continuous monitoring of the environment~\cite{WallsBOOK08,WisemanBOOK09}.
The wave function $\ket{\psi(t)}$ can be interpreted as a single stochastic realization of the dissipative dynamics whose average reproduces the predictions of the Lindblad master equation \eqref{Eq:Liouvillian}.
As discussed in Ref.~\cite{FerrariPRR25}, since the system discussed in this article does not have any weak or strong Liouvillian symmetry, all the possible unravelings are expected to give the same information about steady-state integrability and chaos.
We can therefore assume a diagonal unraveling which we can write down considering the spectral decomposition of the Floquet steady state
\begin{equation}
    \sss^{\rm F} = \sum_k p_k \ket{\psi_k}\bra{\psi_k}.
\end{equation}
Using the spectral decomposition introduced in \eqref{eqs:eigendecomposition}, one can then define
\begin{equation}\label{eq:spectralrho}
\begin{split}
    \hat{\rho}_k =& \ket{\psi_k}\bra{\psi_k}
    = \sum_{j} c_{k,j} (t)  \, \, \hat{\eta}_{j}.
\end{split}
\end{equation}
This procedure allows associating to each eigenvalue $\lambda_j$ the relative spectral weight $c_{k,j}$.
We select the Liouvillian eigenvalues $\lambda_j$, for which $|c_{k,j}(t)|>c_{\rm min}$ \footnote{We set the cutoff $c_{\textrm{min}} = \bar{C}/1000$ where $\bar{C}$ is the average of the spectral weights in \eqref{eq:spectralrho}, as detailed in Ref.~\cite{FerrariPRR25}. 
For the Floquet Liouvillian, we found that some of the $|c_j|$ were very large (order of magnitudes bigger than one). 
As the average procedure of the spectral weights would have been affected by those outliers, we restrict the mean to the ones such that $|c_j|\le 1$. 
Such a choice is justified as a spectral coefficient $|c_j| > 1$ will be for sure chosen with the SSQT protocol, and we get a meaningful $c_{\rm min}$ as in Ref.~\cite{FerrariPRR25}.}.
On each $\hat{\rho}_k$ we perform the spectral analysis by computing, e.g., the complex spacing ratio for the selected eigenvalues $\langle \cos\theta \rangle_{k}$.
We finally obtain $\langle \cos\theta \rangle = \sum_k p_k \langle \cos\theta\rangle_{k}$.

In Fig.~\ref{fig:floquet_chaos} (a), the green curve represents the results of the SSQT criterion. 
Compared to the spectral statistics applied to the full Floquet Liouvillian spectrum, we see a profoundly different behavior of the system as a function of the drive amplitude.
Notably, comparing the results of Fig.~\ref{fig:floquet_chaos} (a) with the purity of $\hat{\rho}^{\textrm{F}}_{\textrm{ss}}$ in Fig.~\ref{fig:floquet_chaos} (b), this time we observe that it drops below $0.1$ only when we enter the steady-state chaotic region.
This result ultimately demonstrates the necessity of the SSQT criterion to correctly interpret the onset of chaos in open quantum systems.

\subsection*{Data availability}
The data used to produce the plots are available on Zenodo with the
identifier https://zenodo.org/records/14883314.

\subsection*{Code availability}
The codes used to analyze the data and produce the plots are available on Zenodo with the identifier https://zenodo.org/records/14883314.

\subsection*{Acknowledgements}
We thank Alberto Mercurio and Sergey Shevchenko for the useful discussion and the insights on the numerical coding.
\add{We are grateful to Guillaume Beaulieu and Davide Sbroggio for their help with the fabrication process.}
M.S. acknowledges support from the EPFL Center for Quantum Science and Engineering postdoctoral fellowship.
F.N. is supported in part by:
Nippon Telegraph and Telephone Corporation (NTT) Research,
the Japan Science and Technology Agency (JST)
[via the Quantum Leap Flagship Program (Q-LEAP), and the Moonshot R\&D Grant Number JPMJMS2061],
the Asian Office of Aerospace Research and Development (AOARD) (via Grant No. FA2386-20-1-4069),
and the Office of Naval Research (ONR) Global (via Grant No. N62909-23-1-2074).
V.S. acknowledges support by the Swiss National Science Foundation through Projects No. 200020\_185015 and 200020\_215172.
P.S. and V.S. acknowledge support from the EPFL Science Seed Fund 2021 and of Swiss State Secretariat for Education, Research and Innovation (SERI) under contract number UeM019-16.
P.S. acknowledges support from the Swiss National Science Foundation through Projects No. 206021\_205335 and Projects No. 200021\_200418, and from the SERI through grant under contract number MB22.00081.

\subsection*{Author Contributions}
L.P., M.S., P.S. designed the experiment.
L.P. fabricated the devices.
M.S. and V.J. performed preliminary measurements.
L.P. performed the measurements presented in the manuscript.
L.P. analyzed the data.
F.M. and F.F. developed the theoretical model and reproduced the experimental data with supervision from V.S.
P.S. supervised the project.
L.P., F.M., M.S., F.F. and P.S. wrote the manuscript with input from all authors.

\subsection*{Competing Interests}
The authors declare no competing interests.


\begin{widetext}
\begin{center}
{\Large\bfseries Supplementary Information}
\end{center}
\vspace{0.2cm}
\end{widetext}

\section{Construction and solution of the Floquet-Liouvillian problem}

The periodically modulated systems described by Eq.~(2) in the main text can be described using a stroboscopic Lindblad master equation of period $T$. 
The equation of motion of such a system is
\begin{equation}
\label{eq:Time_dependent_LME}
\hbar \partial_t \hat{\rho}(t) = \mathcal{L} (t) \hat{\rho}(t), \quad \mathcal{L}(t+T) = \mathcal{L}(t).
\end{equation}
While the temporal dependence of $\LL(t)$ prevents the emergence of a true steady state, one can still reach a \textit{stroboscopic} stationary regime.

\subsection{The average Floquet steady-state}
\label{subsubsec:average_Floquet}

We are interested in the average properties of the system along one modulation period $T = 2 \pi/\Omega$ after a time long enough for the system to have reached a stroboscopic stationary regime. 
To solve this problem, we assume that, for a long enough time, 
\begin{equation}
\hat{\rho}(t)=\sum_{m=-\infty}^{+\infty} \hat{\rho}_m e^{i m \Omega t}.
\end{equation}
One can easily verify that
\begin{equation}
    \frac{1}{T}\int_{t}^{t+T} \hat{\rho}(\tau) \de\tau = \hat{\rho}_0.
\end{equation}
At this point, one has to determine $\hat{\rho}_0$.
A convenient way to find it is to solve it through Fourier analysis (see, e.g., \cite{MaragkouPRB13,MacriPRL22}).

The equation of motion can be recast as
\begin{equation}
\begin{split}
\hbar \frac{\mathrm{d}}{\mathrm{d} t} \hat{\rho}(t) &= \sum_{m=-\infty}^{+\infty} i m \Omega \hat{\rho}_m e^{i m \Omega t} \\
=&\left[\mathcal{L}_0+\mathcal{L}_1 e^{i \Omega t}+\mathcal{L}_{-1} e^{-i \Omega t}\right] \hat{\rho}(t) \\
=&\sum_{m=-\infty}^{+\infty}\left[\mathcal{L}_0+\mathcal{L}_1 e^{i \Omega t}+\mathcal{L}_{-1} e^{-i \Omega t}\right] \hat{\rho}_m e^{i m \Omega t},
\end{split}
\end{equation}
where $\mathcal{L}_0$ is the time-independent part of the Liouvillian in Eq.~(2) of the main text (i.e., $\zeta =0$), while $\mathcal{L}_1$ and $\mathcal{L}_{-1}$ represent the decomposition of the modulation.
Collecting each term evolving with $\Omega$ we have
\begin{equation}
    \sum_{m=-\infty}^{+\infty} \left[ \left(\mathcal{L}_0-i m \Omega \right) \hat{\rho}_m+\mathcal{L}_1 \hat{\rho}_{m-1}+\mathcal{L}_{-1} \hat{\rho}_{m+1} \right] e^{i \Omega t} = 0.
\end{equation}

If we now assume that each term of the sum is stationary, we obtain the recursion relation
\begin{equation}
 \left(\mathcal{L}_0-i m \Omega \right) \hat{\rho}_m+\mathcal{L}_1 \hat{\rho}_{m-1}+\mathcal{L}_{-1} \hat{\rho}_{m+1} = 0.
\end{equation}
By truncating this recursion (i.e., assuming $\hat{\rho}_{m} = 0$ if $m>M$ or $m<-M$), the problem can be then self-consistently solved.

\subsection{Analysis of the Floquet Liouvillian spectrum}
A different approach to solving the Floquet problem consists of constructing the so-called \textit{Floquet evolution superoperator} (a Floquet map for Lindbladian systems).
Indeed, using the time ordering $\mathcal{T}$, we can formally solve \eqref{eq:Time_dependent_LME} as
\begin{equation}\label{eq:solution_Floquet}
\rhot(t) = \mathcal{T} \left[\exp\left(\int_0^{t} \LL(t') dt' /\hbar \right) \right] \hat{\rho}(0)= \mathcal{F}(t, 0)\hat{\rho}(0).
\end{equation}
$\mathcal{F}(t, t_0)$ is the evolution superoperator for the time-dependent Lindblad master equation.
We can then formally introduce the Floquet Liouvillian $\LL_{\rm F}$ as
\begin{equation}\label{Eq:Floquet_Liouvillian}
    \mathcal{F}(T, 0)  = \exp\left( \LL_{\rm F} T/\hbar \right).
\end{equation}
The stroboscopic steady state is the state such that
\begin{equation}
   \LL_{\rm F} \sss^{\rm F} = 0, \quad {\rm or } \quad \mathcal{F} \sss^{\rm F} = \sss^{\rm F}.
\end{equation}

To construct $\mathcal{F}(T, 0)$~\cite{MingantiQuantum22}, let us consider 
\begin{equation}
\mathcal{F}(T, 0) \hat{\rho}_{i,j}, \quad \hat{\rho}_{i,j} = \ket{i}\bra{j}.
\end{equation}
Since $\hat{\rho}_{i,j}$ are an orthonormal basis of the operators space (i.e., any operator can be written as a linear combination of $\hat{\rho}_{i,j}$), we conclude that the matrix form of $\mathcal{F}(T, 0)$ can be obtained as
\begin{equation}\label{Eq:Construction_of_F}
\mathcal{F}_{[m=i \cdot (N+1) + j, :]} = \operatorname{vec}\left[  \hat{\rho}_{i,j}(T)\right],
\end{equation}
where $\mathcal{F}_{[m, :]}$ indicates the $m$th row of the evolution operator in its matrix form, and $\operatorname{vec}\left[  \hat{\rho}_{i,j}(T)\right]$ is the vectorized form of the initial density matrix $\hat{\rho}_{i,j}$ evolved for a time $T$.

\section{SQUID arrays as frequency-tunable Kerr resonators}
\label{Appendix:SQUID_array_model}

An array of $N$ SQUIDs results in $N$ nonlinear bosonic modes whose dispersion relation can be obtained numerically from the linearized Lagrangian of an effective lumped LC model~\cite{maslukMicrowave2012a}.
The dispersion is linear for the lower frequency modes, and saturates at a high-frequency cutoff close to the plasma frequency of the junctions.
Introducing the lowest order nonlinear terms as a perturbation to the previous linear model yields the self- and cross-Kerr terms of each mode~\cite{weisslKerr2015}.
This approximation is valid if the mode frequency $\omega_i$ is much smaller than the Josephson energy $E_J/\hbar$ of individual junctions. 
Increasing the number of SQUIDs in the array reduces both the frequency of the lower modes and the nonlinearity of all modes.
The scaling of the frequency and nonlinearities depends on the circuit parameters and the boundary conditions.

The SQUID arrays considered in this work have between $N=10$ and $N=32$ SQUIDs with nominally identical Josephson junctions.
One end of the array is shorted to ground, while the other end of the array is left open with a capacitance to ground and to a readout waveguide.
Throughout this study we only use the fundamental mode of the SQUID arrays.
The frequency of the second mode of the array is approximately twice that of the first mode.
Consequently, we can safely neglect the second mode, along with all higher modes of the array.
The Josephson inductance of all junctions in the array can be tuned with an external magnetic flux, and the frequency of the first mode the array follows approximately
\begin{equation}
\omega(\Phi_x) = \omega_c \sqrt{|\cos\left( \pi \Phi_x/\Phi_0\right)|},
\label{eq:supp_freq_sq_flux}
\end{equation}
with $\omega_c$ the zero-flux frequency, $\Phi_x$ the flux threading the SQUIDs loop and $\Phi_0=h/2e$ the magnetic flux quantum.
For the large ratio $E_J/E_C$ of the SQUID's junctions, the Kerr nonlinearity depends weakly on the flux $\Phi$.
For the two devices presented here, we choose the flux operating point $\Phi_\mathrm{wp}$ such that the frequency of the first mode of the arrays is similar, with $\omega_\mathrm{wp}/2\pi=4.5$ GHz (4.3 GHz) for the $N=10$ ($N=32$) device.

Retaining only the first mode of the SQUID array, and keeping only the first-order nonlinearity, we arrive at the Kerr resonator Hamiltonian in the lab frame,
\begin{equation}
\mathcal{H} / \hbar = \omega_{\rm wp} \hat{a}^\dagger \hat{a} + \chi \hat{a}^\dagger \hat{a}^\dagger \hat{a} \hat{a} \, ,
\end{equation}
with $\chi$ the Kerr nonlinearity.
After adding a drive $F(\hat a e^{-i\omega_dt} + \hat a^\dagger e^{i\omega_dt})$ and a frequency modulation $\zeta \cos(\Omega t) \hat a ^\dagger \hat a $, and moving to a frame rotating at the pump frequency $\omega_d$, we obtain the Hamiltonian of Eq.~(1) in the main text. 
We note that to accurately model the Kerr multiphoton resonances of the $N=10$ device, we had to include a higher-order nonlinearity~\footnote{
The resonator exhibits deviation from the pure Kerr nonlinearity prediction due to non-negligible effects of higher nonlinearities, e.g., terms of the form $\chi^{(5)} (\hat{a}^\dagger)^3 \hat{a}^3$. We estimate $\chi^{(5)}/2\pi \approx -1.1\,$MHz$\simeq 5\% \chi/2\pi$.
As such, although the $\chi^{(5)}$ term produces small changes compared to the model in Eq.~(1) of the main text, all the relevant physical features of the LZSM interference are captured by the Kerr model.
We also note additional nonlinear effects due to the large values of flux modulation used to obtain the wanted $\zeta$.}.

\section{Derivation of an effective model for the study of the nonlinear modulated resonators}
\label{Appendix:Effective_model}

\begin{figure*}[ht]
    \includegraphics[width=1.0\textwidth]{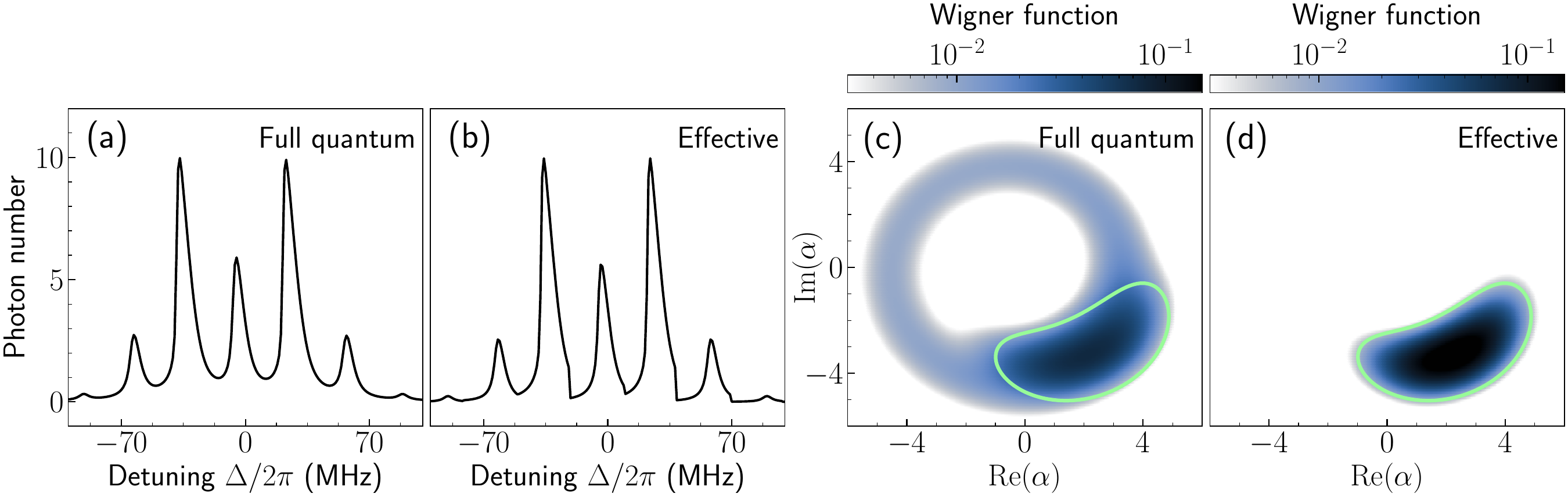}     
    \caption{\label{fig:Bessel_vs_Floquet}
    Comparison between the full Floquet analysis and the effective model derived in Appendix \ref{Appendix:Effective_model}. 
    We show the photon number $n$ computed with the Floquet steady state [panel (a)] and with the effective Hamiltonian Eq.~(3) of the main text [panel (b)].  
    The full quantum solution and the effective model exhibit a good agreement. 
    We plot in panels (c) and (d) the Wigner function $W(\alpha)= 2/\pi \cdot \operatorname{Tr}\left[\hat{D}_{\alpha} e^{i \pi \hat{a}^{\dagger} \hat{a}} \hat{D}_{\alpha}^{\dagger} \hat{\rho}\right]$, with $\hat{D}_{\alpha}=\exp(\alpha \hat{a}^{\dagger}-\alpha^* \hat{a})$, of the Floquet steady state and the effective model respectively.
    The green lines in panels (c) and (d) encircles the region where $W(\alpha) > 7\times 10^{-3}$ according to the effective model. 
    We conclude that the effective model is reliable when computing $\langle\hat{a}^\dagger \hat{a}\rangle$, i.e., the distance from the center of the distribution $W(\alpha)$. 
    On the contrary, $\langle\hat{a}\rangle$ can not be captured by the simple treatment presented in Appendix \ref{Appendix:Effective_model}, as evident from the different angular distribution of $W(\alpha)$ obtained from the full and effective model.
    All physical parameters as in Fig.~(6)~(d) of main text.
    We set $\zeta/\Omega \approx 1.67$ and $\Delta/\Omega = -1.1$.
    }
\end{figure*}

To simplify the equation of motion Eq. (2) from the main text, we want to eliminate the frequency modulation.
To do this, we use the interaction picture $\trhot = \hat{U}^\dagger(t) \rhot \hat{U}(t)$, where 
\begin{equation}
\begin{split}
\hat{U}(t) &= \mathcal T \exp\left[\int_{0}^{t} \,  -i \, dt' \,  \zeta \cos(\Omega \, t')  \hat{a}^\dagger \hat{a} \right] \\
& = \exp\left[-i \frac{\zeta}{\Omega} \sin(\Omega \, t)   \, \hat{a}^\dagger \hat{a}\right] \,.
\end{split}
\end{equation}
We obtain
\begin{equation}
\label{Eq:Liouvillian_interaction_picture}
\hbar \partial_t \trhot = - i [\hat{\tilde{H}}, \trhot] + \kappa \DD{\hat{a}} \trhot + \kappa_{\phi} \DD{\hat{a}^\dagger \hat{a}} \trhot \,,
\end{equation}
where
\begin{equation}
\label{Eq:Hamiltonian_interaction_picture}
\begin{split}
\hat{\tilde{H}}/\hbar = \Delta \hat{a}^\dagger \hat{a} & + \chi \hat{a}^\dagger \hat{a}^\dagger \hat{a} \hat{a} \\ 
 + & F \left\{ \hat{a} \exp\left[-i \frac{\zeta}{\Omega} \sin(\Omega \, t) \right] + {\rm h.c.} \right\} \,.
\end{split}
\end{equation}
Equations (\ref{Eq:Liouvillian_interaction_picture}) and (\ref{Eq:Hamiltonian_interaction_picture}) can be straightforwardly derived thanks to  
\begin{equation}
\hat{U}^\dagger(t) \, \hat{a} \, \hat{U}(t) = \hat{a} \exp\left[-i \frac{\zeta}{\Omega} \sin(\Omega \, t)\right].
\end{equation}

We finally use the Jacobi-Anger expansion, reading
\begin{equation}
e^{i z \sin \theta} \equiv \sum_{m=-\infty}^{\infty}       J_m(z)\, e^{i m \theta} \, ,
\end{equation}
where $J_m(z)$ is the $m$th Bessel function of the first kind, to obtain (up to a phase)
\begin{equation}
\begin{split}
\hat{\tilde{H}}/\hbar  & = \Delta \hat{a}^\dagger \hat{a} + \chi \hat{a}^\dagger \hat{a}^\dagger \hat{a} \hat{a}  \\ & +
\sum_{m=-\infty}^{\infty} 
F  \Bessel{m}{\frac{\zeta}{\Omega}} \left[\hat{a} e^{-i m {\Omega} t} + \hat{a}^\dagger e^{i m {\Omega} t} \right] \, .
\end{split}
\end{equation}
Notice that both dissipation and Kerr nonlinearity remain unchanged by this set of transformations.

Up to this point, no approximations have been made.
For the small-drive amplitudes considered in Fig.~3 of the main text, however, we can assume that only one of the driving frequencies is relevant, and discard the fast-rotating terms. Namely, we select only those frequencies around which $\Delta_{\bar m} = \Delta  - {\bar m} \Omega \simeq 0$, finally obtaining Eq.~(3) of the main text.

In Fig.~\ref{fig:Bessel_vs_Floquet} we benchmark the validity of the effective Hamiltonian given by Eq.~(2) of the main text for the $N=32$ device in the Duffing regime at intermediate input power. 
All the physical parameters have been chosen as in Fig.~6~(d) of the main text.
We compare the photon number $n$ computed with the Floquet steady state [Fig.~\ref{fig:Bessel_vs_Floquet} (a)] and with the effective model [Fig.~\ref{fig:Bessel_vs_Floquet} (b)] showing that the two approaches exhibit a good agreement.
While the approximation is remarkably predictive in determining the photon number, this is not the case for the coherence $\langle\hat{a}\rangle$.
In Figs.~\ref{fig:Bessel_vs_Floquet} (c-d) we compute the Wigner functions obtained from the full quantum simulation of the Floquet steady state and that obtained according to the effective model.  
While the effective model nicely reproduces the radial distribution of the Wigner function (and thus the photon number), it completely misses the phase, which remains accessible only within the full Floquet-Lindbald treatment described in Appendix \ref{subsubsec:average_Floquet}.
In both Figs.~\ref{fig:Bessel_vs_Floquet} (c) and (d) we report the contour of the effective Wigner function, showing that $W(\alpha)$ of the full quantum model contains the effective Wigner function, but the phase coherence is reduced with respect to the effective model.
We argue that these dephasing-like effects are due to higher-order processes not accounted for in the effective model, emerging from the combination of Hamiltonian and dissipative terms, and treating them would require higher-order time-dependent perturbation theories such as the Floquet-Magnus expansion.

\section{Device characterization}
\label{Appendix:device_charac}

\begin{figure*}[]
    \includegraphics[width=.9\textwidth]{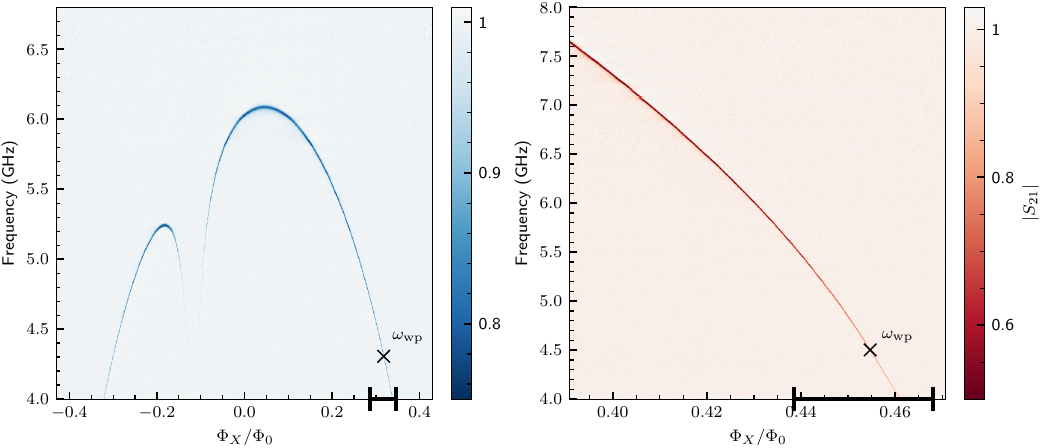} 
    \caption{\label{fig:flux_sweep}
    Measurement of the magnitude of the transmission $|\SParam|$ as a function of the flux $\Phi_X$ in the SQUID arrays for the $N=32$ Duffing device (left, blue) and the $N=10$ Kerr device (right, red).
    The flux working points used throughout the paper are indicated by cross markers.
    The black horizontal segments approximately denote the maximum flux modulation reported is in this work with respectively $\pm 0.4$ GHz and $\pm 1.0$ GHz for the Duffing and Kerr devices.
    }
\end{figure*}

The flux-dependence of the SQUID array frequency is reported for both devices in \figref{fig:flux_sweep}.
The value of the flux $\Phi_X$ is controlled by applying a direct current to the external coil.
We convert the current applied to the flux threading the SQUIDs by fitting a larger flux modulation sweep over more than one period.
The two devices are made of SQUIDs with identical junctions and have a similar total capacitance. As a consequence their maximum frequency differs due to the total number of SQUID $N$ in the two arrays.

We observe an unexplained dip in the flux modulation of device $N=32$ (blue).
This feature is periodically repeated for $\Phi_0$ increments of the flux $\Phi_X$, and we observe no hysteretic effect.
This spurious dip was observed across several cooldowns at the same position.
A similar device with $N=46$ SQUIDs located on the same chip does not show a similar dip.
Cross markers in \figref{fig:flux_sweep} indicate the flux operating point of both devices, and the segment on the x-axes show the maximum flux modulation performed in this work.
The $N=32$ SQUID array is always operated far from the unexpected feature which thus does not impact the results of the experiment.

Because of the nonlinear flux dependence of the frequency of the resonators, the applied frequency modulation is not exactly sinusoidal, $\zeta \cos (\Omega t)$, as stated in the Hamiltonian Eq.~(1) of the main text. 
Instead, we apply a sinuoisidal modulation of the flux threading the SQUID loop as
\begin{equation}
    \Phi_x(t)=\Phi_x^0 + A \cos(\Omega t),
\end{equation}
with $\Phi_x^0$ the static flux, $\Omega$ the frequency of the modulation and $A$ the amplitude of the modulation in $\Phi_0$ unit.
Following Eq.~\ref{eq:supp_freq_sq_flux}, the exact frequency modulation is thus given by 
\begin{equation}
    \omega(\Phi_x(t))=\omega_c\, \sqrt{ | \cos \left[ \pi(\Phi_x^0 + A \cos(\Omega t))/\Phi_0 \right] |}.
\end{equation}
Consequently, for large modulation strength $\zeta$, the frequency modulation is not symmetric around the value without modulation $\omega(\Phi_x^0)=\omega_{\rm wp}$.
This results in a deviation of the LZSM resonances $\bar m$ away from $\Delta_{\bar m}$.
We observe this deviation in our measurements, most clearly in Fig.~4 where $\zeta$ is as large as $\SI{1.0}{\GHz}$.
It is also apparent in Fig.~6~(d-f) of the main text where the data are systematically shifted to negative frequencies compared to the superimposed numerical simulations. 
The deviation from the LZ mode position expected for an ideal modulation is towards negative detuning because of the curvature of the flux dependence of the frequency.
The deviation increases when the flux operating point is brought closer to zero flux where the curvature is more important.
This phenomenon is reported and explained in Ref.~\cite{wuLandau2019}.

From the measurement of the room-temperature normal-state resistance, we estimate the single junction Josephson energy to $E_J/h \approx \SI{170}{\GHz}$.
The frequencies and Kerr non-linearities of the SQUID array modes can be simulated using a lumped-model and assuming $E_J\gg E_C$~\cite{weisslKerr2015}.
From this model, we estimate the plasma frequency of the junctions to $\hbar\omega_P = \sqrt{8 E_J E_C} \approx h \times \SI{39}{\GHz}$.
The zero-flux frequency of the Kerr $N=10$ device is out of our measurement bandwidth of 4-8 GHz, but we estimate it to be approximately $\SI{13}{\GHz}$ from the lumped model discussed above.
We find a single junction charging energy $E_C=e^2/2C_J \approx h\times\SI{1.1}{GHz}$.
Even for the flux operating point $\Phi_X/\Phi_0\approx 0.455$ of the $N=10$ device, the effective Josephson energy of the SQUID remains much larger than the charging energy, ensuring the validity of the Kerr approximation of the Josephson Hamiltonian.

\section{Additional experimental data}
\label{Appendix:Additional_data}

\begin{figure*}[]
    \includegraphics[width=.95\textwidth]{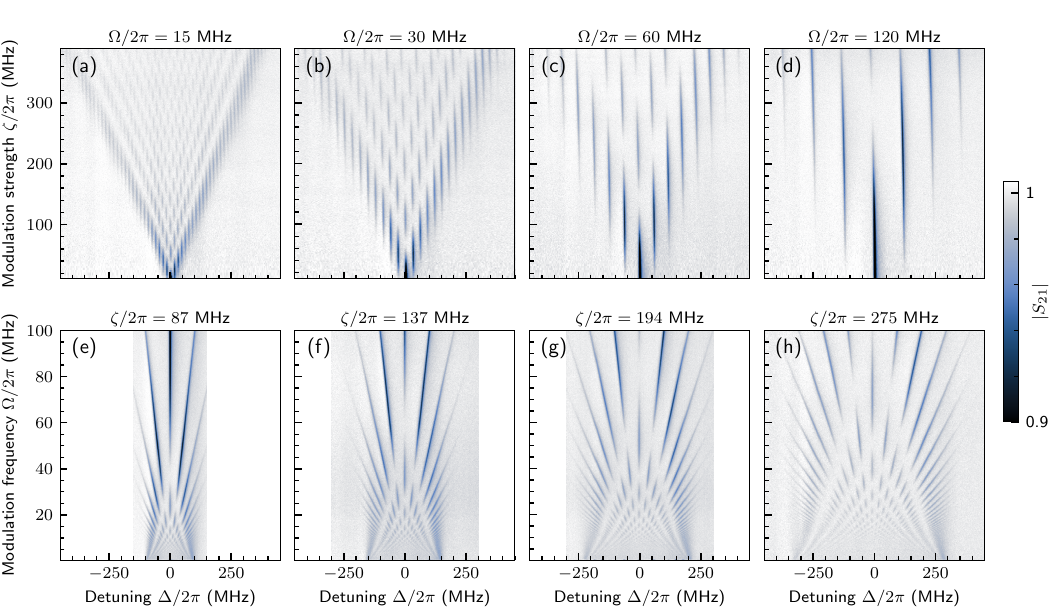} 
    \caption{\label{fig:extended_linear}
     Extended data of LZSM interference patterns in the linear regime. The measurements are performed on the device N = 32 with the same weak drive power as in Fig. 3 (d-f) of the main text. In panels (a-d), the same sweep of modulation strength $\zeta$ is repeated for increasing modulation frequencies $\Omega$. In panels (e-h), the same sweep of modulation frequency $\Omega$ is repeated for increasing modulation strengths $\zeta$. These measurements highlight the exquisite control over both the frequency spacing and the number of resonances offered by the platform, with for instance LZSM resonances up to $m = \pm 25$ visible in (a).
    }
\end{figure*}

\begin{figure*}[]
    \includegraphics[width=1\textwidth]{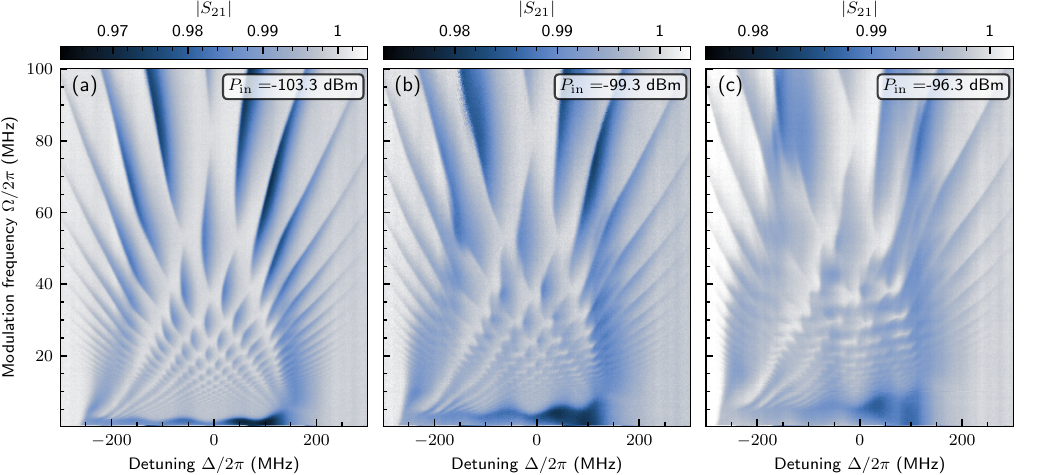} 
    \caption{\label{fig:extended_Duffing_Omega}
    Extended data of LZSM interferometry in the Duffing regime measured with the $N=32$ device.
    The magnitude of $\SParam$ is measured versus $\Delta$ and $\Omega$ for increasing drive power $P_{\rm in}$.
    The modulation strength is fixed to $\zeta/2\pi=206$ MHz.
    }
\end{figure*}

In this section, we report additional measurements performed with the $N=32$ Duffing device in the linear regime.
In \figref{fig:extended_linear} we repeat the linear regime LZSM interferometry measurements of Fig.~3~(d-f) of the main text for different values of $\Omega$ and $\zeta$.
The drive power is set to the same low value to remain in the linear regime with a low photon occupation number. 
In panels (a-d), we sweep the modulation strength $\zeta$ for increasing values of modulation frequency $\Omega$.
LZSM resonances are visible for $\zeta \geq |\Delta| $, irrespective of the value of $\Omega$.
As expected from Eq.~(3), the spacing between LZSM resonances is equal to $\Omega$. 
In panels (e-h), we sweep the modulation frequency $\Omega$ for increasing modulation strengths $\zeta$.
We observe more and more LZSM resonances as $\zeta$ is increased, and again the extension of the resonances is approximately confined to $|\Delta|<\zeta$.
These measurements highlight the superb control offered by the platform on the position and number of modes, as for instance in panel (a) we observe clearly LZSM resonances up to mode $\bar m=25$.

Next, in \figref{fig:extended_Duffing_Omega} we repeat the Duffing regime LZSM interferometry of Fig.~6 of the main text, but this time sweeping the modulation frequency $\Omega$ at fixed $\zeta$. 
We repeat this measurement for three increasing values of drive power $P_{\rm in}$.
For the lowest drive shown in panel (a), individual LZSM resonances remain mostly isolated.
However, when compared to the linear regime of weak drive, LZSM resonances appear distorted with a rounded shape.
This rounding is a combination of the Kerr nonlinearity bending the peak to negative frequencies, and the modulation of the effective drives $F_{\bar m}$ that controls the peak bendings.
For increasing drive power, as shown in panels (b) and (c), the interference pattern gets more distorted and individual resonances start merging together.
In this regime, the effective model of Eq.~(3) is no longer valid and a full Floquet-Lindblad treatment is required.  
As we have theoretically shown, the broadening and distortion of the LZSM interference pattern in the Duffing regime is associated to a dissipative quantum chaotic phase.


\begin{thebibliography}{131}%
\makeatletter
\providecommand \@ifxundefined [1]{%
 \@ifx{#1\undefined}
}%
\providecommand \@ifnum [1]{%
 \ifnum #1\expandafter \@firstoftwo
 \else \expandafter \@secondoftwo
 \fi
}%
\providecommand \@ifx [1]{%
 \ifx #1\expandafter \@firstoftwo
 \else \expandafter \@secondoftwo
 \fi
}%
\providecommand \natexlab [1]{#1}%
\providecommand \enquote  [1]{``#1''}%
\providecommand \bibnamefont  [1]{#1}%
\providecommand \bibfnamefont [1]{#1}%
\providecommand \citenamefont [1]{#1}%
\providecommand \href@noop [0]{\@secondoftwo}%
\providecommand \href [0]{\begingroup \@sanitize@url \@href}%
\providecommand \@href[1]{\@@startlink{#1}\@@href}%
\providecommand \@@href[1]{\endgroup#1\@@endlink}%
\providecommand \@sanitize@url [0]{\catcode `\\12\catcode `\$12\catcode
  `\&12\catcode `\#12\catcode `\^12\catcode `\_12\catcode `\%12\relax}%
\providecommand \@@startlink[1]{}%
\providecommand \@@endlink[0]{}%
\providecommand \url  [0]{\begingroup\@sanitize@url \@url }%
\providecommand \@url [1]{\endgroup\@href {#1}{\urlprefix }}%
\providecommand \urlprefix  [0]{URL }%
\providecommand \Eprint [0]{\href }%
\providecommand \doibase [0]{http://dx.doi.org/}%
\providecommand \selectlanguage [0]{\@gobble}%
\providecommand \bibinfo  [0]{\@secondoftwo}%
\providecommand \bibfield  [0]{\@secondoftwo}%
\providecommand \translation [1]{[#1]}%
\providecommand \BibitemOpen [0]{}%
\providecommand \bibitemStop [0]{}%
\providecommand \bibitemNoStop [0]{.\EOS\space}%
\providecommand \EOS [0]{\spacefactor3000\relax}%
\providecommand \BibitemShut  [1]{\csname bibitem#1\endcsname}%
\let\auto@bib@innerbib\@empty
\bibitem [{\citenamefont {Blais}\ \emph {et~al.}(2021)\citenamefont {Blais},
  \citenamefont {Grimsmo}, \citenamefont {Girvin},\ and\ \citenamefont
  {Wallraff}}]{BlaisRMP21}%
  \BibitemOpen
  \bibfield  {author} {\bibinfo {author} {\bibfnamefont {Alexandre}\
  \bibnamefont {Blais}}, \bibinfo {author} {\bibfnamefont {Arne~L.}\
  \bibnamefont {Grimsmo}}, \bibinfo {author} {\bibfnamefont {S.~M.}\
  \bibnamefont {Girvin}}, \ and\ \bibinfo {author} {\bibfnamefont {Andreas}\
  \bibnamefont {Wallraff}},\ }\bibfield  {title} {\enquote {\bibinfo {title}
  {Circuit quantum electrodynamics},}\ }\href {\doibase
  10.1103/RevModPhys.93.025005} {\bibfield  {journal} {\bibinfo  {journal}
  {Rev. Mod. Phys.}\ }\textbf {\bibinfo {volume} {93}},\ \bibinfo {pages}
  {025005} (\bibinfo {year} {2021})}\BibitemShut {NoStop}%
\bibitem [{\citenamefont {Altman}\ \emph {et~al.}(2021)\citenamefont {Altman},
  \citenamefont {Brown}, \citenamefont {Carleo}, \citenamefont {Carr},
  \citenamefont {Demler}, \citenamefont {Chin}, \citenamefont {DeMarco},
  \citenamefont {Economou}, \citenamefont {Eriksson}, \citenamefont {Fu},
  \citenamefont {Greiner}, \citenamefont {Hazzard}, \citenamefont {Hulet},
  \citenamefont {Koll\'ar}, \citenamefont {Lev}, \citenamefont {Lukin},
  \citenamefont {Ma}, \citenamefont {Mi}, \citenamefont {Misra}, \citenamefont
  {Monroe}, \citenamefont {Murch}, \citenamefont {Nazario}, \citenamefont {Ni},
  \citenamefont {Potter}, \citenamefont {Roushan}, \citenamefont {Saffman},
  \citenamefont {Schleier-Smith}, \citenamefont {Siddiqi}, \citenamefont
  {Simmonds}, \citenamefont {Singh}, \citenamefont {Spielman}, \citenamefont
  {Temme}, \citenamefont {Weiss}, \citenamefont {Vu\ifmmode \check{c}\else
  \v{c}\fi{}kovi\ifmmode~\acute{c}\else \'{c}\fi{}}, \citenamefont
  {Vuleti\ifmmode~\acute{c}\else \'{c}\fi{}}, \citenamefont {Ye},\ and\
  \citenamefont {Zwierlein}}]{AltmanPRXQ21}%
  \BibitemOpen
  \bibfield  {author} {\bibinfo {author} {\bibfnamefont {Ehud}\ \bibnamefont
  {Altman}}, \bibinfo {author} {\bibfnamefont {Kenneth~R.}\ \bibnamefont
  {Brown}}, \bibinfo {author} {\bibfnamefont {Giuseppe}\ \bibnamefont
  {Carleo}}, \bibinfo {author} {\bibfnamefont {Lincoln~D.}\ \bibnamefont
  {Carr}}, \bibinfo {author} {\bibfnamefont {Eugene}\ \bibnamefont {Demler}},
  \bibinfo {author} {\bibfnamefont {Cheng}\ \bibnamefont {Chin}}, \bibinfo
  {author} {\bibfnamefont {Brian}\ \bibnamefont {DeMarco}}, \bibinfo {author}
  {\bibfnamefont {Sophia~E.}\ \bibnamefont {Economou}}, \bibinfo {author}
  {\bibfnamefont {Mark~A.}\ \bibnamefont {Eriksson}}, \bibinfo {author}
  {\bibfnamefont {Kai-Mei~C.}\ \bibnamefont {Fu}}, \bibinfo {author}
  {\bibfnamefont {Markus}\ \bibnamefont {Greiner}}, \bibinfo {author}
  {\bibfnamefont {Kaden~R.A.}\ \bibnamefont {Hazzard}}, \bibinfo {author}
  {\bibfnamefont {Randall~G.}\ \bibnamefont {Hulet}}, \bibinfo {author}
  {\bibfnamefont {Alicia~J.}\ \bibnamefont {Koll\'ar}}, \bibinfo {author}
  {\bibfnamefont {Benjamin~L.}\ \bibnamefont {Lev}}, \bibinfo {author}
  {\bibfnamefont {Mikhail~D.}\ \bibnamefont {Lukin}}, \bibinfo {author}
  {\bibfnamefont {Ruichao}\ \bibnamefont {Ma}}, \bibinfo {author}
  {\bibfnamefont {Xiao}\ \bibnamefont {Mi}}, \bibinfo {author} {\bibfnamefont
  {Shashank}\ \bibnamefont {Misra}}, \bibinfo {author} {\bibfnamefont
  {Christopher}\ \bibnamefont {Monroe}}, \bibinfo {author} {\bibfnamefont
  {Kater}\ \bibnamefont {Murch}}, \bibinfo {author} {\bibfnamefont {Zaira}\
  \bibnamefont {Nazario}}, \bibinfo {author} {\bibfnamefont {Kang-Kuen}\
  \bibnamefont {Ni}}, \bibinfo {author} {\bibfnamefont {Andrew~C.}\
  \bibnamefont {Potter}}, \bibinfo {author} {\bibfnamefont {Pedram}\
  \bibnamefont {Roushan}}, \bibinfo {author} {\bibfnamefont {Mark}\
  \bibnamefont {Saffman}}, \bibinfo {author} {\bibfnamefont {Monika}\
  \bibnamefont {Schleier-Smith}}, \bibinfo {author} {\bibfnamefont {Irfan}\
  \bibnamefont {Siddiqi}}, \bibinfo {author} {\bibfnamefont {Raymond}\
  \bibnamefont {Simmonds}}, \bibinfo {author} {\bibfnamefont {Meenakshi}\
  \bibnamefont {Singh}}, \bibinfo {author} {\bibfnamefont {I.B.}\ \bibnamefont
  {Spielman}}, \bibinfo {author} {\bibfnamefont {Kristan}\ \bibnamefont
  {Temme}}, \bibinfo {author} {\bibfnamefont {David~S.}\ \bibnamefont {Weiss}},
  \bibinfo {author} {\bibfnamefont {Jelena}\ \bibnamefont {Vu\ifmmode
  \check{c}\else \v{c}\fi{}kovi\ifmmode~\acute{c}\else \'{c}\fi{}}}, \bibinfo
  {author} {\bibfnamefont {Vladan}\ \bibnamefont {Vuleti\ifmmode~\acute{c}\else
  \'{c}\fi{}}}, \bibinfo {author} {\bibfnamefont {Jun}\ \bibnamefont {Ye}}, \
  and\ \bibinfo {author} {\bibfnamefont {Martin}\ \bibnamefont {Zwierlein}},\
  }\bibfield  {title} {\enquote {\bibinfo {title} {Quantum simulators:
  Architectures and opportunities},}\ }\href {\doibase
  10.1103/PRXQuantum.2.017003} {\bibfield  {journal} {\bibinfo  {journal} {PRX
  Quantum}\ }\textbf {\bibinfo {volume} {2}},\ \bibinfo {pages} {017003}
  (\bibinfo {year} {2021})}\BibitemShut {NoStop}%
\bibitem [{\citenamefont {Landau}(1932)}]{landau1932theorie}%
  \BibitemOpen
  \bibfield  {author} {\bibinfo {author} {\bibfnamefont {Lev}\ \bibnamefont
  {Landau}},\ }\bibfield  {title} {\enquote {\bibinfo {title} {Zur theorie der
  energieubertragung. ii},}\ }\href@noop {} {\bibfield  {journal} {\bibinfo
  {journal} {Physikalische Zeitschrift der Sowjetunion}\ }\textbf {\bibinfo
  {volume} {2}},\ \bibinfo {pages} {46} (\bibinfo {year} {1932})}\BibitemShut
  {NoStop}%
\bibitem [{\citenamefont {Zener}(1932)}]{zener1932non}%
  \BibitemOpen
  \bibfield  {author} {\bibinfo {author} {\bibfnamefont {Clarence}\
  \bibnamefont {Zener}},\ }\bibfield  {title} {\enquote {\bibinfo {title}
  {Non-adiabatic crossing of energy levels},}\ }\href@noop {} {\bibfield
  {journal} {\bibinfo  {journal} {Proceedings of the Royal Society of London.
  Series A, Containing Papers of a Mathematical and Physical Character}\
  }\textbf {\bibinfo {volume} {137}},\ \bibinfo {pages} {696--702} (\bibinfo
  {year} {1932})}\BibitemShut {NoStop}%
\bibitem [{\citenamefont {St{\"u}ckelberg}(1932)}]{stuckelberg1932theorie}%
  \BibitemOpen
  \bibfield  {author} {\bibinfo {author} {\bibfnamefont {ECG}\ \bibnamefont
  {St{\"u}ckelberg}},\ }\bibfield  {title} {\enquote {\bibinfo {title} {Theorie
  der unelastischen st{\"o}sse zwischen atomen},}\ }\href@noop {} {\bibfield
  {journal} {\bibinfo  {journal} {Helv. Phys. Acta}\ }\textbf {\bibinfo
  {volume} {5}},\ \bibinfo {pages} {369} (\bibinfo {year} {1932})}\BibitemShut
  {NoStop}%
\bibitem [{\citenamefont {Majorana}(1932)}]{majorana1932atomi}%
  \BibitemOpen
  \bibfield  {author} {\bibinfo {author} {\bibfnamefont {Ettore}\ \bibnamefont
  {Majorana}},\ }\bibfield  {title} {\enquote {\bibinfo {title} {Atomi
  orientati in campo magnetico variabile},}\ }\href@noop {} {\bibfield
  {journal} {\bibinfo  {journal} {Il Nuovo Cimento (1924-1942)}\ }\textbf
  {\bibinfo {volume} {9}},\ \bibinfo {pages} {43--50} (\bibinfo {year}
  {1932})}\BibitemShut {NoStop}%
\bibitem [{\citenamefont {Ivakhnenko}\ \emph {et~al.}(2023)\citenamefont
  {Ivakhnenko}, \citenamefont {Shevchenko},\ and\ \citenamefont
  {Nori}}]{IvakhnenkoPHYSREP23}%
  \BibitemOpen
  \bibfield  {author} {\bibinfo {author} {\bibfnamefont {Oleh~V.}\ \bibnamefont
  {Ivakhnenko}}, \bibinfo {author} {\bibfnamefont {Sergey~N.}\ \bibnamefont
  {Shevchenko}}, \ and\ \bibinfo {author} {\bibfnamefont {Franco}\ \bibnamefont
  {Nori}},\ }\bibfield  {title} {\enquote {\bibinfo {title} {Nonadiabatic
  landau–zener–st\"{u}ckelberg–majorana transitions, dynamics, and
  interference},}\ }\href {\doibase 10.1016/j.physrep.2022.10.002} {\bibfield
  {journal} {\bibinfo  {journal} {Physics Reports}\ }\textbf {\bibinfo {volume}
  {995}},\ \bibinfo {pages} {1–89} (\bibinfo {year} {2023})}\BibitemShut
  {NoStop}%
\bibitem [{\citenamefont {Oliver}\ \emph {et~al.}(2005)\citenamefont {Oliver},
  \citenamefont {Yu}, \citenamefont {Lee}, \citenamefont {Berggren},
  \citenamefont {Levitov},\ and\ \citenamefont {Orlando}}]{OliverSCIENCE05}%
  \BibitemOpen
  \bibfield  {author} {\bibinfo {author} {\bibfnamefont {William~D.}\
  \bibnamefont {Oliver}}, \bibinfo {author} {\bibfnamefont {Yang}\ \bibnamefont
  {Yu}}, \bibinfo {author} {\bibfnamefont {Janice~C.}\ \bibnamefont {Lee}},
  \bibinfo {author} {\bibfnamefont {Karl~K.}\ \bibnamefont {Berggren}},
  \bibinfo {author} {\bibfnamefont {Leonid~S.}\ \bibnamefont {Levitov}}, \ and\
  \bibinfo {author} {\bibfnamefont {Terry~P.}\ \bibnamefont {Orlando}},\
  }\bibfield  {title} {\enquote {\bibinfo {title} {Mach-{Z}ehnder
  interferometry in a strongly driven superconducting qubit},}\ }\href
  {\doibase 10.1126/science.1119678} {\bibfield  {journal} {\bibinfo  {journal}
  {Science}\ }\textbf {\bibinfo {volume} {310}},\ \bibinfo {pages}
  {1653–1657} (\bibinfo {year} {2005})}\BibitemShut {NoStop}%
\bibitem [{\citenamefont {Sillanp{\"a}{\"a}}\ \emph {et~al.}(2006)\citenamefont
  {Sillanp{\"a}{\"a}}, \citenamefont {Lehtinen}, \citenamefont {Paila},
  \citenamefont {Makhlin},\ and\ \citenamefont
  {Hakonen}}]{sillanpaaContinuousTime2006}%
  \BibitemOpen
  \bibfield  {author} {\bibinfo {author} {\bibfnamefont {Mika}\ \bibnamefont
  {Sillanp{\"a}{\"a}}}, \bibinfo {author} {\bibfnamefont {Teijo}\ \bibnamefont
  {Lehtinen}}, \bibinfo {author} {\bibfnamefont {Antti}\ \bibnamefont {Paila}},
  \bibinfo {author} {\bibfnamefont {Yuriy}\ \bibnamefont {Makhlin}}, \ and\
  \bibinfo {author} {\bibfnamefont {Pertti}\ \bibnamefont {Hakonen}},\
  }\bibfield  {title} {\enquote {\bibinfo {title} {Continuous-{{Time
  Monitoring}} of {{Landau-Zener Interference}} in a {{Cooper-Pair Box}}},}\
  }\href {\doibase 10.1103/PhysRevLett.96.187002} {\bibfield  {journal}
  {\bibinfo  {journal} {Physical Review Letters}\ }\textbf {\bibinfo {volume}
  {96}},\ \bibinfo {pages} {187002} (\bibinfo {year} {2006})}\BibitemShut
  {NoStop}%
\bibitem [{\citenamefont {Stehlik}\ \emph {et~al.}(2012)\citenamefont
  {Stehlik}, \citenamefont {Dovzhenko}, \citenamefont {Petta}, \citenamefont
  {Johansson}, \citenamefont {Nori}, \citenamefont {Lu},\ and\ \citenamefont
  {Gossard}}]{stehlikLandauZenerSt2012}%
  \BibitemOpen
  \bibfield  {author} {\bibinfo {author} {\bibfnamefont {J.}~\bibnamefont
  {Stehlik}}, \bibinfo {author} {\bibfnamefont {Y.}~\bibnamefont {Dovzhenko}},
  \bibinfo {author} {\bibfnamefont {J.~R.}\ \bibnamefont {Petta}}, \bibinfo
  {author} {\bibfnamefont {J.~R.}\ \bibnamefont {Johansson}}, \bibinfo {author}
  {\bibfnamefont {F.}~\bibnamefont {Nori}}, \bibinfo {author} {\bibfnamefont
  {H.}~\bibnamefont {Lu}}, \ and\ \bibinfo {author} {\bibfnamefont {A.~C.}\
  \bibnamefont {Gossard}},\ }\bibfield  {title} {\enquote {\bibinfo {title}
  {Landau-{Z}ener-{S}t{\"u}ckelberg interferometry of a single electron charge
  qubit},}\ }\href {\doibase 10.1103/PhysRevB.86.121303} {\bibfield  {journal}
  {\bibinfo  {journal} {Physical Review B}\ }\textbf {\bibinfo {volume} {86}},\
  \bibinfo {pages} {121303} (\bibinfo {year} {2012})}\BibitemShut {NoStop}%
\bibitem [{\citenamefont {Forster}\ \emph {et~al.}(2014)\citenamefont
  {Forster}, \citenamefont {Petersen}, \citenamefont {Manus}, \citenamefont
  {H{\"a}nggi}, \citenamefont {Schuh}, \citenamefont {Wegscheider},
  \citenamefont {Kohler},\ and\ \citenamefont
  {Ludwig}}]{forsterCharacterization2014}%
  \BibitemOpen
  \bibfield  {author} {\bibinfo {author} {\bibfnamefont {F.}~\bibnamefont
  {Forster}}, \bibinfo {author} {\bibfnamefont {G.}~\bibnamefont {Petersen}},
  \bibinfo {author} {\bibfnamefont {S.}~\bibnamefont {Manus}}, \bibinfo
  {author} {\bibfnamefont {P.}~\bibnamefont {H{\"a}nggi}}, \bibinfo {author}
  {\bibfnamefont {D.}~\bibnamefont {Schuh}}, \bibinfo {author} {\bibfnamefont
  {W.}~\bibnamefont {Wegscheider}}, \bibinfo {author} {\bibfnamefont
  {S.}~\bibnamefont {Kohler}}, \ and\ \bibinfo {author} {\bibfnamefont
  {S.}~\bibnamefont {Ludwig}},\ }\bibfield  {title} {\enquote {\bibinfo {title}
  {Characterization of {{Qubit Dephasing}} by
  {{Landau-Zener-St}}{\"u}ckelberg-{{Majorana Interferometry}}},}\ }\href
  {\doibase 10.1103/PhysRevLett.112.116803} {\bibfield  {journal} {\bibinfo
  {journal} {Physical Review Letters}\ }\textbf {\bibinfo {volume} {112}},\
  \bibinfo {pages} {116803} (\bibinfo {year} {2014})}\BibitemShut {NoStop}%
\bibitem [{\citenamefont {Childress}\ and\ \citenamefont
  {McIntyre}(2010)}]{childressMultifrequency2010}%
  \BibitemOpen
  \bibfield  {author} {\bibinfo {author} {\bibfnamefont {Lilian}\ \bibnamefont
  {Childress}}\ and\ \bibinfo {author} {\bibfnamefont {Jean}\ \bibnamefont
  {McIntyre}},\ }\bibfield  {title} {\enquote {\bibinfo {title} {Multifrequency
  spin resonance in diamond},}\ }\href {\doibase 10.1103/PhysRevA.82.033839}
  {\bibfield  {journal} {\bibinfo  {journal} {Physical Review A}\ }\textbf
  {\bibinfo {volume} {82}},\ \bibinfo {pages} {033839} (\bibinfo {year}
  {2010})}\BibitemShut {NoStop}%
\bibitem [{\citenamefont {Niepce}\ \emph {et~al.}(2021)\citenamefont {Niepce},
  \citenamefont {Burnett}, \citenamefont {Kudra}, \citenamefont {Cole},\ and\
  \citenamefont {Bylander}}]{niepce_stability_2021}%
  \BibitemOpen
  \bibfield  {author} {\bibinfo {author} {\bibfnamefont {David}\ \bibnamefont
  {Niepce}}, \bibinfo {author} {\bibfnamefont {Jonathan~J.}\ \bibnamefont
  {Burnett}}, \bibinfo {author} {\bibfnamefont {Marina}\ \bibnamefont {Kudra}},
  \bibinfo {author} {\bibfnamefont {Jared~H.}\ \bibnamefont {Cole}}, \ and\
  \bibinfo {author} {\bibfnamefont {Jonas}\ \bibnamefont {Bylander}},\
  }\bibfield  {title} {\enquote {\bibinfo {title} {Stability of superconducting
  resonators: Motional narrowing and the role of {L}andau-{Z}ener driving of
  two-level defects},}\ }\href {\doibase 10.1126/sciadv.abh0462} {\bibfield
  {journal} {\bibinfo  {journal} {Science Advances}\ }\textbf {\bibinfo
  {volume} {7}} (\bibinfo {year} {2021}),\ 10.1126/sciadv.abh0462}\BibitemShut
  {NoStop}%
\bibitem [{\citenamefont {{Dupont-Ferrier}}\ \emph {et~al.}(2013)\citenamefont
  {{Dupont-Ferrier}}, \citenamefont {Roche}, \citenamefont {Voisin},
  \citenamefont {Jehl}, \citenamefont {Wacquez}, \citenamefont {Vinet},
  \citenamefont {Sanquer},\ and\ \citenamefont
  {De~Franceschi}}]{dupont-ferrierCoherent2013}%
  \BibitemOpen
  \bibfield  {author} {\bibinfo {author} {\bibfnamefont {E.}~\bibnamefont
  {{Dupont-Ferrier}}}, \bibinfo {author} {\bibfnamefont {B.}~\bibnamefont
  {Roche}}, \bibinfo {author} {\bibfnamefont {B.}~\bibnamefont {Voisin}},
  \bibinfo {author} {\bibfnamefont {X.}~\bibnamefont {Jehl}}, \bibinfo {author}
  {\bibfnamefont {R.}~\bibnamefont {Wacquez}}, \bibinfo {author} {\bibfnamefont
  {M.}~\bibnamefont {Vinet}}, \bibinfo {author} {\bibfnamefont
  {M.}~\bibnamefont {Sanquer}}, \ and\ \bibinfo {author} {\bibfnamefont
  {S.}~\bibnamefont {De~Franceschi}},\ }\bibfield  {title} {\enquote {\bibinfo
  {title} {Coherent {{Coupling}} of {{Two Dopants}} in a {{Silicon Nanowire
  Probed}} by {{Landau-Zener-St}}{\"u}ckelberg {{Interferometry}}},}\ }\href
  {\doibase 10.1103/PhysRevLett.110.136802} {\bibfield  {journal} {\bibinfo
  {journal} {Physical Review Letters}\ }\textbf {\bibinfo {volume} {110}},\
  \bibinfo {pages} {136802} (\bibinfo {year} {2013})}\BibitemShut {NoStop}%
\bibitem [{\citenamefont {He}\ \emph {et~al.}(2024)\citenamefont {He},
  \citenamefont {Pan}, \citenamefont {Liu}, \citenamefont {Lyu}, \citenamefont
  {Jia}, \citenamefont {Yang}, \citenamefont {Zhu}, \citenamefont {Liu},
  \citenamefont {Shen}, \citenamefont {Shevchenko}, \citenamefont {Nori},
  \citenamefont {Zhao}, \citenamefont {Lu},\ and\ \citenamefont
  {Qu}}]{heQuantifying2024}%
  \BibitemOpen
  \bibfield  {author} {\bibinfo {author} {\bibfnamefont {Jiangbo}\ \bibnamefont
  {He}}, \bibinfo {author} {\bibfnamefont {Dong}\ \bibnamefont {Pan}}, \bibinfo
  {author} {\bibfnamefont {Mingli}\ \bibnamefont {Liu}}, \bibinfo {author}
  {\bibfnamefont {Zhaozheng}\ \bibnamefont {Lyu}}, \bibinfo {author}
  {\bibfnamefont {Zhongmou}\ \bibnamefont {Jia}}, \bibinfo {author}
  {\bibfnamefont {Guang}\ \bibnamefont {Yang}}, \bibinfo {author}
  {\bibfnamefont {Shang}\ \bibnamefont {Zhu}}, \bibinfo {author} {\bibfnamefont
  {Guangtong}\ \bibnamefont {Liu}}, \bibinfo {author} {\bibfnamefont {Jie}\
  \bibnamefont {Shen}}, \bibinfo {author} {\bibfnamefont {Sergey~N.}\
  \bibnamefont {Shevchenko}}, \bibinfo {author} {\bibfnamefont {Franco}\
  \bibnamefont {Nori}}, \bibinfo {author} {\bibfnamefont {Jianhua}\
  \bibnamefont {Zhao}}, \bibinfo {author} {\bibfnamefont {Li}~\bibnamefont
  {Lu}}, \ and\ \bibinfo {author} {\bibfnamefont {Fanming}\ \bibnamefont
  {Qu}},\ }\bibfield  {title} {\enquote {\bibinfo {title} {Quantifying quantum
  coherence of multiple-charge states in tunable {{Josephson}} junctions},}\
  }\href {\doibase 10.1038/s41534-023-00798-2} {\bibfield  {journal} {\bibinfo
  {journal} {npj Quantum Information}\ }\textbf {\bibinfo {volume} {10}},\
  \bibinfo {pages} {1--8} (\bibinfo {year} {2024})}\BibitemShut {NoStop}%
\bibitem [{\citenamefont {Cao}\ \emph {et~al.}(2013)\citenamefont {Cao},
  \citenamefont {Li}, \citenamefont {Tu}, \citenamefont {Wang}, \citenamefont
  {Zhou}, \citenamefont {Xiao}, \citenamefont {Guo}, \citenamefont {Jiang},\
  and\ \citenamefont {Guo}}]{caoUltrafast2013}%
  \BibitemOpen
  \bibfield  {author} {\bibinfo {author} {\bibfnamefont {Gang}\ \bibnamefont
  {Cao}}, \bibinfo {author} {\bibfnamefont {Hai-Ou}\ \bibnamefont {Li}},
  \bibinfo {author} {\bibfnamefont {Tao}\ \bibnamefont {Tu}}, \bibinfo {author}
  {\bibfnamefont {Li}~\bibnamefont {Wang}}, \bibinfo {author} {\bibfnamefont
  {Cheng}\ \bibnamefont {Zhou}}, \bibinfo {author} {\bibfnamefont {Ming}\
  \bibnamefont {Xiao}}, \bibinfo {author} {\bibfnamefont {Guang-Can}\
  \bibnamefont {Guo}}, \bibinfo {author} {\bibfnamefont {Hong-Wen}\
  \bibnamefont {Jiang}}, \ and\ \bibinfo {author} {\bibfnamefont {Guo-Ping}\
  \bibnamefont {Guo}},\ }\bibfield  {title} {\enquote {\bibinfo {title}
  {Ultrafast universal quantum control of a quantum-dot charge qubit using
  {{Landau}}--{{Zener}}--{{St{\"u}ckelberg}} interference},}\ }\href {\doibase
  10.1038/ncomms2412} {\bibfield  {journal} {\bibinfo  {journal} {Nature
  Communications}\ }\textbf {\bibinfo {volume} {4}},\ \bibinfo {pages} {1401}
  (\bibinfo {year} {2013})}\BibitemShut {NoStop}%
\bibitem [{\citenamefont {Chatterjee}\ \emph {et~al.}(2018)\citenamefont
  {Chatterjee}, \citenamefont {Shevchenko}, \citenamefont {Barraud},
  \citenamefont {Otxoa}, \citenamefont {Nori}, \citenamefont {Morton},\ and\
  \citenamefont {{Gonzalez-Zalba}}}]{chatterjeesiliconbased2018}%
  \BibitemOpen
  \bibfield  {author} {\bibinfo {author} {\bibfnamefont {Anasua}\ \bibnamefont
  {Chatterjee}}, \bibinfo {author} {\bibfnamefont {Sergey~N.}\ \bibnamefont
  {Shevchenko}}, \bibinfo {author} {\bibfnamefont {Sylvain}\ \bibnamefont
  {Barraud}}, \bibinfo {author} {\bibfnamefont {Rub{\'e}n~M.}\ \bibnamefont
  {Otxoa}}, \bibinfo {author} {\bibfnamefont {Franco}\ \bibnamefont {Nori}},
  \bibinfo {author} {\bibfnamefont {John J.~L.}\ \bibnamefont {Morton}}, \ and\
  \bibinfo {author} {\bibfnamefont {M.~Fernando}\ \bibnamefont
  {{Gonzalez-Zalba}}},\ }\bibfield  {title} {\enquote {\bibinfo {title} {A
  silicon-based single-electron interferometer coupled to a fermionic sea},}\
  }\href {\doibase 10.1103/PhysRevB.97.045405} {\bibfield  {journal} {\bibinfo
  {journal} {Physical Review B}\ }\textbf {\bibinfo {volume} {97}},\ \bibinfo
  {pages} {045405} (\bibinfo {year} {2018})}\BibitemShut {NoStop}%
\bibitem [{\citenamefont {Bogan}\ \emph {et~al.}(2018)\citenamefont {Bogan},
  \citenamefont {Studenikin}, \citenamefont {Korkusinski}, \citenamefont
  {Gaudreau}, \citenamefont {Zawadzki}, \citenamefont {Sachrajda},
  \citenamefont {Tracy}, \citenamefont {Reno},\ and\ \citenamefont
  {Hargett}}]{boganLandauZenerSt2018}%
  \BibitemOpen
  \bibfield  {author} {\bibinfo {author} {\bibfnamefont {Alex}\ \bibnamefont
  {Bogan}}, \bibinfo {author} {\bibfnamefont {Sergei}\ \bibnamefont
  {Studenikin}}, \bibinfo {author} {\bibfnamefont {Marek}\ \bibnamefont
  {Korkusinski}}, \bibinfo {author} {\bibfnamefont {Louis}\ \bibnamefont
  {Gaudreau}}, \bibinfo {author} {\bibfnamefont {Piotr}\ \bibnamefont
  {Zawadzki}}, \bibinfo {author} {\bibfnamefont {Andy~S.}\ \bibnamefont
  {Sachrajda}}, \bibinfo {author} {\bibfnamefont {Lisa}\ \bibnamefont {Tracy}},
  \bibinfo {author} {\bibfnamefont {John}\ \bibnamefont {Reno}}, \ and\
  \bibinfo {author} {\bibfnamefont {Terry}\ \bibnamefont {Hargett}},\
  }\bibfield  {title} {\enquote {\bibinfo {title}
  {Landau-{{Zener-St}}{\textbackslash}"uckelberg-{{Majorana Interferometry}} of
  a {{Single Hole}}},}\ }\href {\doibase 10.1103/PhysRevLett.120.207701}
  {\bibfield  {journal} {\bibinfo  {journal} {Physical Review Letters}\
  }\textbf {\bibinfo {volume} {120}},\ \bibinfo {pages} {207701} (\bibinfo
  {year} {2018})}\BibitemShut {NoStop}%
\bibitem [{\citenamefont {Khomitsky}\ and\ \citenamefont
  {Studenikin}(2022)}]{khomitskySinglespin2022}%
  \BibitemOpen
  \bibfield  {author} {\bibinfo {author} {\bibfnamefont {D.~V.}\ \bibnamefont
  {Khomitsky}}\ and\ \bibinfo {author} {\bibfnamefont {S.~A.}\ \bibnamefont
  {Studenikin}},\ }\bibfield  {title} {\enquote {\bibinfo {title} {Single-spin
  {{Landau-Zener-St}}{\textbackslash}"uckelberg-{{Majorana}} interferometry of
  {{Zeeman-split}} states with strong spin-orbit interaction in a double
  quantum dot},}\ }\href {\doibase 10.1103/PhysRevB.106.195414} {\bibfield
  {journal} {\bibinfo  {journal} {Physical Review B}\ }\textbf {\bibinfo
  {volume} {106}},\ \bibinfo {pages} {195414} (\bibinfo {year}
  {2022})}\BibitemShut {NoStop}%
\bibitem [{\citenamefont {Khomitsky}\ \emph {et~al.}(2023)\citenamefont
  {Khomitsky}, \citenamefont {Bastrakova}, \citenamefont {Munyaev},
  \citenamefont {Zaprudnov},\ and\ \citenamefont
  {Studenikin}}]{khomitskyControllable2023}%
  \BibitemOpen
  \bibfield  {author} {\bibinfo {author} {\bibfnamefont {D.~V.}\ \bibnamefont
  {Khomitsky}}, \bibinfo {author} {\bibfnamefont {M.~V.}\ \bibnamefont
  {Bastrakova}}, \bibinfo {author} {\bibfnamefont {V.~O.}\ \bibnamefont
  {Munyaev}}, \bibinfo {author} {\bibfnamefont {N.~A.}\ \bibnamefont
  {Zaprudnov}}, \ and\ \bibinfo {author} {\bibfnamefont {S.~A.}\ \bibnamefont
  {Studenikin}},\ }\bibfield  {title} {\enquote {\bibinfo {title} {Controllable
  single-spin evolution at subharmonics of electric dipole spin resonance
  enhanced by four-level
  {{Landau-Zener-St}}{\textbackslash}"uckelberg-{{Majorana}} interference},}\
  }\href {\doibase 10.1103/PhysRevB.108.205404} {\bibfield  {journal} {\bibinfo
   {journal} {Physical Review B}\ }\textbf {\bibinfo {volume} {108}},\ \bibinfo
  {pages} {205404} (\bibinfo {year} {2023})}\BibitemShut {NoStop}%
\bibitem [{\citenamefont {Munyaev}\ and\ \citenamefont
  {Bastrakova}(2021)}]{munyaevControl2021}%
  \BibitemOpen
  \bibfield  {author} {\bibinfo {author} {\bibfnamefont {V.~O.}\ \bibnamefont
  {Munyaev}}\ and\ \bibinfo {author} {\bibfnamefont {M.~V.}\ \bibnamefont
  {Bastrakova}},\ }\bibfield  {title} {\enquote {\bibinfo {title} {Control of
  spectroscopic features of multiphoton transitions in two coupled qubits by
  driving fields},}\ }\href {\doibase 10.1103/PhysRevA.104.012613} {\bibfield
  {journal} {\bibinfo  {journal} {Physical Review A}\ }\textbf {\bibinfo
  {volume} {104}},\ \bibinfo {pages} {012613} (\bibinfo {year}
  {2021})}\BibitemShut {NoStop}%
\bibitem [{\citenamefont {Kervinen}\ \emph {et~al.}(2019)\citenamefont
  {Kervinen}, \citenamefont {{Ram{\'i}rez-Mu{\~n}oz}}, \citenamefont
  {V{\"a}limaa},\ and\ \citenamefont
  {Sillanp{\"a}{\"a}}}]{kervinenLandauZenerSt2019}%
  \BibitemOpen
  \bibfield  {author} {\bibinfo {author} {\bibfnamefont {Mikael}\ \bibnamefont
  {Kervinen}}, \bibinfo {author} {\bibfnamefont {Jhon~E.}\ \bibnamefont
  {{Ram{\'i}rez-Mu{\~n}oz}}}, \bibinfo {author} {\bibfnamefont {Alpo}\
  \bibnamefont {V{\"a}limaa}}, \ and\ \bibinfo {author} {\bibfnamefont
  {Mika~A.}\ \bibnamefont {Sillanp{\"a}{\"a}}},\ }\bibfield  {title} {\enquote
  {\bibinfo {title} {Landau-{{Zener-St}}{\"u}ckelberg {{Interference}} in a
  {{Multimode Electromechanical System}} in the {{Quantum Regime}}},}\ }\href
  {\doibase 10.1103/PhysRevLett.123.240401} {\bibfield  {journal} {\bibinfo
  {journal} {Physical Review Letters}\ }\textbf {\bibinfo {volume} {123}},\
  \bibinfo {pages} {240401} (\bibinfo {year} {2019})}\BibitemShut {NoStop}%
\bibitem [{\citenamefont {Yang}\ \emph {et~al.}(2017)\citenamefont {Yang},
  \citenamefont {Pang},\ and\ \citenamefont {Jordan}}]{yangQuantum2017}%
  \BibitemOpen
  \bibfield  {author} {\bibinfo {author} {\bibfnamefont {Jing}\ \bibnamefont
  {Yang}}, \bibinfo {author} {\bibfnamefont {Shengshi}\ \bibnamefont {Pang}}, \
  and\ \bibinfo {author} {\bibfnamefont {Andrew~N.}\ \bibnamefont {Jordan}},\
  }\bibfield  {title} {\enquote {\bibinfo {title} {Quantum parameter estimation
  with the {{Landau-Zener}} transition},}\ }\href {\doibase
  10.1103/PhysRevA.96.020301} {\bibfield  {journal} {\bibinfo  {journal}
  {Physical Review A}\ }\textbf {\bibinfo {volume} {96}},\ \bibinfo {pages}
  {020301} (\bibinfo {year} {2017})}\BibitemShut {NoStop}%
\bibitem [{\citenamefont {Wen}\ \emph {et~al.}(2020)\citenamefont {Wen},
  \citenamefont {Ivakhnenko}, \citenamefont {Nakonechnyi}, \citenamefont
  {Suri}, \citenamefont {Lin}, \citenamefont {Lin}, \citenamefont {Chen},
  \citenamefont {Shevchenko}, \citenamefont {Nori},\ and\ \citenamefont
  {Hoi}}]{wenLandauZenerStuckelbergMajorana2020}%
  \BibitemOpen
  \bibfield  {author} {\bibinfo {author} {\bibfnamefont {P.~Y.}\ \bibnamefont
  {Wen}}, \bibinfo {author} {\bibfnamefont {O.~V.}\ \bibnamefont {Ivakhnenko}},
  \bibinfo {author} {\bibfnamefont {M.~A.}\ \bibnamefont {Nakonechnyi}},
  \bibinfo {author} {\bibfnamefont {B.}~\bibnamefont {Suri}}, \bibinfo {author}
  {\bibfnamefont {J.-J.}\ \bibnamefont {Lin}}, \bibinfo {author} {\bibfnamefont
  {W.-J.}\ \bibnamefont {Lin}}, \bibinfo {author} {\bibfnamefont {J.~C.}\
  \bibnamefont {Chen}}, \bibinfo {author} {\bibfnamefont {S.~N.}\ \bibnamefont
  {Shevchenko}}, \bibinfo {author} {\bibfnamefont {Franco}\ \bibnamefont
  {Nori}}, \ and\ \bibinfo {author} {\bibfnamefont {I.-C.}\ \bibnamefont
  {Hoi}},\ }\bibfield  {title} {\enquote {\bibinfo {title}
  {Landau-{{Zener-St{\"u}ckelberg-Majorana}} interferometry of a
  superconducting qubit in front of a mirror},}\ }\href {\doibase
  10.1103/PhysRevB.102.075448} {\bibfield  {journal} {\bibinfo  {journal}
  {Physical Review B}\ }\textbf {\bibinfo {volume} {102}},\ \bibinfo {pages}
  {075448} (\bibinfo {year} {2020})}\BibitemShut {NoStop}%
\bibitem [{\citenamefont {Chang}\ \emph {et~al.}(2022)\citenamefont {Chang},
  \citenamefont {Dubyna}, \citenamefont {Chien}, \citenamefont {Chen},
  \citenamefont {Wu},\ and\ \citenamefont {Kuo}}]{changCircuit2022}%
  \BibitemOpen
  \bibfield  {author} {\bibinfo {author} {\bibfnamefont {Yu-Han}\ \bibnamefont
  {Chang}}, \bibinfo {author} {\bibfnamefont {Dmytro}\ \bibnamefont {Dubyna}},
  \bibinfo {author} {\bibfnamefont {Wei-Chen}\ \bibnamefont {Chien}}, \bibinfo
  {author} {\bibfnamefont {Chien-Han}\ \bibnamefont {Chen}}, \bibinfo {author}
  {\bibfnamefont {Cen-Shawn}\ \bibnamefont {Wu}}, \ and\ \bibinfo {author}
  {\bibfnamefont {Watson}\ \bibnamefont {Kuo}},\ }\bibfield  {title} {\enquote
  {\bibinfo {title} {Circuit quantum electrodynamics with dressed states of a
  superconducting artificial atom},}\ }\href {\doibase
  10.1038/s41598-022-26828-1} {\bibfield  {journal} {\bibinfo  {journal}
  {Scientific Reports}\ }\textbf {\bibinfo {volume} {12}},\ \bibinfo {pages}
  {22308} (\bibinfo {year} {2022})}\BibitemShut {NoStop}%
\bibitem [{\citenamefont {Lidal}\ and\ \citenamefont
  {Danon}(2020)}]{lidalGeneration2020}%
  \BibitemOpen
  \bibfield  {author} {\bibinfo {author} {\bibfnamefont {Jonas}\ \bibnamefont
  {Lidal}}\ and\ \bibinfo {author} {\bibfnamefont {Jeroen}\ \bibnamefont
  {Danon}},\ }\bibfield  {title} {\enquote {\bibinfo {title} {Generation of
  {{Schr}}{\"o}dinger-cat states through photon-assisted
  {{Landau-Zener-St}}{\"u}ckelberg interferometry},}\ }\href {\doibase
  10.1103/PhysRevA.102.043717} {\bibfield  {journal} {\bibinfo  {journal}
  {Physical Review A}\ }\textbf {\bibinfo {volume} {102}},\ \bibinfo {pages}
  {043717} (\bibinfo {year} {2020})}\BibitemShut {NoStop}%
\bibitem [{\citenamefont {Wang}\ \emph {et~al.}(2021)\citenamefont {Wang},
  \citenamefont {Zheng}, \citenamefont {Wang}, \citenamefont {Gro{\ss}mann},\
  and\ \citenamefont {Zhao}}]{wangSchrodingerCat2021}%
  \BibitemOpen
  \bibfield  {author} {\bibinfo {author} {\bibfnamefont {Lu}~\bibnamefont
  {Wang}}, \bibinfo {author} {\bibfnamefont {Fulu}\ \bibnamefont {Zheng}},
  \bibinfo {author} {\bibfnamefont {Jiaming}\ \bibnamefont {Wang}}, \bibinfo
  {author} {\bibfnamefont {Frank}\ \bibnamefont {Gro{\ss}mann}}, \ and\
  \bibinfo {author} {\bibfnamefont {Yang}\ \bibnamefont {Zhao}},\ }\bibfield
  {title} {\enquote {\bibinfo {title} {Schr{\"o}dinger-{{Cat States}} in
  {{Landau}}--{{Zener}}--{{St{\"u}ckelberg}}--{{Majorana Interferometry}}: {{A
  Multiple Davydov Ansatz Approach}}},}\ }\href {\doibase
  10.1021/acs.jpcb.1c00887} {\bibfield  {journal} {\bibinfo  {journal} {The
  Journal of Physical Chemistry B}\ }\textbf {\bibinfo {volume} {125}},\
  \bibinfo {pages} {3184--3196} (\bibinfo {year} {2021})}\BibitemShut {NoStop}%
\bibitem [{\citenamefont {Ivakhnenko}\ \emph {et~al.}(2018)\citenamefont
  {Ivakhnenko}, \citenamefont {Shevchenko},\ and\ \citenamefont
  {Nori}}]{ivakhnenkoSimulating2018}%
  \BibitemOpen
  \bibfield  {author} {\bibinfo {author} {\bibfnamefont {O.~V.}\ \bibnamefont
  {Ivakhnenko}}, \bibinfo {author} {\bibfnamefont {S.~N.}\ \bibnamefont
  {Shevchenko}}, \ and\ \bibinfo {author} {\bibfnamefont {Franco}\ \bibnamefont
  {Nori}},\ }\bibfield  {title} {\enquote {\bibinfo {title} {Simulating quantum
  dynamical phenomena using classical oscillators:
  {{Landau-Zener-St{\"u}ckelberg-Majorana}} interferometry, latching
  modulation, and motional averaging},}\ }\href {\doibase
  10.1038/s41598-018-28993-8} {\bibfield  {journal} {\bibinfo  {journal}
  {Scientific Reports}\ }\textbf {\bibinfo {volume} {8}},\ \bibinfo {pages}
  {12218} (\bibinfo {year} {2018})}\BibitemShut {NoStop}%
\bibitem [{\citenamefont {Zhou}\ \emph {et~al.}(2019)\citenamefont {Zhou},
  \citenamefont {Zhao}, \citenamefont {Xiao}, \citenamefont {Sun},
  \citenamefont {Sobreviela}, \citenamefont {Gerrard}, \citenamefont {Chen},
  \citenamefont {Flader}, \citenamefont {Kenny}, \citenamefont {Wu},\ and\
  \citenamefont {Seshia}}]{zhouDynamic2019}%
  \BibitemOpen
  \bibfield  {author} {\bibinfo {author} {\bibfnamefont {Xin}\ \bibnamefont
  {Zhou}}, \bibinfo {author} {\bibfnamefont {Chun}\ \bibnamefont {Zhao}},
  \bibinfo {author} {\bibfnamefont {Dingbang}\ \bibnamefont {Xiao}}, \bibinfo
  {author} {\bibfnamefont {Jiangkun}\ \bibnamefont {Sun}}, \bibinfo {author}
  {\bibfnamefont {Guillermo}\ \bibnamefont {Sobreviela}}, \bibinfo {author}
  {\bibfnamefont {Dustin~D.}\ \bibnamefont {Gerrard}}, \bibinfo {author}
  {\bibfnamefont {Yunhan}\ \bibnamefont {Chen}}, \bibinfo {author}
  {\bibfnamefont {Ian}\ \bibnamefont {Flader}}, \bibinfo {author}
  {\bibfnamefont {Thomas~W.}\ \bibnamefont {Kenny}}, \bibinfo {author}
  {\bibfnamefont {Xuezhong}\ \bibnamefont {Wu}}, \ and\ \bibinfo {author}
  {\bibfnamefont {Ashwin~A.}\ \bibnamefont {Seshia}},\ }\bibfield  {title}
  {\enquote {\bibinfo {title} {Dynamic modulation of modal coupling in
  microelectromechanical gyroscopic ring resonators},}\ }\href {\doibase
  10.1038/s41467-019-12796-0} {\bibfield  {journal} {\bibinfo  {journal}
  {Nature Communications}\ }\textbf {\bibinfo {volume} {10}},\ \bibinfo {pages}
  {4980} (\bibinfo {year} {2019})}\BibitemShut {NoStop}%
\bibitem [{\citenamefont {Lorenz}\ \emph {et~al.}(2023)\citenamefont {Lorenz},
  \citenamefont {Kohler}, \citenamefont {Parafilo}, \citenamefont {Kiselev},\
  and\ \citenamefont {Ludwig}}]{lorenzClassical2023}%
  \BibitemOpen
  \bibfield  {author} {\bibinfo {author} {\bibfnamefont {Heribert}\
  \bibnamefont {Lorenz}}, \bibinfo {author} {\bibfnamefont {Sigmund}\
  \bibnamefont {Kohler}}, \bibinfo {author} {\bibfnamefont {Anton}\
  \bibnamefont {Parafilo}}, \bibinfo {author} {\bibfnamefont {Mikhail}\
  \bibnamefont {Kiselev}}, \ and\ \bibinfo {author} {\bibfnamefont {Stefan}\
  \bibnamefont {Ludwig}},\ }\bibfield  {title} {\enquote {\bibinfo {title}
  {Classical analogue to driven quantum bits based on macroscopic pendula},}\
  }\href {\doibase 10.1038/s41598-023-45118-y} {\bibfield  {journal} {\bibinfo
  {journal} {Scientific Reports}\ }\textbf {\bibinfo {volume} {13}},\ \bibinfo
  {pages} {18386} (\bibinfo {year} {2023})}\BibitemShut {NoStop}%
\bibitem [{\citenamefont {Bernazzani}\ and\ \citenamefont
  {Burkard}(2024)}]{bernazzaniFluctuating2024}%
  \BibitemOpen
  \bibfield  {author} {\bibinfo {author} {\bibfnamefont {Lorenzo}\ \bibnamefont
  {Bernazzani}}\ and\ \bibinfo {author} {\bibfnamefont {Guido}\ \bibnamefont
  {Burkard}},\ }\bibfield  {title} {\enquote {\bibinfo {title} {Fluctuating
  parametric drive of coupled classical oscillators can simulate dissipative
  qubits},}\ }\href {\doibase 10.1103/PhysRevResearch.6.013284} {\bibfield
  {journal} {\bibinfo  {journal} {Physical Review Research}\ }\textbf {\bibinfo
  {volume} {6}},\ \bibinfo {pages} {013284} (\bibinfo {year}
  {2024})}\BibitemShut {NoStop}%
\bibitem [{\citenamefont {Carusotto}\ \emph {et~al.}(2020)\citenamefont
  {Carusotto}, \citenamefont {Houck}, \citenamefont {Kollár}, \citenamefont
  {Roushan}, \citenamefont {Schuster},\ and\ \citenamefont
  {Simon}}]{CarusottoNATPHYS20}%
  \BibitemOpen
  \bibfield  {author} {\bibinfo {author} {\bibfnamefont {Iacopo}\ \bibnamefont
  {Carusotto}}, \bibinfo {author} {\bibfnamefont {Andrew~A.}\ \bibnamefont
  {Houck}}, \bibinfo {author} {\bibfnamefont {Alicia~J.}\ \bibnamefont
  {Kollár}}, \bibinfo {author} {\bibfnamefont {Pedram}\ \bibnamefont
  {Roushan}}, \bibinfo {author} {\bibfnamefont {David~I.}\ \bibnamefont
  {Schuster}}, \ and\ \bibinfo {author} {\bibfnamefont {Jonathan}\ \bibnamefont
  {Simon}},\ }\bibfield  {title} {\enquote {\bibinfo {title} {Photonic
  materials in circuit quantum electrodynamics},}\ }\href {\doibase
  10.1038/s41567-020-0815-y} {\bibfield  {journal} {\bibinfo  {journal} {Nature
  Physics}\ }\textbf {\bibinfo {volume} {16}},\ \bibinfo {pages} {268–279}
  (\bibinfo {year} {2020})}\BibitemShut {NoStop}%
\bibitem [{\citenamefont {Carusotto}\ and\ \citenamefont
  {Ciuti}(2013)}]{CarusottoRMP13}%
  \BibitemOpen
  \bibfield  {author} {\bibinfo {author} {\bibfnamefont {Iacopo}\ \bibnamefont
  {Carusotto}}\ and\ \bibinfo {author} {\bibfnamefont {Cristiano}\ \bibnamefont
  {Ciuti}},\ }\bibfield  {title} {\enquote {\bibinfo {title} {Quantum fluids of
  light},}\ }\href {\doibase 10.1103/RevModPhys.85.299} {\bibfield  {journal}
  {\bibinfo  {journal} {Rev. Mod. Phys.}\ }\textbf {\bibinfo {volume} {85}},\
  \bibinfo {pages} {299--366} (\bibinfo {year} {2013})}\BibitemShut {NoStop}%
\bibitem [{\citenamefont {Huber}\ \emph {et~al.}(2020)\citenamefont {Huber},
  \citenamefont {Rastelli}, \citenamefont {Seitner}, \citenamefont {K\"olbl},
  \citenamefont {Belzig}, \citenamefont {Dykman},\ and\ \citenamefont
  {Weig}}]{HuberPRX20}%
  \BibitemOpen
  \bibfield  {author} {\bibinfo {author} {\bibfnamefont {J.~S.}\ \bibnamefont
  {Huber}}, \bibinfo {author} {\bibfnamefont {G.}~\bibnamefont {Rastelli}},
  \bibinfo {author} {\bibfnamefont {M.~J.}\ \bibnamefont {Seitner}}, \bibinfo
  {author} {\bibfnamefont {J.}~\bibnamefont {K\"olbl}}, \bibinfo {author}
  {\bibfnamefont {W.}~\bibnamefont {Belzig}}, \bibinfo {author} {\bibfnamefont
  {M.~I.}\ \bibnamefont {Dykman}}, \ and\ \bibinfo {author} {\bibfnamefont
  {E.~M.}\ \bibnamefont {Weig}},\ }\bibfield  {title} {\enquote {\bibinfo
  {title} {Spectral evidence of squeezing of a weakly damped driven
  nanomechanical mode},}\ }\href {\doibase 10.1103/PhysRevX.10.021066}
  {\bibfield  {journal} {\bibinfo  {journal} {Phys. Rev. X}\ }\textbf {\bibinfo
  {volume} {10}},\ \bibinfo {pages} {021066} (\bibinfo {year}
  {2020})}\BibitemShut {NoStop}%
\bibitem [{\citenamefont {Ding}\ \emph {et~al.}(2017)\citenamefont {Ding},
  \citenamefont {Maslennikov}, \citenamefont {Habl\"utzel},\ and\ \citenamefont
  {Matsukevich}}]{DingPRL17}%
  \BibitemOpen
  \bibfield  {author} {\bibinfo {author} {\bibfnamefont {Shiqian}\ \bibnamefont
  {Ding}}, \bibinfo {author} {\bibfnamefont {Gleb}\ \bibnamefont
  {Maslennikov}}, \bibinfo {author} {\bibfnamefont {Roland}\ \bibnamefont
  {Habl\"utzel}}, \ and\ \bibinfo {author} {\bibfnamefont {Dzmitry}\
  \bibnamefont {Matsukevich}},\ }\bibfield  {title} {\enquote {\bibinfo {title}
  {Cross-{K}err nonlinearity for phonon counting},}\ }\href {\doibase
  10.1103/PhysRevLett.119.193602} {\bibfield  {journal} {\bibinfo  {journal}
  {Phys. Rev. Lett.}\ }\textbf {\bibinfo {volume} {119}},\ \bibinfo {pages}
  {193602} (\bibinfo {year} {2017})}\BibitemShut {NoStop}%
\bibitem [{\citenamefont {Andersen}\ \emph {et~al.}(2020)\citenamefont
  {Andersen}, \citenamefont {Kamal}, \citenamefont {Masluk}, \citenamefont
  {Pop}, \citenamefont {Blais},\ and\ \citenamefont
  {Devoret}}]{andersenQuantum2020}%
  \BibitemOpen
  \bibfield  {author} {\bibinfo {author} {\bibfnamefont {Christian~Kraglund}\
  \bibnamefont {Andersen}}, \bibinfo {author} {\bibfnamefont {Archana}\
  \bibnamefont {Kamal}}, \bibinfo {author} {\bibfnamefont {Nicholas~A.}\
  \bibnamefont {Masluk}}, \bibinfo {author} {\bibfnamefont {Ioan~M.}\
  \bibnamefont {Pop}}, \bibinfo {author} {\bibfnamefont {Alexandre}\
  \bibnamefont {Blais}}, \ and\ \bibinfo {author} {\bibfnamefont {Michel~H.}\
  \bibnamefont {Devoret}},\ }\bibfield  {title} {\enquote {\bibinfo {title}
  {Quantum {{Versus Classical Switching Dynamics}} of {{Driven Dissipative Kerr
  Resonators}}},}\ }\href {\doibase 10.1103/PhysRevApplied.13.044017}
  {\bibfield  {journal} {\bibinfo  {journal} {Physical Review Applied}\
  }\textbf {\bibinfo {volume} {13}},\ \bibinfo {pages} {044017} (\bibinfo
  {year} {2020})}\BibitemShut {NoStop}%
\bibitem [{\citenamefont {Yamaji}\ \emph {et~al.}(2022)\citenamefont {Yamaji},
  \citenamefont {Kagami}, \citenamefont {Yamaguchi}, \citenamefont {Satoh},
  \citenamefont {Koshino}, \citenamefont {Goto}, \citenamefont {Lin},
  \citenamefont {Nakamura},\ and\ \citenamefont {Yamamoto}}]{YamajiPRA22}%
  \BibitemOpen
  \bibfield  {author} {\bibinfo {author} {\bibfnamefont {T.}~\bibnamefont
  {Yamaji}}, \bibinfo {author} {\bibfnamefont {S.}~\bibnamefont {Kagami}},
  \bibinfo {author} {\bibfnamefont {A.}~\bibnamefont {Yamaguchi}}, \bibinfo
  {author} {\bibfnamefont {T.}~\bibnamefont {Satoh}}, \bibinfo {author}
  {\bibfnamefont {K.}~\bibnamefont {Koshino}}, \bibinfo {author} {\bibfnamefont
  {H.}~\bibnamefont {Goto}}, \bibinfo {author} {\bibfnamefont {Z.~R.}\
  \bibnamefont {Lin}}, \bibinfo {author} {\bibfnamefont {Y.}~\bibnamefont
  {Nakamura}}, \ and\ \bibinfo {author} {\bibfnamefont {T.}~\bibnamefont
  {Yamamoto}},\ }\bibfield  {title} {\enquote {\bibinfo {title} {Spectroscopic
  observation of the crossover from a classical {Duffing} oscillator to a
  {K}err parametric oscillator},}\ }\href {\doibase
  10.1103/PhysRevA.105.023519} {\bibfield  {journal} {\bibinfo  {journal}
  {Phys. Rev. A}\ }\textbf {\bibinfo {volume} {105}},\ \bibinfo {pages}
  {023519} (\bibinfo {year} {2022})}\BibitemShut {NoStop}%
\bibitem [{\citenamefont {Winkel}\ \emph {et~al.}(2020)\citenamefont {Winkel},
  \citenamefont {Borisov}, \citenamefont {Gr\"unhaupt}, \citenamefont {Rieger},
  \citenamefont {Spiecker}, \citenamefont {Valenti}, \citenamefont {Ustinov},
  \citenamefont {Wernsdorfer},\ and\ \citenamefont {Pop}}]{WinkelPRX20}%
  \BibitemOpen
  \bibfield  {author} {\bibinfo {author} {\bibfnamefont {Patrick}\ \bibnamefont
  {Winkel}}, \bibinfo {author} {\bibfnamefont {Kiril}\ \bibnamefont {Borisov}},
  \bibinfo {author} {\bibfnamefont {Lukas}\ \bibnamefont {Gr\"unhaupt}},
  \bibinfo {author} {\bibfnamefont {Dennis}\ \bibnamefont {Rieger}}, \bibinfo
  {author} {\bibfnamefont {Martin}\ \bibnamefont {Spiecker}}, \bibinfo {author}
  {\bibfnamefont {Francesco}\ \bibnamefont {Valenti}}, \bibinfo {author}
  {\bibfnamefont {Alexey~V.}\ \bibnamefont {Ustinov}}, \bibinfo {author}
  {\bibfnamefont {Wolfgang}\ \bibnamefont {Wernsdorfer}}, \ and\ \bibinfo
  {author} {\bibfnamefont {Ioan~M.}\ \bibnamefont {Pop}},\ }\bibfield  {title}
  {\enquote {\bibinfo {title} {Implementation of a transmon qubit using
  superconducting granular aluminum},}\ }\href {\doibase
  10.1103/PhysRevX.10.031032} {\bibfield  {journal} {\bibinfo  {journal} {Phys.
  Rev. X}\ }\textbf {\bibinfo {volume} {10}},\ \bibinfo {pages} {031032}
  (\bibinfo {year} {2020})}\BibitemShut {NoStop}%
\bibitem [{Note1()}]{Note1}%
  \BibitemOpen
  \bibinfo {note} {Note that here multiphoton resonance refers to the fact that
  absorbing $n$ photons leads to the $n$th excited state of the resonator. This
  is not the multiphoton Rabi resonance, where $n$ driving photons are absorbed
  to populate the excited level of the qubit.}\BibitemShut {Stop}%
\bibitem [{\citenamefont {Dykman}(2012)}]{dykmanFluctuating2012}%
  \BibitemOpen
  \bibfield  {author} {\bibinfo {author} {\bibfnamefont {Mark~I.}\ \bibnamefont
  {Dykman}},\ }\href {\doibase 10.1093/acprof:oso/9780199691388.001.0001}
  {\emph {\bibinfo {title} {Fluctuating {{Nonlinear Oscillators}}: {{From
  Nanomechanics}} to {{Quantum Superconducting Circuits}}}}}\ (\bibinfo
  {publisher} {Oxford University Press},\ \bibinfo {year} {2012})\BibitemShut
  {NoStop}%
\bibitem [{\citenamefont {Chen}\ \emph {et~al.}(2023)\citenamefont {Chen},
  \citenamefont {Fischer}, \citenamefont {Nojiri}, \citenamefont {Renger},
  \citenamefont {Xie}, \citenamefont {Partanen}, \citenamefont {Pogorzalek},
  \citenamefont {Fedorov}, \citenamefont {Marx}, \citenamefont {Deppe},\ and\
  \citenamefont {Gross}}]{ChenNATCOM23}%
  \BibitemOpen
  \bibfield  {author} {\bibinfo {author} {\bibfnamefont {Qi-Ming}\ \bibnamefont
  {Chen}}, \bibinfo {author} {\bibfnamefont {Michael}\ \bibnamefont {Fischer}},
  \bibinfo {author} {\bibfnamefont {Yuki}\ \bibnamefont {Nojiri}}, \bibinfo
  {author} {\bibfnamefont {Michael}\ \bibnamefont {Renger}}, \bibinfo {author}
  {\bibfnamefont {Edwar}\ \bibnamefont {Xie}}, \bibinfo {author} {\bibfnamefont
  {Matti}\ \bibnamefont {Partanen}}, \bibinfo {author} {\bibfnamefont {Stefan}\
  \bibnamefont {Pogorzalek}}, \bibinfo {author} {\bibfnamefont {Kirill~G.}\
  \bibnamefont {Fedorov}}, \bibinfo {author} {\bibfnamefont {Achim}\
  \bibnamefont {Marx}}, \bibinfo {author} {\bibfnamefont {Frank}\ \bibnamefont
  {Deppe}}, \ and\ \bibinfo {author} {\bibfnamefont {Rudolf}\ \bibnamefont
  {Gross}},\ }\bibfield  {title} {\enquote {\bibinfo {title} {Quantum behavior
  of the {Duffing} oscillator at the dissipative phase transition},}\ }\href
  {\doibase 10.1038/s41467-023-38217-x} {\bibfield  {journal} {\bibinfo
  {journal} {Nature Communications}\ }\textbf {\bibinfo {volume} {14}}
  (\bibinfo {year} {2023}),\ 10.1038/s41467-023-38217-x}\BibitemShut {NoStop}%
\bibitem [{\citenamefont {Beaulieu}\ \emph {et~al.}(2023)\citenamefont
  {Beaulieu}, \citenamefont {Minganti}, \citenamefont {Frasca}, \citenamefont
  {Savona}, \citenamefont {Felicetti}, \citenamefont {Di~Candia},\ and\
  \citenamefont {Scarlino}}]{BeaulieuARXIV23}%
  \BibitemOpen
  \bibfield  {author} {\bibinfo {author} {\bibfnamefont {Guillaume}\
  \bibnamefont {Beaulieu}}, \bibinfo {author} {\bibfnamefont {Fabrizio}\
  \bibnamefont {Minganti}}, \bibinfo {author} {\bibfnamefont {Simone}\
  \bibnamefont {Frasca}}, \bibinfo {author} {\bibfnamefont {Vincenzo}\
  \bibnamefont {Savona}}, \bibinfo {author} {\bibfnamefont {Simone}\
  \bibnamefont {Felicetti}}, \bibinfo {author} {\bibfnamefont {Roberto}\
  \bibnamefont {Di~Candia}}, \ and\ \bibinfo {author} {\bibfnamefont
  {Pasquale}\ \bibnamefont {Scarlino}},\ }\bibfield  {title} {\enquote
  {\bibinfo {title} {Observation of first- and second-order dissipative phase
  transitions in a two-photon driven {K}err resonator},}\ }\href {\doibase
  10.48550/ARXIV.2310.13636} {\  (\bibinfo {year} {2023}),\
  10.48550/ARXIV.2310.13636}\BibitemShut {NoStop}%
\bibitem [{\citenamefont {{Foss-Feig}}\ \emph {et~al.}(2017)\citenamefont
  {{Foss-Feig}}, \citenamefont {Niroula}, \citenamefont {Young}, \citenamefont
  {Hafezi}, \citenamefont {Gorshkov}, \citenamefont {Wilson},\ and\
  \citenamefont {Maghrebi}}]{foss-feigEmergent2017}%
  \BibitemOpen
  \bibfield  {author} {\bibinfo {author} {\bibfnamefont {M.}~\bibnamefont
  {{Foss-Feig}}}, \bibinfo {author} {\bibfnamefont {P.}~\bibnamefont
  {Niroula}}, \bibinfo {author} {\bibfnamefont {J.~T.}\ \bibnamefont {Young}},
  \bibinfo {author} {\bibfnamefont {M.}~\bibnamefont {Hafezi}}, \bibinfo
  {author} {\bibfnamefont {A.~V.}\ \bibnamefont {Gorshkov}}, \bibinfo {author}
  {\bibfnamefont {R.~M.}\ \bibnamefont {Wilson}}, \ and\ \bibinfo {author}
  {\bibfnamefont {M.~F.}\ \bibnamefont {Maghrebi}},\ }\bibfield  {title}
  {\enquote {\bibinfo {title} {Emergent equilibrium in many-body optical
  bistability},}\ }\href {\doibase 10.1103/PhysRevA.95.043826} {\bibfield
  {journal} {\bibinfo  {journal} {Physical Review A}\ }\textbf {\bibinfo
  {volume} {95}},\ \bibinfo {pages} {043826} (\bibinfo {year}
  {2017})}\BibitemShut {NoStop}%
\bibitem [{\citenamefont {Vicentini}\ \emph {et~al.}(2018)\citenamefont
  {Vicentini}, \citenamefont {Minganti}, \citenamefont {Rota}, \citenamefont
  {Orso},\ and\ \citenamefont {Ciuti}}]{VicentiniPRA18}%
  \BibitemOpen
  \bibfield  {author} {\bibinfo {author} {\bibfnamefont {Filippo}\ \bibnamefont
  {Vicentini}}, \bibinfo {author} {\bibfnamefont {Fabrizio}\ \bibnamefont
  {Minganti}}, \bibinfo {author} {\bibfnamefont {Riccardo}\ \bibnamefont
  {Rota}}, \bibinfo {author} {\bibfnamefont {Giuliano}\ \bibnamefont {Orso}}, \
  and\ \bibinfo {author} {\bibfnamefont {Cristiano}\ \bibnamefont {Ciuti}},\
  }\bibfield  {title} {\enquote {\bibinfo {title} {Critical slowing down in
  driven-dissipative {Bose-Hubbard} lattices},}\ }\href {\doibase
  10.1103/PhysRevA.97.013853} {\bibfield  {journal} {\bibinfo  {journal} {Phys.
  Rev. A}\ }\textbf {\bibinfo {volume} {97}},\ \bibinfo {pages} {013853}
  (\bibinfo {year} {2018})}\BibitemShut {NoStop}%
\bibitem [{\citenamefont {Li}\ \emph {et~al.}(2022)\citenamefont {Li},
  \citenamefont {Claude}, \citenamefont {Boulier}, \citenamefont {Giacobino},
  \citenamefont {Glorieux}, \citenamefont {Bramati},\ and\ \citenamefont
  {Ciuti}}]{LiPRL22}%
  \BibitemOpen
  \bibfield  {author} {\bibinfo {author} {\bibfnamefont {Zejian}\ \bibnamefont
  {Li}}, \bibinfo {author} {\bibfnamefont {Ferdinand}\ \bibnamefont {Claude}},
  \bibinfo {author} {\bibfnamefont {Thomas}\ \bibnamefont {Boulier}}, \bibinfo
  {author} {\bibfnamefont {Elisabeth}\ \bibnamefont {Giacobino}}, \bibinfo
  {author} {\bibfnamefont {Quentin}\ \bibnamefont {Glorieux}}, \bibinfo
  {author} {\bibfnamefont {Alberto}\ \bibnamefont {Bramati}}, \ and\ \bibinfo
  {author} {\bibfnamefont {Cristiano}\ \bibnamefont {Ciuti}},\ }\bibfield
  {title} {\enquote {\bibinfo {title} {Dissipative phase transition with
  driving-controlled spatial dimension and diffusive boundary conditions},}\
  }\href {\doibase 10.1103/PhysRevLett.128.093601} {\bibfield  {journal}
  {\bibinfo  {journal} {Phys. Rev. Lett.}\ }\textbf {\bibinfo {volume} {128}},\
  \bibinfo {pages} {093601} (\bibinfo {year} {2022})}\BibitemShut {NoStop}%
\bibitem [{\citenamefont {Ferrari}\ \emph {et~al.}(2025)\citenamefont
  {Ferrari}, \citenamefont {Gravina}, \citenamefont {Eeltink}, \citenamefont
  {Scarlino}, \citenamefont {Savona},\ and\ \citenamefont
  {Minganti}}]{FerrariPRR25}%
  \BibitemOpen
  \bibfield  {author} {\bibinfo {author} {\bibfnamefont {Filippo}\ \bibnamefont
  {Ferrari}}, \bibinfo {author} {\bibfnamefont {Luca}\ \bibnamefont {Gravina}},
  \bibinfo {author} {\bibfnamefont {Debbie}\ \bibnamefont {Eeltink}}, \bibinfo
  {author} {\bibfnamefont {Pasquale}\ \bibnamefont {Scarlino}}, \bibinfo
  {author} {\bibfnamefont {Vincenzo}\ \bibnamefont {Savona}}, \ and\ \bibinfo
  {author} {\bibfnamefont {Fabrizio}\ \bibnamefont {Minganti}},\ }
  \bibfield  {title} {\enquote {\bibinfo {title} {Dissipative quantum chaos
  unveiled by stochastic quantum trajectories},}\ }
  \href{https://doi.org/10.1103/PhysRevResearch.7.013276}{%
  \bibfield  {journal} {\bibinfo {journal} {Phys. Rev. Res.} }\textbf
  {\bibinfo {volume} {7}}, \bibinfo {pages} {013276} (\bibinfo {year} {2025})}%
  \BibitemShut {NoStop}%
\bibitem [{\citenamefont {Dahan}\ \emph {et~al.}(2022)\citenamefont {Dahan},
  \citenamefont {Arwas},\ and\ \citenamefont {Grosfeld}}]{DahanNPJ22}%
  \BibitemOpen
  \bibfield  {author} {\bibinfo {author} {\bibfnamefont {Daniel}\ \bibnamefont
  {Dahan}}, \bibinfo {author} {\bibfnamefont {Geva}\ \bibnamefont {Arwas}}, \
  and\ \bibinfo {author} {\bibfnamefont {Eytan}\ \bibnamefont {Grosfeld}},\
  }\bibfield  {title} {\enquote {\bibinfo {title} {Classical and quantum chaos
  in chirally-driven, dissipative {Bose-Hubbard} systems},}\ }\href {\doibase
  10.1038/s41534-022-00518-2} {\bibfield  {journal} {\bibinfo  {journal} {npj
  Quantum Information}\ }\textbf {\bibinfo {volume} {8}} (\bibinfo {year}
  {2022}),\ 10.1038/s41534-022-00518-2}\BibitemShut {NoStop}%
\bibitem [{\citenamefont {Cohen}\ \emph {et~al.}(2023)\citenamefont {Cohen},
  \citenamefont {Petrescu}, \citenamefont {Shillito},\ and\ \citenamefont
  {Blais}}]{CohenPRXQ23}%
  \BibitemOpen
  \bibfield  {author} {\bibinfo {author} {\bibfnamefont {Joachim}\ \bibnamefont
  {Cohen}}, \bibinfo {author} {\bibfnamefont {Alexandru}\ \bibnamefont
  {Petrescu}}, \bibinfo {author} {\bibfnamefont {Ross}\ \bibnamefont
  {Shillito}}, \ and\ \bibinfo {author} {\bibfnamefont {Alexandre}\
  \bibnamefont {Blais}},\ }\bibfield  {title} {\enquote {\bibinfo {title}
  {Reminiscence of classical chaos in driven transmons},}\ }\href {\doibase
  10.1103/PRXQuantum.4.020312} {\bibfield  {journal} {\bibinfo  {journal} {PRX
  Quantum}\ }\textbf {\bibinfo {volume} {4}},\ \bibinfo {pages} {020312}
  (\bibinfo {year} {2023})}\BibitemShut {NoStop}%
\bibitem [{\citenamefont {Nguyen}\ \emph {et~al.}(2024)\citenamefont {Nguyen},
  \citenamefont {Kim}, \citenamefont {Hashim}, \citenamefont {Goss},
  \citenamefont {Marinelli}, \citenamefont {Bhandari}, \citenamefont {Das},
  \citenamefont {Naik}, \citenamefont {Kreikebaum}, \citenamefont {Jordan},
  \citenamefont {Santiago},\ and\ \citenamefont
  {Siddiqi}}]{nguyenProgrammable2024}%
  \BibitemOpen
  \bibfield  {author} {\bibinfo {author} {\bibfnamefont {Long~B.}\ \bibnamefont
  {Nguyen}}, \bibinfo {author} {\bibfnamefont {Yosep}\ \bibnamefont {Kim}},
  \bibinfo {author} {\bibfnamefont {Akel}\ \bibnamefont {Hashim}}, \bibinfo
  {author} {\bibfnamefont {Noah}\ \bibnamefont {Goss}}, \bibinfo {author}
  {\bibfnamefont {Brian}\ \bibnamefont {Marinelli}}, \bibinfo {author}
  {\bibfnamefont {Bibek}\ \bibnamefont {Bhandari}}, \bibinfo {author}
  {\bibfnamefont {Debmalya}\ \bibnamefont {Das}}, \bibinfo {author}
  {\bibfnamefont {Ravi~K.}\ \bibnamefont {Naik}}, \bibinfo {author}
  {\bibfnamefont {John~Mark}\ \bibnamefont {Kreikebaum}}, \bibinfo {author}
  {\bibfnamefont {Andrew~N.}\ \bibnamefont {Jordan}}, \bibinfo {author}
  {\bibfnamefont {David~I.}\ \bibnamefont {Santiago}}, \ and\ \bibinfo {author}
  {\bibfnamefont {Irfan}\ \bibnamefont {Siddiqi}},\ }\bibfield  {title}
  {\enquote {\bibinfo {title} {Programmable {{Heisenberg}} interactions between
  {{Floquet}} qubits},}\ }\href {\doibase 10.1038/s41567-023-02326-7}
  {\bibfield  {journal} {\bibinfo  {journal} {Nature Physics}\ }\textbf
  {\bibinfo {volume} {20}},\ \bibinfo {pages} {240--246} (\bibinfo {year}
  {2024})}\BibitemShut {NoStop}%
\bibitem [{\citenamefont {Gandon}\ \emph {et~al.}(2022)\citenamefont {Gandon},
  \citenamefont {Le~Calonnec}, \citenamefont {Shillito}, \citenamefont
  {Petrescu},\ and\ \citenamefont {Blais}}]{gandonEngineering2022}%
  \BibitemOpen
  \bibfield  {author} {\bibinfo {author} {\bibfnamefont {Anthony}\ \bibnamefont
  {Gandon}}, \bibinfo {author} {\bibfnamefont {Camille}\ \bibnamefont
  {Le~Calonnec}}, \bibinfo {author} {\bibfnamefont {Ross}\ \bibnamefont
  {Shillito}}, \bibinfo {author} {\bibfnamefont {Alexandru}\ \bibnamefont
  {Petrescu}}, \ and\ \bibinfo {author} {\bibfnamefont {Alexandre}\
  \bibnamefont {Blais}},\ }\bibfield  {title} {\enquote {\bibinfo {title}
  {Engineering, {{Control}}, and {{Longitudinal Readout}} of {{Floquet
  Qubits}}},}\ }\href {\doibase 10.1103/PhysRevApplied.17.064006} {\bibfield
  {journal} {\bibinfo  {journal} {Physical Review Applied}\ }\textbf {\bibinfo
  {volume} {17}},\ \bibinfo {pages} {064006} (\bibinfo {year}
  {2022})}\BibitemShut {NoStop}%
\bibitem [{\citenamefont {Rudner}\ \emph {et~al.}(2013)\citenamefont {Rudner},
  \citenamefont {Lindner}, \citenamefont {Berg},\ and\ \citenamefont
  {Levin}}]{rudnerAnomalous2013}%
  \BibitemOpen
  \bibfield  {author} {\bibinfo {author} {\bibfnamefont {Mark~S.}\ \bibnamefont
  {Rudner}}, \bibinfo {author} {\bibfnamefont {Netanel~H.}\ \bibnamefont
  {Lindner}}, \bibinfo {author} {\bibfnamefont {Erez}\ \bibnamefont {Berg}}, \
  and\ \bibinfo {author} {\bibfnamefont {Michael}\ \bibnamefont {Levin}},\
  }\bibfield  {title} {\enquote {\bibinfo {title} {Anomalous {{Edge States}}
  and the {{Bulk-Edge Correspondence}} for {{Periodically Driven
  Two-Dimensional Systems}}},}\ }\href {\doibase 10.1103/PhysRevX.3.031005}
  {\bibfield  {journal} {\bibinfo  {journal} {Physical Review X}\ }\textbf
  {\bibinfo {volume} {3}},\ \bibinfo {pages} {031005} (\bibinfo {year}
  {2013})}\BibitemShut {NoStop}%
\bibitem [{\citenamefont {Maczewsky}\ \emph {et~al.}(2017)\citenamefont
  {Maczewsky}, \citenamefont {Zeuner}, \citenamefont {Nolte},\ and\
  \citenamefont {Szameit}}]{maczewskyObservation2017}%
  \BibitemOpen
  \bibfield  {author} {\bibinfo {author} {\bibfnamefont {Lukas~J.}\
  \bibnamefont {Maczewsky}}, \bibinfo {author} {\bibfnamefont {Julia~M.}\
  \bibnamefont {Zeuner}}, \bibinfo {author} {\bibfnamefont {Stefan}\
  \bibnamefont {Nolte}}, \ and\ \bibinfo {author} {\bibfnamefont {Alexander}\
  \bibnamefont {Szameit}},\ }\bibfield  {title} {\enquote {\bibinfo {title}
  {Observation of photonic anomalous {{Floquet}} topological insulators},}\
  }\href {\doibase 10.1038/ncomms13756} {\bibfield  {journal} {\bibinfo
  {journal} {Nature Communications}\ }\textbf {\bibinfo {volume} {8}},\
  \bibinfo {pages} {13756} (\bibinfo {year} {2017})}\BibitemShut {NoStop}%
\bibitem [{\citenamefont {Oka}\ and\ \citenamefont
  {Kitamura}(2019)}]{okaFloquet2019}%
  \BibitemOpen
  \bibfield  {author} {\bibinfo {author} {\bibfnamefont {Takashi}\ \bibnamefont
  {Oka}}\ and\ \bibinfo {author} {\bibfnamefont {Sota}\ \bibnamefont
  {Kitamura}},\ }\bibfield  {title} {\enquote {\bibinfo {title} {Floquet
  {{Engineering}} of {{Quantum Materials}}},}\ }\href {\doibase
  10.1146/annurev-conmatphys-031218-013423} {\bibfield  {journal} {\bibinfo
  {journal} {Annual Review of Condensed Matter Physics}\ }\textbf {\bibinfo
  {volume} {10}},\ \bibinfo {pages} {387--408} (\bibinfo {year}
  {2019})}\BibitemShut {NoStop}%
\bibitem [{\citenamefont {Weitenberg}\ and\ \citenamefont
  {Simonet}(2021)}]{weitenbergTailoring2021}%
  \BibitemOpen
  \bibfield  {author} {\bibinfo {author} {\bibfnamefont {Christof}\
  \bibnamefont {Weitenberg}}\ and\ \bibinfo {author} {\bibfnamefont {Juliette}\
  \bibnamefont {Simonet}},\ }\bibfield  {title} {\enquote {\bibinfo {title}
  {Tailoring quantum gases by {{Floquet}} engineering},}\ }\href {\doibase
  10.1038/s41567-021-01316-x} {\bibfield  {journal} {\bibinfo  {journal}
  {Nature Physics}\ }\textbf {\bibinfo {volume} {17}},\ \bibinfo {pages}
  {1342--1348} (\bibinfo {year} {2021})}\BibitemShut {NoStop}%
\bibitem [{\citenamefont {Ozawa}\ and\ \citenamefont
  {Price}(2019)}]{ozawaTopological2019}%
  \BibitemOpen
  \bibfield  {author} {\bibinfo {author} {\bibfnamefont {Tomoki}\ \bibnamefont
  {Ozawa}}\ and\ \bibinfo {author} {\bibfnamefont {Hannah~M.}\ \bibnamefont
  {Price}},\ }\bibfield  {title} {\enquote {\bibinfo {title} {Topological
  quantum matter in synthetic dimensions},}\ }\href {\doibase
  10.1038/s42254-019-0045-3} {\bibfield  {journal} {\bibinfo  {journal} {Nature
  Reviews Physics}\ }\textbf {\bibinfo {volume} {1}},\ \bibinfo {pages}
  {349--357} (\bibinfo {year} {2019})}\BibitemShut {NoStop}%
\bibitem [{\citenamefont {Meier}\ \emph {et~al.}(2019)\citenamefont {Meier},
  \citenamefont {Ang'ong'a}, \citenamefont {An},\ and\ \citenamefont
  {Gadway}}]{meierExploring2019}%
  \BibitemOpen
  \bibfield  {author} {\bibinfo {author} {\bibfnamefont {Eric~J.}\ \bibnamefont
  {Meier}}, \bibinfo {author} {\bibfnamefont {Jackson}\ \bibnamefont
  {Ang'ong'a}}, \bibinfo {author} {\bibfnamefont {Fangzhao~Alex}\ \bibnamefont
  {An}}, \ and\ \bibinfo {author} {\bibfnamefont {Bryce}\ \bibnamefont
  {Gadway}},\ }\bibfield  {title} {\enquote {\bibinfo {title} {Exploring
  quantum signatures of chaos on a {{Floquet}} synthetic lattice},}\ }\href
  {\doibase 10.1103/PhysRevA.100.013623} {\bibfield  {journal} {\bibinfo
  {journal} {Physical Review A}\ }\textbf {\bibinfo {volume} {100}},\ \bibinfo
  {pages} {013623} (\bibinfo {year} {2019})}\BibitemShut {NoStop}%
\bibitem [{\citenamefont {Arnal}\ \emph {et~al.}(2020)\citenamefont {Arnal},
  \citenamefont {Chatelain}, \citenamefont {Martinez}, \citenamefont {Dupont},
  \citenamefont {Giraud}, \citenamefont {Ullmo}, \citenamefont {Georgeot},
  \citenamefont {Lemari{\'e}}, \citenamefont {Billy},\ and\ \citenamefont
  {{Gu{\'e}ry-Odelin}}}]{arnalChaosassisted2020}%
  \BibitemOpen
  \bibfield  {author} {\bibinfo {author} {\bibfnamefont {M.}~\bibnamefont
  {Arnal}}, \bibinfo {author} {\bibfnamefont {G.}~\bibnamefont {Chatelain}},
  \bibinfo {author} {\bibfnamefont {M.}~\bibnamefont {Martinez}}, \bibinfo
  {author} {\bibfnamefont {N.}~\bibnamefont {Dupont}}, \bibinfo {author}
  {\bibfnamefont {O.}~\bibnamefont {Giraud}}, \bibinfo {author} {\bibfnamefont
  {D.}~\bibnamefont {Ullmo}}, \bibinfo {author} {\bibfnamefont
  {B.}~\bibnamefont {Georgeot}}, \bibinfo {author} {\bibfnamefont
  {G.}~\bibnamefont {Lemari{\'e}}}, \bibinfo {author} {\bibfnamefont
  {J.}~\bibnamefont {Billy}}, \ and\ \bibinfo {author} {\bibfnamefont
  {D.}~\bibnamefont {{Gu{\'e}ry-Odelin}}},\ }\bibfield  {title} {\enquote
  {\bibinfo {title} {Chaos-assisted tunneling resonances in a synthetic
  {{Floquet}} superlattice},}\ }\href {\doibase 10.1126/sciadv.abc4886}
  {\bibfield  {journal} {\bibinfo  {journal} {Science Advances}\ }\textbf
  {\bibinfo {volume} {6}},\ \bibinfo {pages} {eabc4886} (\bibinfo {year}
  {2020})}\BibitemShut {NoStop}%
\bibitem [{\citenamefont {Ikeda}\ \emph {et~al.}(2022)\citenamefont {Ikeda},
  \citenamefont {Tanaka},\ and\ \citenamefont
  {Kayanuma}}]{ikedaFloquetLandauZener2022}%
  \BibitemOpen
  \bibfield  {author} {\bibinfo {author} {\bibfnamefont {Tatsuhiko~N.}\
  \bibnamefont {Ikeda}}, \bibinfo {author} {\bibfnamefont {Satoshi}\
  \bibnamefont {Tanaka}}, \ and\ \bibinfo {author} {\bibfnamefont {Yosuke}\
  \bibnamefont {Kayanuma}},\ }\bibfield  {title} {\enquote {\bibinfo {title}
  {Floquet-{{Landau-Zener}} interferometry: {{Usefulness}} of the {{Floquet}}
  theory in pulse-laser-driven systems},}\ }\href {\doibase
  10.1103/PhysRevResearch.4.033075} {\bibfield  {journal} {\bibinfo  {journal}
  {Physical Review Research}\ }\textbf {\bibinfo {volume} {4}},\ \bibinfo
  {pages} {033075} (\bibinfo {year} {2022})}\BibitemShut {NoStop}%
\bibitem [{\citenamefont {Wang}\ \emph {et~al.}(2023)\citenamefont {Wang},
  \citenamefont {Qin}, \citenamefont {Zhao}, \citenamefont {Ye}, \citenamefont
  {Longhi}, \citenamefont {Lu},\ and\ \citenamefont {Wang}}]{wangPhotonic2023}%
  \BibitemOpen
  \bibfield  {author} {\bibinfo {author} {\bibfnamefont {Shulin}\ \bibnamefont
  {Wang}}, \bibinfo {author} {\bibfnamefont {Chengzhi}\ \bibnamefont {Qin}},
  \bibinfo {author} {\bibfnamefont {Lange}\ \bibnamefont {Zhao}}, \bibinfo
  {author} {\bibfnamefont {Han}\ \bibnamefont {Ye}}, \bibinfo {author}
  {\bibfnamefont {Stefano}\ \bibnamefont {Longhi}}, \bibinfo {author}
  {\bibfnamefont {Peixiang}\ \bibnamefont {Lu}}, \ and\ \bibinfo {author}
  {\bibfnamefont {Bing}\ \bibnamefont {Wang}},\ }\bibfield  {title} {\enquote
  {\bibinfo {title} {Photonic {{Floquet Landau-Zener}} tunneling and temporal
  beam splitters},}\ }\href {\doibase 10.1126/sciadv.adh0415} {\bibfield
  {journal} {\bibinfo  {journal} {Science Advances}\ }\textbf {\bibinfo
  {volume} {9}},\ \bibinfo {pages} {eadh0415} (\bibinfo {year}
  {2023})}\BibitemShut {NoStop}%
\bibitem [{\citenamefont {Sato}\ \emph {et~al.}(2020)\citenamefont {Sato},
  \citenamefont {Giovannini}, \citenamefont {Aeschlimann}, \citenamefont
  {Gierz}, \citenamefont {H{\"u}bener},\ and\ \citenamefont
  {Rubio}}]{satoFloquet2020}%
  \BibitemOpen
  \bibfield  {author} {\bibinfo {author} {\bibfnamefont {S.~A.}\ \bibnamefont
  {Sato}}, \bibinfo {author} {\bibfnamefont {U.~De}\ \bibnamefont
  {Giovannini}}, \bibinfo {author} {\bibfnamefont {S.}~\bibnamefont
  {Aeschlimann}}, \bibinfo {author} {\bibfnamefont {I.}~\bibnamefont {Gierz}},
  \bibinfo {author} {\bibfnamefont {H.}~\bibnamefont {H{\"u}bener}}, \ and\
  \bibinfo {author} {\bibfnamefont {A.}~\bibnamefont {Rubio}},\ }\bibfield
  {title} {\enquote {\bibinfo {title} {Floquet states in dissipative open
  quantum systems},}\ }\href {\doibase 10.1088/1361-6455/abb127} {\bibfield
  {journal} {\bibinfo  {journal} {Journal of Physics B: Atomic, Molecular and
  Optical Physics}\ }\textbf {\bibinfo {volume} {53}},\ \bibinfo {pages}
  {225601} (\bibinfo {year} {2020})}\BibitemShut {NoStop}%
\bibitem [{\citenamefont {Mori}(2023)}]{moriFloquet2023}%
  \BibitemOpen
  \bibfield  {author} {\bibinfo {author} {\bibfnamefont {Takashi}\ \bibnamefont
  {Mori}},\ }\bibfield  {title} {\enquote {\bibinfo {title} {Floquet {{States}}
  in {{Open Quantum Systems}}},}\ }\href {\doibase
  10.1146/annurev-conmatphys-040721-015537} {\bibfield  {journal} {\bibinfo
  {journal} {Annual Review of Condensed Matter Physics}\ }\textbf {\bibinfo
  {volume} {14}},\ \bibinfo {pages} {35--56} (\bibinfo {year}
  {2023})}\BibitemShut {NoStop}%
\bibitem [{\citenamefont {Shan}\ \emph {et~al.}(2021)\citenamefont {Shan},
  \citenamefont {Ye}, \citenamefont {Chu}, \citenamefont {Lee}, \citenamefont
  {Park}, \citenamefont {Balents},\ and\ \citenamefont
  {Hsieh}}]{shanGiant2021}%
  \BibitemOpen
  \bibfield  {author} {\bibinfo {author} {\bibfnamefont {Jun-Yi}\ \bibnamefont
  {Shan}}, \bibinfo {author} {\bibfnamefont {M.}~\bibnamefont {Ye}}, \bibinfo
  {author} {\bibfnamefont {H.}~\bibnamefont {Chu}}, \bibinfo {author}
  {\bibfnamefont {Sungmin}\ \bibnamefont {Lee}}, \bibinfo {author}
  {\bibfnamefont {Je-Geun}\ \bibnamefont {Park}}, \bibinfo {author}
  {\bibfnamefont {L.}~\bibnamefont {Balents}}, \ and\ \bibinfo {author}
  {\bibfnamefont {D.}~\bibnamefont {Hsieh}},\ }\bibfield  {title} {\enquote
  {\bibinfo {title} {Giant modulation of optical nonlinearity by {{Floquet}}
  engineering},}\ }\href {\doibase 10.1038/s41586-021-04051-8} {\bibfield
  {journal} {\bibinfo  {journal} {Nature}\ }\textbf {\bibinfo {volume} {600}},\
  \bibinfo {pages} {235--239} (\bibinfo {year} {2021})}\BibitemShut {NoStop}%
\bibitem [{\citenamefont {Mukherjee}\ and\ \citenamefont
  {Rechtsman}(2020)}]{mukherjeeObservation2020}%
  \BibitemOpen
  \bibfield  {author} {\bibinfo {author} {\bibfnamefont {Sebabrata}\
  \bibnamefont {Mukherjee}}\ and\ \bibinfo {author} {\bibfnamefont {Mikael~C.}\
  \bibnamefont {Rechtsman}},\ }\bibfield  {title} {\enquote {\bibinfo {title}
  {Observation of {{Floquet}} solitons in a topological bandgap},}\ }\href
  {\doibase 10.1126/science.aba8725} {\bibfield  {journal} {\bibinfo  {journal}
  {Science}\ }\textbf {\bibinfo {volume} {368}},\ \bibinfo {pages} {856--859}
  (\bibinfo {year} {2020})}\BibitemShut {NoStop}%
\bibitem [{\citenamefont {Lu}\ \emph {et~al.}(2021)\citenamefont {Lu},
  \citenamefont {He}, \citenamefont {Addison}, \citenamefont {Mele},\ and\
  \citenamefont {Zhen}}]{luFloquet2021}%
  \BibitemOpen
  \bibfield  {author} {\bibinfo {author} {\bibfnamefont {Jian}\ \bibnamefont
  {Lu}}, \bibinfo {author} {\bibfnamefont {Li}~\bibnamefont {He}}, \bibinfo
  {author} {\bibfnamefont {Zachariah}\ \bibnamefont {Addison}}, \bibinfo
  {author} {\bibfnamefont {Eugene~J.}\ \bibnamefont {Mele}}, \ and\ \bibinfo
  {author} {\bibfnamefont {Bo}~\bibnamefont {Zhen}},\ }\bibfield  {title}
  {\enquote {\bibinfo {title} {Floquet {{Topological Phases}} in
  {{One-Dimensional Nonlinear Photonic Crystals}}},}\ }\href {\doibase
  10.1103/PhysRevLett.126.113901} {\bibfield  {journal} {\bibinfo  {journal}
  {Physical Review Letters}\ }\textbf {\bibinfo {volume} {126}},\ \bibinfo
  {pages} {113901} (\bibinfo {year} {2021})}\BibitemShut {NoStop}%
\bibitem [{\citenamefont {Goldman}\ \emph {et~al.}(2023)\citenamefont
  {Goldman}, \citenamefont {Diessel}, \citenamefont {Barbiero}, \citenamefont
  {Pr{\"u}fer}, \citenamefont {Di~Liberto},\ and\ \citenamefont
  {Peralta~Gavensky}}]{goldmanFloquetEngineered2023a}%
  \BibitemOpen
  \bibfield  {author} {\bibinfo {author} {\bibfnamefont {N.}~\bibnamefont
  {Goldman}}, \bibinfo {author} {\bibfnamefont {O.K.}\ \bibnamefont {Diessel}},
  \bibinfo {author} {\bibfnamefont {L.}~\bibnamefont {Barbiero}}, \bibinfo
  {author} {\bibfnamefont {M.}~\bibnamefont {Pr{\"u}fer}}, \bibinfo {author}
  {\bibfnamefont {M.}~\bibnamefont {Di~Liberto}}, \ and\ \bibinfo {author}
  {\bibfnamefont {L.}~\bibnamefont {Peralta~Gavensky}},\ }\bibfield  {title}
  {\enquote {\bibinfo {title} {Floquet-{{Engineered Nonlinearities}} and
  {{Controllable Pair-Hopping Processes}}: {{From Optical Kerr Cavities}} to
  {{Correlated Quantum Matter}}},}\ }\href {\doibase
  10.1103/PRXQuantum.4.040327} {\bibfield  {journal} {\bibinfo  {journal} {PRX
  Quantum}\ }\textbf {\bibinfo {volume} {4}},\ \bibinfo {pages} {040327}
  (\bibinfo {year} {2023})}\BibitemShut {NoStop}%
\bibitem [{\citenamefont {Masluk}\ \emph {et~al.}(2012)\citenamefont {Masluk},
  \citenamefont {Pop}, \citenamefont {Kamal}, \citenamefont {Minev},\ and\
  \citenamefont {Devoret}}]{maslukMicrowave2012a}%
  \BibitemOpen
  \bibfield  {author} {\bibinfo {author} {\bibfnamefont {Nicholas~A.}\
  \bibnamefont {Masluk}}, \bibinfo {author} {\bibfnamefont {Ioan~M.}\
  \bibnamefont {Pop}}, \bibinfo {author} {\bibfnamefont {Archana}\ \bibnamefont
  {Kamal}}, \bibinfo {author} {\bibfnamefont {Zlatko~K.}\ \bibnamefont
  {Minev}}, \ and\ \bibinfo {author} {\bibfnamefont {Michel~H.}\ \bibnamefont
  {Devoret}},\ }\bibfield  {title} {\enquote {\bibinfo {title} {Microwave
  {{Characterization}} of {{Josephson Junction Arrays}}: {{Implementing}} a
  {{Low Loss Superinductance}}},}\ }\href {\doibase
  10.1103/PhysRevLett.109.137002} {\bibfield  {journal} {\bibinfo  {journal}
  {Physical Review Letters}\ }\textbf {\bibinfo {volume} {109}},\ \bibinfo
  {pages} {137002} (\bibinfo {year} {2012})}\BibitemShut {NoStop}%
\bibitem [{\citenamefont {Wei{\ss}l}\ \emph {et~al.}(2015)\citenamefont
  {Wei{\ss}l}, \citenamefont {K{\"u}ng}, \citenamefont {Dumur}, \citenamefont
  {Feofanov}, \citenamefont {Matei}, \citenamefont {Naud}, \citenamefont
  {Buisson}, \citenamefont {Hekking},\ and\ \citenamefont
  {Guichard}}]{weisslKerr2015}%
  \BibitemOpen
  \bibfield  {author} {\bibinfo {author} {\bibfnamefont {T.}~\bibnamefont
  {Wei{\ss}l}}, \bibinfo {author} {\bibfnamefont {B.}~\bibnamefont {K{\"u}ng}},
  \bibinfo {author} {\bibfnamefont {E.}~\bibnamefont {Dumur}}, \bibinfo
  {author} {\bibfnamefont {A.~K.}\ \bibnamefont {Feofanov}}, \bibinfo {author}
  {\bibfnamefont {I.}~\bibnamefont {Matei}}, \bibinfo {author} {\bibfnamefont
  {C.}~\bibnamefont {Naud}}, \bibinfo {author} {\bibfnamefont {O.}~\bibnamefont
  {Buisson}}, \bibinfo {author} {\bibfnamefont {F.~W.~J.}\ \bibnamefont
  {Hekking}}, \ and\ \bibinfo {author} {\bibfnamefont {W.}~\bibnamefont
  {Guichard}},\ }\bibfield  {title} {\enquote {\bibinfo {title} {Kerr
  coefficients of plasma resonances in {{Josephson}} junction chains},}\ }\href
  {\doibase 10.1103/PhysRevB.92.104508} {\bibfield  {journal} {\bibinfo
  {journal} {Physical Review B}\ }\textbf {\bibinfo {volume} {92}},\ \bibinfo
  {pages} {104508} (\bibinfo {year} {2015})}\BibitemShut {NoStop}%
\bibitem [{\citenamefont {Krupko}\ \emph {et~al.}(2018)\citenamefont {Krupko},
  \citenamefont {Nguyen}, \citenamefont {Wei{\ss}l}, \citenamefont {Dumur},
  \citenamefont {Puertas}, \citenamefont {Dassonneville}, \citenamefont {Naud},
  \citenamefont {Hekking}, \citenamefont {Basko}, \citenamefont {Buisson},
  \citenamefont {Roch},\ and\ \citenamefont
  {{Hasch-Guichard}}}]{krupkoKerr2018}%
  \BibitemOpen
  \bibfield  {author} {\bibinfo {author} {\bibfnamefont {{\relax
  Yu}.}~\bibnamefont {Krupko}}, \bibinfo {author} {\bibfnamefont {V.~D.}\
  \bibnamefont {Nguyen}}, \bibinfo {author} {\bibfnamefont {T.}~\bibnamefont
  {Wei{\ss}l}}, \bibinfo {author} {\bibfnamefont {{\'E}.}~\bibnamefont
  {Dumur}}, \bibinfo {author} {\bibfnamefont {J.}~\bibnamefont {Puertas}},
  \bibinfo {author} {\bibfnamefont {R.}~\bibnamefont {Dassonneville}}, \bibinfo
  {author} {\bibfnamefont {C.}~\bibnamefont {Naud}}, \bibinfo {author}
  {\bibfnamefont {F.~W.~J.}\ \bibnamefont {Hekking}}, \bibinfo {author}
  {\bibfnamefont {D.~M.}\ \bibnamefont {Basko}}, \bibinfo {author}
  {\bibfnamefont {O.}~\bibnamefont {Buisson}}, \bibinfo {author} {\bibfnamefont
  {N.}~\bibnamefont {Roch}}, \ and\ \bibinfo {author} {\bibfnamefont
  {W.}~\bibnamefont {{Hasch-Guichard}}},\ }\bibfield  {title} {\enquote
  {\bibinfo {title} {Kerr nonlinearity in a superconducting {{Josephson}}
  metamaterial},}\ }\href {\doibase 10.1103/PhysRevB.98.094516} {\bibfield
  {journal} {\bibinfo  {journal} {Physical Review B}\ }\textbf {\bibinfo
  {volume} {98}},\ \bibinfo {pages} {094516} (\bibinfo {year}
  {2018})}\BibitemShut {NoStop}%
\bibitem [{\citenamefont {Sivak}\ \emph {et~al.}(2020)\citenamefont {Sivak},
  \citenamefont {Shankar}, \citenamefont {Liu}, \citenamefont {Aumentado},\
  and\ \citenamefont {Devoret}}]{sivakJosephson2020}%
  \BibitemOpen
  \bibfield  {author} {\bibinfo {author} {\bibfnamefont {V.~V.}\ \bibnamefont
  {Sivak}}, \bibinfo {author} {\bibfnamefont {S.}~\bibnamefont {Shankar}},
  \bibinfo {author} {\bibfnamefont {G.}~\bibnamefont {Liu}}, \bibinfo {author}
  {\bibfnamefont {J.}~\bibnamefont {Aumentado}}, \ and\ \bibinfo {author}
  {\bibfnamefont {M.~H.}\ \bibnamefont {Devoret}},\ }\bibfield  {title}
  {\enquote {\bibinfo {title} {Josephson {{Array-Mode Parametric
  Amplifier}}},}\ }\href {\doibase 10.1103/PhysRevApplied.13.024014} {\bibfield
   {journal} {\bibinfo  {journal} {Physical Review Applied}\ }\textbf {\bibinfo
  {volume} {13}},\ \bibinfo {pages} {024014} (\bibinfo {year}
  {2020})}\BibitemShut {NoStop}%
\bibitem [{\citenamefont {Lidar}(2019)}]{LidarARXIV19}%
  \BibitemOpen
  \bibfield  {author} {\bibinfo {author} {\bibfnamefont {Daniel~A.}\
  \bibnamefont {Lidar}},\ }\bibfield  {title} {\enquote {\bibinfo {title}
  {Lecture notes on the theory of open quantum systems},}\ }\href {\doibase
  10.48550/ARXIV.1902.00967} {\  (\bibinfo {year} {2019}),\
  10.48550/ARXIV.1902.00967}\BibitemShut {NoStop}%
\bibitem [{Note2()}]{Note2}%
  \BibitemOpen
  \bibinfo {note} {The resonator exhibits deviation from the pure Kerr
  nonlinearity prediction due to non-negligible effects of higher
  nonlinearities, e.g., terms of the form $\chi ^{(5)} (\protect \hat
  {a}^\dagger )^3 \protect \hat {a}^3$. We estimate $\chi ^{(5)}/2\pi \approx
  -1.1\protect \,$MHz$\simeq 5\% \chi /2\pi $. As such, although the $\chi
  ^{(5)}$ term produces small changes compared to the model in \protect \mbox
  {Eq.~(\ref {Eq:Hamiltonian})}, all the relevant physical features of the LZSM
  interference are captured by the Kerr model. We also note additional
  nonlinear effects due to the large values of flux modulation used to obtain
  the wanted $\zeta $.}\BibitemShut {Stop}%
\bibitem [{Note3()}]{Note3}%
  \BibitemOpen
  \bibinfo {note} {One shows that, assuming at most two-photon in the system,
  the maximum of the two-photon population occurs at multiphoton resonance
  $\Delta = \chi $, where $$ \left \langle 2\left | \protect \hat {\rho
  }_{\protect \rm ss} \right | 2\right \rangle =\protect \frac {2 F^4}{9 F^4+2
  \kappa ^2 \left [2 \left (\kappa ^2+\chi ^2\right )-5 F^2\right ]} \simeq
  \protect \frac { F^4}{2 \kappa ^2 \chi ^2}. $$ It follows that $F^2 \ll |\chi
  |\kappa $ ensures the validity of the qubit approximation}\BibitemShut
  {NoStop}%
\bibitem [{Note4()}]{Note4}%
  \BibitemOpen
  \bibinfo {note} {To justify this approximation, consider the semiclassical
  (coherent state approximation) $\protect \hat {\rho }_{\protect \rm ss}=
  \left | \alpha \rangle \langle \alpha \right |$. One finds that the photon
  number $n=|\alpha |^2$ satisfies \begin {equation*} \begin {split} &\left [
  \protect \frac {\kappa ^2}{4} +(\Delta +n \chi )^2 \right ] n -F^2 \\ &\hskip
  1em\relax \simeq 2 \Delta n^2 \chi +n \left (\Delta ^2+\protect \frac {\kappa
  ^2}{4}\right )-F^2=0. \end {split} \end {equation*} The solution to this
  equation can be expanded in powers of $\chi $ as $$ n = n_0 \left (1 - n_0
  \protect \frac {8 \Delta \chi }{4 \Delta ^2 + \kappa ^2} \right ), \protect
  \, \protect \text { with } \protect \, n_0 = \protect \frac {F^2}{\Delta ^2 +
  \kappa ^2/4}. $$ The deviation from the linear regime $\delta n = 1- n / n_0
  $ is then maximal for $\Delta = \kappa / (2\protect \sqrt {3})$, and leading
  to $$ \delta n = \protect \frac {3 \protect \sqrt {3} F^2 \chi }{\kappa ^3}.
  $$ As we are interested in $\delta n \ll 1$, we find back the formula in the
  main text.}\BibitemShut {Stop}%
\bibitem [{\citenamefont {Lee}\ \emph {et~al.}(2020)\citenamefont {Lee},
  \citenamefont {Pechal}, \citenamefont {Wollack}, \citenamefont
  {{Arrangoiz-Arriola}}, \citenamefont {Wang},\ and\ \citenamefont
  {{Safavi-Naeni}}}]{leePropagation2020}%
  \BibitemOpen
  \bibfield  {author} {\bibinfo {author} {\bibfnamefont {Nathan R.~A.}\
  \bibnamefont {Lee}}, \bibinfo {author} {\bibfnamefont {Marek}\ \bibnamefont
  {Pechal}}, \bibinfo {author} {\bibfnamefont {E.~Alex}\ \bibnamefont
  {Wollack}}, \bibinfo {author} {\bibfnamefont {Patricio}\ \bibnamefont
  {{Arrangoiz-Arriola}}}, \bibinfo {author} {\bibfnamefont {Zhaoyou}\
  \bibnamefont {Wang}}, \ and\ \bibinfo {author} {\bibfnamefont {Amir~H.}\
  \bibnamefont {{Safavi-Naeni}}},\ }\bibfield  {title} {\enquote {\bibinfo
  {title} {Propagation of microwave photons along a synthetic dimension},}\
  }\href {\doibase 10.1103/PhysRevA.101.053807} {\bibfield  {journal} {\bibinfo
   {journal} {Physical Review A}\ }\textbf {\bibinfo {volume} {101}},\ \bibinfo
  {pages} {053807} (\bibinfo {year} {2020})}\BibitemShut {NoStop}%
\bibitem [{\citenamefont {Lecocq}\ \emph {et~al.}(2017)\citenamefont {Lecocq},
  \citenamefont {Ranzani}, \citenamefont {Peterson}, \citenamefont {Cicak},
  \citenamefont {Simmonds}, \citenamefont {Teufel},\ and\ \citenamefont
  {Aumentado}}]{lecocqNonreciprocal2017}%
  \BibitemOpen
  \bibfield  {author} {\bibinfo {author} {\bibfnamefont {F.}~\bibnamefont
  {Lecocq}}, \bibinfo {author} {\bibfnamefont {L.}~\bibnamefont {Ranzani}},
  \bibinfo {author} {\bibfnamefont {G.~A.}\ \bibnamefont {Peterson}}, \bibinfo
  {author} {\bibfnamefont {K.}~\bibnamefont {Cicak}}, \bibinfo {author}
  {\bibfnamefont {R.~W.}\ \bibnamefont {Simmonds}}, \bibinfo {author}
  {\bibfnamefont {J.~D.}\ \bibnamefont {Teufel}}, \ and\ \bibinfo {author}
  {\bibfnamefont {J.}~\bibnamefont {Aumentado}},\ }\bibfield  {title} {\enquote
  {\bibinfo {title} {Nonreciprocal microwave signal processing with a
  {{Field-Programmable Josephson Amplifier}}},}\ }\href {\doibase
  10.1103/PhysRevApplied.7.024028} {\bibfield  {journal} {\bibinfo  {journal}
  {Physical Review Applied}\ }\textbf {\bibinfo {volume} {7}},\ \bibinfo
  {pages} {024028} (\bibinfo {year} {2017})},\ \Eprint
  {http://arxiv.org/abs/1612.01438} {arXiv:1612.01438 [quant-ph]} \BibitemShut
  {NoStop}%
\bibitem [{\citenamefont {Zagoskin}\ \emph {et~al.}(2008)\citenamefont
  {Zagoskin}, \citenamefont {Il'ichev}, \citenamefont {McCutcheon},
  \citenamefont {Young},\ and\ \citenamefont {Nori}}]{zagoskinControlled2008}%
  \BibitemOpen
  \bibfield  {author} {\bibinfo {author} {\bibfnamefont {A.~M.}\ \bibnamefont
  {Zagoskin}}, \bibinfo {author} {\bibfnamefont {E.}~\bibnamefont {Il'ichev}},
  \bibinfo {author} {\bibfnamefont {M.~W.}\ \bibnamefont {McCutcheon}},
  \bibinfo {author} {\bibfnamefont {Jeff~F.}\ \bibnamefont {Young}}, \ and\
  \bibinfo {author} {\bibfnamefont {Franco}\ \bibnamefont {Nori}},\ }\bibfield
  {title} {\enquote {\bibinfo {title} {Controlled {{Generation}} of {{Squeezed
  States}} of {{Microwave Radiation}} in a {{Superconducting Resonant
  Circuit}}},}\ }\href {\doibase 10.1103/PhysRevLett.101.253602} {\bibfield
  {journal} {\bibinfo  {journal} {Physical Review Letters}\ }\textbf {\bibinfo
  {volume} {101}},\ \bibinfo {pages} {253602} (\bibinfo {year}
  {2008})}\BibitemShut {NoStop}%
\bibitem [{\citenamefont {Silveri}\ \emph {et~al.}(2017)\citenamefont
  {Silveri}, \citenamefont {Tuorila}, \citenamefont {Thuneberg},\ and\
  \citenamefont {Paraoanu}}]{silveriQuantum2017}%
  \BibitemOpen
  \bibfield  {author} {\bibinfo {author} {\bibfnamefont {M.~P.}\ \bibnamefont
  {Silveri}}, \bibinfo {author} {\bibfnamefont {J.~A.}\ \bibnamefont
  {Tuorila}}, \bibinfo {author} {\bibfnamefont {E.~V.}\ \bibnamefont
  {Thuneberg}}, \ and\ \bibinfo {author} {\bibfnamefont {G.~S.}\ \bibnamefont
  {Paraoanu}},\ }\bibfield  {title} {\enquote {\bibinfo {title} {Quantum
  systems under frequency modulation},}\ }\href {\doibase
  10.1088/1361-6633/aa5170} {\bibfield  {journal} {\bibinfo  {journal} {Reports
  on Progress in Physics}\ }\textbf {\bibinfo {volume} {80}},\ \bibinfo {pages}
  {056002} (\bibinfo {year} {2017})}\BibitemShut {NoStop}%
\bibitem [{\citenamefont {Le~Boit{\'e}}(2015)}]{leboiteTHESIS15}%
  \BibitemOpen
  \bibfield  {author} {\bibinfo {author} {\bibfnamefont {Alexandre}\
  \bibnamefont {Le~Boit{\'e}}},\ }\emph {\bibinfo {title} {{Strongly correlated
  photons in arrays of nonlinear cavities}}},\ \href
  {https://theses.hal.science/tel-01172202} {\bibinfo {type} {Theses}},\
  \bibinfo  {school} {{Universit{\'e} Paris Diderot - Paris 7}} (\bibinfo
  {year} {2015})\BibitemShut {NoStop}%
\bibitem [{\citenamefont {Bj{\"o}rkman}\ \emph {et~al.}(2024)\citenamefont
  {Bj{\"o}rkman}, \citenamefont {Kuzmanovi{\'c}},\ and\ \citenamefont
  {Paraoanu}}]{bjorkmanObservation2024}%
  \BibitemOpen
  \bibfield  {author} {\bibinfo {author} {\bibfnamefont {Isak}\ \bibnamefont
  {Bj{\"o}rkman}}, \bibinfo {author} {\bibfnamefont {Marko}\ \bibnamefont
  {Kuzmanovi{\'c}}}, \ and\ \bibinfo {author} {\bibfnamefont {Gheorghe~Sorin}\
  \bibnamefont {Paraoanu}},\ }\bibfield  {title} {\enquote {\bibinfo {title}
  {Observation of the two-photon {{Landau-Zener-St}}{\"u}ckelberg-{{Majorana}}
  effect},}\ }\href {\doibase 10.48550/arXiv.2402.10833} {\  (\bibinfo {year}
  {2024}),\ 10.48550/arXiv.2402.10833},\ \Eprint
  {http://arxiv.org/abs/2402.10833} {2402.10833 [cond-mat, physics:quant-ph]}
  \BibitemShut {NoStop}%
\bibitem [{\citenamefont {Strogatz}(2018)}]{strogatz_nonlinear_2018}%
  \BibitemOpen
  \bibfield  {author} {\bibinfo {author} {\bibfnamefont {Steven~H.}\
  \bibnamefont {Strogatz}},\ }\href {\doibase 10.1201/9780429492563} {\emph
  {\bibinfo {title} {Nonlinear {Dynamics} and {Chaos}}}},\ \bibinfo {edition}
  {0th}\ ed.\ (\bibinfo  {publisher} {CRC Press},\ \bibinfo {year}
  {2018})\BibitemShut {NoStop}%
\bibitem [{\citenamefont {Bohigas}\ \emph {et~al.}(1984)\citenamefont
  {Bohigas}, \citenamefont {Giannoni},\ and\ \citenamefont
  {Schmit}}]{bohigas_characterization_1984}%
  \BibitemOpen
  \bibfield  {author} {\bibinfo {author} {\bibfnamefont {O.}~\bibnamefont
  {Bohigas}}, \bibinfo {author} {\bibfnamefont {M.~J.}\ \bibnamefont
  {Giannoni}}, \ and\ \bibinfo {author} {\bibfnamefont {C.}~\bibnamefont
  {Schmit}},\ }\bibfield  {title} {\enquote {\bibinfo {title} {Characterization
  of {Chaotic} {Quantum} {Spectra} and {Universality} of {Level} {Fluctuation}
  {Laws}},}\ }\href {\doibase 10.1103/PhysRevLett.52.1} {\bibfield  {journal}
  {\bibinfo  {journal} {Physical Review Letters}\ }\textbf {\bibinfo {volume}
  {52}},\ \bibinfo {pages} {1--4} (\bibinfo {year} {1984})}\BibitemShut
  {NoStop}%
\bibitem [{\citenamefont {Haake}(2001)}]{haake_quantum_2001}%
  \BibitemOpen
  \bibfield  {author} {\bibinfo {author} {\bibfnamefont {Fritz}\ \bibnamefont
  {Haake}},\ }\href {\doibase 10.1007/978-3-662-04506-0} {\emph {\bibinfo
  {title} {Quantum {Signatures} of {Chaos}}}},\ edited by\ \bibinfo {editor}
  {\bibfnamefont {Hermann}\ \bibnamefont {Haken}},\ \bibinfo {series} {Springer
  {Series} in {Synergetics}}, Vol.~\bibinfo {volume} {54}\ (\bibinfo
  {publisher} {Springer Berlin Heidelberg},\ \bibinfo {address} {Berlin,
  Heidelberg},\ \bibinfo {year} {2001})\BibitemShut {NoStop}%
\bibitem [{\citenamefont {D'Alessio}\ \emph {et~al.}(2016)\citenamefont
  {D'Alessio}, \citenamefont {Kafri}, \citenamefont {Polkovnikov},\ and\
  \citenamefont {Rigol}}]{dalessio_quantum_2016}%
  \BibitemOpen
  \bibfield  {author} {\bibinfo {author} {\bibfnamefont {Luca}\ \bibnamefont
  {D'Alessio}}, \bibinfo {author} {\bibfnamefont {Yariv}\ \bibnamefont
  {Kafri}}, \bibinfo {author} {\bibfnamefont {Anatoli}\ \bibnamefont
  {Polkovnikov}}, \ and\ \bibinfo {author} {\bibfnamefont {Marcos}\
  \bibnamefont {Rigol}},\ }\bibfield  {title} {\enquote {\bibinfo {title} {From
  quantum chaos and eigenstate thermalization to statistical mechanics and
  thermodynamics},}\ }\href {\doibase 10.1080/00018732.2016.1198134} {\bibfield
   {journal} {\bibinfo  {journal} {Advances in Physics}\ }\textbf {\bibinfo
  {volume} {65}},\ \bibinfo {pages} {239--362} (\bibinfo {year}
  {2016})}\BibitemShut {NoStop}%
\bibitem [{\citenamefont {Grobe}\ \emph {et~al.}(1988)\citenamefont {Grobe},
  \citenamefont {Haake},\ and\ \citenamefont {Sommers}}]{GrobePRL88}%
  \BibitemOpen
  \bibfield  {author} {\bibinfo {author} {\bibfnamefont {Rainer}\ \bibnamefont
  {Grobe}}, \bibinfo {author} {\bibfnamefont {Fritz}\ \bibnamefont {Haake}}, \
  and\ \bibinfo {author} {\bibfnamefont {Hans-Jürgen}\ \bibnamefont
  {Sommers}},\ }\bibfield  {title} {\enquote {\bibinfo {title} {Quantum
  {Distinction} of {Regular} and {Chaotic} {Dissipative} {Motion}},}\ }\href
  {\doibase 10.1103/PhysRevLett.61.1899} {\bibfield  {journal} {\bibinfo
  {journal} {Physical Review Letters}\ }\textbf {\bibinfo {volume} {61}},\
  \bibinfo {pages} {1899--1902} (\bibinfo {year} {1988})}\BibitemShut {NoStop}%
\bibitem [{\citenamefont {Sá}\ \emph {et~al.}(2020)\citenamefont {Sá},
  \citenamefont {Ribeiro},\ and\ \citenamefont {Prosen}}]{SaPRX20}%
  \BibitemOpen
  \bibfield  {author} {\bibinfo {author} {\bibfnamefont {Lucas}\ \bibnamefont
  {Sá}}, \bibinfo {author} {\bibfnamefont {Pedro}\ \bibnamefont {Ribeiro}}, \
  and\ \bibinfo {author} {\bibfnamefont {Tomaž}\ \bibnamefont {Prosen}},\
  }\bibfield  {title} {\enquote {\bibinfo {title} {Complex {Spacing} {Ratios}:
  {A} {Signature} of {Dissipative} {Quantum} {Chaos}},}\ }\href {\doibase
  10.1103/PhysRevX.10.021019} {\bibfield  {journal} {\bibinfo  {journal}
  {Physical Review X}\ }\textbf {\bibinfo {volume} {10}},\ \bibinfo {pages}
  {021019} (\bibinfo {year} {2020})}\BibitemShut {NoStop}%
\bibitem [{\citenamefont {Berke}\ \emph {et~al.}(2022)\citenamefont {Berke},
  \citenamefont {Varvelis}, \citenamefont {Trebst}, \citenamefont {Altland},\
  and\ \citenamefont {DiVincenzo}}]{berke_transmon_2022}%
  \BibitemOpen
  \bibfield  {author} {\bibinfo {author} {\bibfnamefont {Christoph}\
  \bibnamefont {Berke}}, \bibinfo {author} {\bibfnamefont {Evangelos}\
  \bibnamefont {Varvelis}}, \bibinfo {author} {\bibfnamefont {Simon}\
  \bibnamefont {Trebst}}, \bibinfo {author} {\bibfnamefont {Alexander}\
  \bibnamefont {Altland}}, \ and\ \bibinfo {author} {\bibfnamefont {David~P.}\
  \bibnamefont {DiVincenzo}},\ }\bibfield  {title} {\enquote {\bibinfo {title}
  {Transmon platform for quantum computing challenged by chaotic
  fluctuations},}\ }\href {\doibase 10.1038/s41467-022-29940-y} {\bibfield
  {journal} {\bibinfo  {journal} {Nature Communications}\ }\textbf {\bibinfo
  {volume} {13}},\ \bibinfo {pages} {2495} (\bibinfo {year}
  {2022})}\BibitemShut {NoStop}%
\bibitem [{\citenamefont {Dumas}\ \emph {et~al.}(2024)\citenamefont {Dumas},
  \citenamefont {{Groleau-Par{\'e}}}, \citenamefont {McDonald}, \citenamefont
  {{Mu{\~n}oz-Arias}}, \citenamefont {Lled{\'o}}, \citenamefont {D'Anjou},\
  and\ \citenamefont {Blais}}]{dumasMeasurementInduced2024}%
  \BibitemOpen
  \bibfield  {author} {\bibinfo {author} {\bibfnamefont
  {Marie~Fr{\'e}d{\'e}rique}\ \bibnamefont {Dumas}}, \bibinfo {author}
  {\bibfnamefont {Benjamin}\ \bibnamefont {{Groleau-Par{\'e}}}}, \bibinfo
  {author} {\bibfnamefont {Alexander}\ \bibnamefont {McDonald}}, \bibinfo
  {author} {\bibfnamefont {Manuel~H.}\ \bibnamefont {{Mu{\~n}oz-Arias}}},
  \bibinfo {author} {\bibfnamefont {Crist{\'o}bal}\ \bibnamefont {Lled{\'o}}},
  \bibinfo {author} {\bibfnamefont {Benjamin}\ \bibnamefont {D'Anjou}}, \ and\
  \bibinfo {author} {\bibfnamefont {Alexandre}\ \bibnamefont {Blais}},\
  }\bibfield  {title} {\enquote {\bibinfo {title} {Measurement-{{Induced
  Transmon Ionization}}},}\ }\href {\doibase 10.1103/PhysRevX.14.041023}
  {\bibfield  {journal} {\bibinfo  {journal} {Physical Review X}\ }\textbf
  {\bibinfo {volume} {14}},\ \bibinfo {pages} {041023} (\bibinfo {year}
  {2024})}\BibitemShut {NoStop}%
\bibitem [{\citenamefont {Wang}\ \emph {et~al.}(2019)\citenamefont {Wang},
  \citenamefont {Pechal}, \citenamefont {Wollack}, \citenamefont
  {{Arrangoiz-Arriola}}, \citenamefont {Gao}, \citenamefont {Lee},\ and\
  \citenamefont {{Safavi-Naeini}}}]{wangQuantum2019}%
  \BibitemOpen
  \bibfield  {author} {\bibinfo {author} {\bibfnamefont {Zhaoyou}\ \bibnamefont
  {Wang}}, \bibinfo {author} {\bibfnamefont {Marek}\ \bibnamefont {Pechal}},
  \bibinfo {author} {\bibfnamefont {E.~Alex}\ \bibnamefont {Wollack}}, \bibinfo
  {author} {\bibfnamefont {Patricio}\ \bibnamefont {{Arrangoiz-Arriola}}},
  \bibinfo {author} {\bibfnamefont {Maodong}\ \bibnamefont {Gao}}, \bibinfo
  {author} {\bibfnamefont {Nathan~R.}\ \bibnamefont {Lee}}, \ and\ \bibinfo
  {author} {\bibfnamefont {Amir~H.}\ \bibnamefont {{Safavi-Naeini}}},\
  }\bibfield  {title} {\enquote {\bibinfo {title} {Quantum {{Dynamics}} of a
  {{Few-Photon Parametric Oscillator}}},}\ }\href {\doibase
  10.1103/PhysRevX.9.021049} {\bibfield  {journal} {\bibinfo  {journal}
  {Physical Review X}\ }\textbf {\bibinfo {volume} {9}},\ \bibinfo {pages}
  {021049} (\bibinfo {year} {2019})}\BibitemShut {NoStop}%
\bibitem [{\citenamefont {Berdou}\ \emph {et~al.}(2023)\citenamefont {Berdou},
  \citenamefont {Murani}, \citenamefont {R{\'e}glade}, \citenamefont {Smith},
  \citenamefont {Villiers}, \citenamefont {Palomo}, \citenamefont {Rosticher},
  \citenamefont {Denis}, \citenamefont {Morfin}, \citenamefont {Delbecq},
  \citenamefont {Kontos}, \citenamefont {Pankratova}, \citenamefont
  {Rautschke}, \citenamefont {Peronnin}, \citenamefont {Sellem}, \citenamefont
  {Rouchon}, \citenamefont {Sarlette}, \citenamefont {Mirrahimi}, \citenamefont
  {{Campagne-Ibarcq}}, \citenamefont {Jezouin}, \citenamefont {Lescanne},\ and\
  \citenamefont {Leghtas}}]{berdouOne2023}%
  \BibitemOpen
  \bibfield  {author} {\bibinfo {author} {\bibfnamefont {C.}~\bibnamefont
  {Berdou}}, \bibinfo {author} {\bibfnamefont {A.}~\bibnamefont {Murani}},
  \bibinfo {author} {\bibfnamefont {U.}~\bibnamefont {R{\'e}glade}}, \bibinfo
  {author} {\bibfnamefont {W.C.}\ \bibnamefont {Smith}}, \bibinfo {author}
  {\bibfnamefont {M.}~\bibnamefont {Villiers}}, \bibinfo {author}
  {\bibfnamefont {J.}~\bibnamefont {Palomo}}, \bibinfo {author} {\bibfnamefont
  {M.}~\bibnamefont {Rosticher}}, \bibinfo {author} {\bibfnamefont
  {A.}~\bibnamefont {Denis}}, \bibinfo {author} {\bibfnamefont
  {P.}~\bibnamefont {Morfin}}, \bibinfo {author} {\bibfnamefont
  {M.}~\bibnamefont {Delbecq}}, \bibinfo {author} {\bibfnamefont
  {T.}~\bibnamefont {Kontos}}, \bibinfo {author} {\bibfnamefont
  {N.}~\bibnamefont {Pankratova}}, \bibinfo {author} {\bibfnamefont
  {F.}~\bibnamefont {Rautschke}}, \bibinfo {author} {\bibfnamefont
  {T.}~\bibnamefont {Peronnin}}, \bibinfo {author} {\bibfnamefont {L.-A.}\
  \bibnamefont {Sellem}}, \bibinfo {author} {\bibfnamefont {P.}~\bibnamefont
  {Rouchon}}, \bibinfo {author} {\bibfnamefont {A.}~\bibnamefont {Sarlette}},
  \bibinfo {author} {\bibfnamefont {M.}~\bibnamefont {Mirrahimi}}, \bibinfo
  {author} {\bibfnamefont {P.}~\bibnamefont {{Campagne-Ibarcq}}}, \bibinfo
  {author} {\bibfnamefont {S.}~\bibnamefont {Jezouin}}, \bibinfo {author}
  {\bibfnamefont {R.}~\bibnamefont {Lescanne}}, \ and\ \bibinfo {author}
  {\bibfnamefont {Z.}~\bibnamefont {Leghtas}},\ }\bibfield  {title} {\enquote
  {\bibinfo {title} {One {{Hundred Second Bit-Flip Time}} in a {{Two-Photon
  Dissipative Oscillator}}},}\ }\href {\doibase 10.1103/PRXQuantum.4.020350}
  {\bibfield  {journal} {\bibinfo  {journal} {PRX Quantum}\ }\textbf {\bibinfo
  {volume} {4}},\ \bibinfo {pages} {020350} (\bibinfo {year}
  {2023})}\BibitemShut {NoStop}%
\bibitem [{\citenamefont {Di~Candia}\ \emph {et~al.}(2023)\citenamefont
  {Di~Candia}, \citenamefont {Minganti}, \citenamefont {Petrovnin},
  \citenamefont {Paraoanu},\ and\ \citenamefont
  {Felicetti}}]{dicandiaCritical2023}%
  \BibitemOpen
  \bibfield  {author} {\bibinfo {author} {\bibfnamefont {R.}~\bibnamefont
  {Di~Candia}}, \bibinfo {author} {\bibfnamefont {F.}~\bibnamefont {Minganti}},
  \bibinfo {author} {\bibfnamefont {K.~V.}\ \bibnamefont {Petrovnin}}, \bibinfo
  {author} {\bibfnamefont {G.~S.}\ \bibnamefont {Paraoanu}}, \ and\ \bibinfo
  {author} {\bibfnamefont {S.}~\bibnamefont {Felicetti}},\ }\bibfield  {title}
  {\enquote {\bibinfo {title} {Critical parametric quantum sensing},}\ }\href
  {\doibase 10.1038/s41534-023-00690-z} {\bibfield  {journal} {\bibinfo
  {journal} {npj Quantum Information}\ }\textbf {\bibinfo {volume} {9}},\
  \bibinfo {pages} {1--9} (\bibinfo {year} {2023})}\BibitemShut {NoStop}%
\bibitem [{\citenamefont {Montenegro}\ \emph {et~al.}(2024)\citenamefont
  {Montenegro}, \citenamefont {Mukhopadhyay}, \citenamefont {Yousefjani},
  \citenamefont {Sarkar}, \citenamefont {Mishra}, \citenamefont {Paris},\ and\
  \citenamefont {Bayat}}]{montenegroReview2024}%
  \BibitemOpen
  \bibfield  {author} {\bibinfo {author} {\bibfnamefont {Victor}\ \bibnamefont
  {Montenegro}}, \bibinfo {author} {\bibfnamefont {Chiranjib}\ \bibnamefont
  {Mukhopadhyay}}, \bibinfo {author} {\bibfnamefont {Rozhin}\ \bibnamefont
  {Yousefjani}}, \bibinfo {author} {\bibfnamefont {Saubhik}\ \bibnamefont
  {Sarkar}}, \bibinfo {author} {\bibfnamefont {Utkarsh}\ \bibnamefont
  {Mishra}}, \bibinfo {author} {\bibfnamefont {Matteo G.~A.}\ \bibnamefont
  {Paris}}, \ and\ \bibinfo {author} {\bibfnamefont {Abolfazl}\ \bibnamefont
  {Bayat}},\ }\href {\doibase 10.48550/arXiv.2408.15323} {\enquote {\bibinfo
  {title} {Review: {{Quantum Metrology}} and {{Sensing}} with {{Many-Body
  Systems}}},}\ } (\bibinfo {year} {2024}),\ \Eprint
  {http://arxiv.org/abs/2408.15323} {arXiv:2408.15323 [quant-ph]} \BibitemShut
  {NoStop}%
\bibitem [{\citenamefont {Li}\ \emph {et~al.}(2017)\citenamefont {Li},
  \citenamefont {Fan}, \citenamefont {Wang}, \citenamefont {Ye}, \citenamefont
  {Zeng}, \citenamefont {Zhai}, \citenamefont {Peng},\ and\ \citenamefont
  {Du}}]{liMeasuring2017}%
  \BibitemOpen
  \bibfield  {author} {\bibinfo {author} {\bibfnamefont {Jun}\ \bibnamefont
  {Li}}, \bibinfo {author} {\bibfnamefont {Ruihua}\ \bibnamefont {Fan}},
  \bibinfo {author} {\bibfnamefont {Hengyan}\ \bibnamefont {Wang}}, \bibinfo
  {author} {\bibfnamefont {Bingtian}\ \bibnamefont {Ye}}, \bibinfo {author}
  {\bibfnamefont {Bei}\ \bibnamefont {Zeng}}, \bibinfo {author} {\bibfnamefont
  {Hui}\ \bibnamefont {Zhai}}, \bibinfo {author} {\bibfnamefont {Xinhua}\
  \bibnamefont {Peng}}, \ and\ \bibinfo {author} {\bibfnamefont {Jiangfeng}\
  \bibnamefont {Du}},\ }\bibfield  {title} {\enquote {\bibinfo {title}
  {Measuring {{Out-of-Time-Order Correlators}} on a {{Nuclear Magnetic
  Resonance Quantum Simulator}}},}\ }\href {\doibase 10.1103/PhysRevX.7.031011}
  {\bibfield  {journal} {\bibinfo  {journal} {Physical Review X}\ }\textbf
  {\bibinfo {volume} {7}},\ \bibinfo {pages} {031011} (\bibinfo {year}
  {2017})}\BibitemShut {NoStop}%
\bibitem [{\citenamefont {Braum{\"u}ller}\ \emph {et~al.}(2022)\citenamefont
  {Braum{\"u}ller}, \citenamefont {Karamlou}, \citenamefont {Yanay},
  \citenamefont {Kannan}, \citenamefont {Kim}, \citenamefont {Kjaergaard},
  \citenamefont {Melville}, \citenamefont {Niedzielski}, \citenamefont {Sung},
  \citenamefont {Veps{\"a}l{\"a}inen}, \citenamefont {Winik}, \citenamefont
  {Yoder}, \citenamefont {Orlando}, \citenamefont {Gustavsson}, \citenamefont
  {Tahan},\ and\ \citenamefont {Oliver}}]{braumullerProbing2022}%
  \BibitemOpen
  \bibfield  {author} {\bibinfo {author} {\bibfnamefont {Jochen}\ \bibnamefont
  {Braum{\"u}ller}}, \bibinfo {author} {\bibfnamefont {Amir~H.}\ \bibnamefont
  {Karamlou}}, \bibinfo {author} {\bibfnamefont {Yariv}\ \bibnamefont {Yanay}},
  \bibinfo {author} {\bibfnamefont {Bharath}\ \bibnamefont {Kannan}}, \bibinfo
  {author} {\bibfnamefont {David}\ \bibnamefont {Kim}}, \bibinfo {author}
  {\bibfnamefont {Morten}\ \bibnamefont {Kjaergaard}}, \bibinfo {author}
  {\bibfnamefont {Alexander}\ \bibnamefont {Melville}}, \bibinfo {author}
  {\bibfnamefont {Bethany~M.}\ \bibnamefont {Niedzielski}}, \bibinfo {author}
  {\bibfnamefont {Youngkyu}\ \bibnamefont {Sung}}, \bibinfo {author}
  {\bibfnamefont {Antti}\ \bibnamefont {Veps{\"a}l{\"a}inen}}, \bibinfo
  {author} {\bibfnamefont {Roni}\ \bibnamefont {Winik}}, \bibinfo {author}
  {\bibfnamefont {Jonilyn~L.}\ \bibnamefont {Yoder}}, \bibinfo {author}
  {\bibfnamefont {Terry~P.}\ \bibnamefont {Orlando}}, \bibinfo {author}
  {\bibfnamefont {Simon}\ \bibnamefont {Gustavsson}}, \bibinfo {author}
  {\bibfnamefont {Charles}\ \bibnamefont {Tahan}}, \ and\ \bibinfo {author}
  {\bibfnamefont {William~D.}\ \bibnamefont {Oliver}},\ }\bibfield  {title}
  {\enquote {\bibinfo {title} {Probing quantum information propagation with
  out-of-time-ordered correlators},}\ }\href {\doibase
  10.1038/s41567-021-01430-w} {\bibfield  {journal} {\bibinfo  {journal}
  {Nature Physics}\ }\textbf {\bibinfo {volume} {18}},\ \bibinfo {pages}
  {172--178} (\bibinfo {year} {2022})}\BibitemShut {NoStop}%
\bibitem [{\citenamefont {Xu}\ and\ \citenamefont
  {Swingle}(2024)}]{xuScrambling2024}%
  \BibitemOpen
  \bibfield  {author} {\bibinfo {author} {\bibfnamefont {Shenglong}\
  \bibnamefont {Xu}}\ and\ \bibinfo {author} {\bibfnamefont {Brian}\
  \bibnamefont {Swingle}},\ }\bibfield  {title} {\enquote {\bibinfo {title}
  {Scrambling {{Dynamics}} and {{Out-of-Time-Ordered Correlators}} in {{Quantum
  Many-Body Systems}}},}\ }\href {\doibase 10.1103/PRXQuantum.5.010201}
  {\bibfield  {journal} {\bibinfo  {journal} {PRX Quantum}\ }\textbf {\bibinfo
  {volume} {5}},\ \bibinfo {pages} {010201} (\bibinfo {year}
  {2024})}\BibitemShut {NoStop}%
\bibitem [{\citenamefont {Hung}\ \emph {et~al.}(2021)\citenamefont {Hung},
  \citenamefont {Busnaina}, \citenamefont {Chang}, \citenamefont {Vadiraj},
  \citenamefont {Nsanzineza}, \citenamefont {Solano}, \citenamefont {Alaeian},
  \citenamefont {Rico},\ and\ \citenamefont {Wilson}}]{hungQuantum2021}%
  \BibitemOpen
  \bibfield  {author} {\bibinfo {author} {\bibfnamefont {Jimmy S.~C.}\
  \bibnamefont {Hung}}, \bibinfo {author} {\bibfnamefont {J.~H.}\ \bibnamefont
  {Busnaina}}, \bibinfo {author} {\bibfnamefont {C.~W.~Sandbo}\ \bibnamefont
  {Chang}}, \bibinfo {author} {\bibfnamefont {A.~M.}\ \bibnamefont {Vadiraj}},
  \bibinfo {author} {\bibfnamefont {I.}~\bibnamefont {Nsanzineza}}, \bibinfo
  {author} {\bibfnamefont {E.}~\bibnamefont {Solano}}, \bibinfo {author}
  {\bibfnamefont {H.}~\bibnamefont {Alaeian}}, \bibinfo {author} {\bibfnamefont
  {E.}~\bibnamefont {Rico}}, \ and\ \bibinfo {author} {\bibfnamefont {C.~M.}\
  \bibnamefont {Wilson}},\ }\bibfield  {title} {\enquote {\bibinfo {title}
  {Quantum {{Simulation}} of the {{Bosonic Creutz Ladder}} with a {{Parametric
  Cavity}}},}\ }\href {\doibase 10.1103/PhysRevLett.127.100503} {\bibfield
  {journal} {\bibinfo  {journal} {Physical Review Letters}\ }\textbf {\bibinfo
  {volume} {127}},\ \bibinfo {pages} {100503} (\bibinfo {year}
  {2021})}\BibitemShut {NoStop}%
\bibitem [{\citenamefont {{G{\'o}mez-Le{\'o}n}}\ and\ \citenamefont
  {Platero}(2013)}]{gomez-leonFloquetBloch2013}%
  \BibitemOpen
  \bibfield  {author} {\bibinfo {author} {\bibfnamefont {A.}~\bibnamefont
  {{G{\'o}mez-Le{\'o}n}}}\ and\ \bibinfo {author} {\bibfnamefont
  {G.}~\bibnamefont {Platero}},\ }\bibfield  {title} {\enquote {\bibinfo
  {title} {Floquet-{{Bloch Theory}} and {{Topology}} in {{Periodically Driven
  Lattices}}},}\ }\href {\doibase 10.1103/PhysRevLett.110.200403} {\bibfield
  {journal} {\bibinfo  {journal} {Physical Review Letters}\ }\textbf {\bibinfo
  {volume} {110}},\ \bibinfo {pages} {200403} (\bibinfo {year}
  {2013})}\BibitemShut {NoStop}%
\bibitem [{\citenamefont {{Le{\'o}n-Montiel}}\ \emph
  {et~al.}(2018)\citenamefont {{Le{\'o}n-Montiel}}, \citenamefont
  {{Quiroz-Ju{\'a}rez}}, \citenamefont {{Dom{\'i}nguez-Ju{\'a}rez}},
  \citenamefont {{Quintero-Torres}}, \citenamefont {Arag{\'o}n}, \citenamefont
  {Harter},\ and\ \citenamefont {Joglekar}}]{leon-montielObservation2018a}%
  \BibitemOpen
  \bibfield  {author} {\bibinfo {author} {\bibfnamefont {Roberto de~J.}\
  \bibnamefont {{Le{\'o}n-Montiel}}}, \bibinfo {author} {\bibfnamefont
  {Mario~A.}\ \bibnamefont {{Quiroz-Ju{\'a}rez}}}, \bibinfo {author}
  {\bibfnamefont {Jorge~L.}\ \bibnamefont {{Dom{\'i}nguez-Ju{\'a}rez}}},
  \bibinfo {author} {\bibfnamefont {Rafael}\ \bibnamefont {{Quintero-Torres}}},
  \bibinfo {author} {\bibfnamefont {Jos{\'e}~L.}\ \bibnamefont {Arag{\'o}n}},
  \bibinfo {author} {\bibfnamefont {Andrew~K.}\ \bibnamefont {Harter}}, \ and\
  \bibinfo {author} {\bibfnamefont {Yogesh~N.}\ \bibnamefont {Joglekar}},\
  }\bibfield  {title} {\enquote {\bibinfo {title} {Observation of slowly
  decaying eigenmodes without exceptional points in {{Floquet}} dissipative
  synthetic circuits},}\ }\href {\doibase 10.1038/s42005-018-0087-3} {\bibfield
   {journal} {\bibinfo  {journal} {Communications Physics}\ }\textbf {\bibinfo
  {volume} {1}},\ \bibinfo {pages} {1--11} (\bibinfo {year}
  {2018})}\BibitemShut {NoStop}%
\bibitem [{\citenamefont {Yamaji}\ \emph {et~al.}(2023)\citenamefont {Yamaji},
  \citenamefont {Masuda}, \citenamefont {Yamaguchi}, \citenamefont {Satoh},
  \citenamefont {Morioka}, \citenamefont {Igarashi}, \citenamefont {Shirane},\
  and\ \citenamefont {Yamamoto}}]{yamajiCorrelated2023}%
  \BibitemOpen
  \bibfield  {author} {\bibinfo {author} {\bibfnamefont {T.}~\bibnamefont
  {Yamaji}}, \bibinfo {author} {\bibfnamefont {S.}~\bibnamefont {Masuda}},
  \bibinfo {author} {\bibfnamefont {A.}~\bibnamefont {Yamaguchi}}, \bibinfo
  {author} {\bibfnamefont {T.}~\bibnamefont {Satoh}}, \bibinfo {author}
  {\bibfnamefont {A.}~\bibnamefont {Morioka}}, \bibinfo {author} {\bibfnamefont
  {Y.}~\bibnamefont {Igarashi}}, \bibinfo {author} {\bibfnamefont
  {M.}~\bibnamefont {Shirane}}, \ and\ \bibinfo {author} {\bibfnamefont
  {T.}~\bibnamefont {Yamamoto}},\ }\bibfield  {title} {\enquote {\bibinfo
  {title} {Correlated {{Oscillations}} in {{Kerr Parametric Oscillators}} with
  {{Tunable Effective Coupling}}},}\ }\href {\doibase
  10.1103/PhysRevApplied.20.014057} {\bibfield  {journal} {\bibinfo  {journal}
  {Physical Review Applied}\ }\textbf {\bibinfo {volume} {20}},\ \bibinfo
  {pages} {014057} (\bibinfo {year} {2023})}\BibitemShut {NoStop}%
\bibitem [{\citenamefont {Heugel}\ \emph {et~al.}(2023)\citenamefont {Heugel},
  \citenamefont {Eichler}, \citenamefont {Chitra},\ and\ \citenamefont
  {Zilberberg}}]{heugelrole2023a}%
  \BibitemOpen
  \bibfield  {author} {\bibinfo {author} {\bibfnamefont {Toni~Louis}\
  \bibnamefont {Heugel}}, \bibinfo {author} {\bibfnamefont {Alexander}\
  \bibnamefont {Eichler}}, \bibinfo {author} {\bibfnamefont {Ramasubramanian}\
  \bibnamefont {Chitra}}, \ and\ \bibinfo {author} {\bibfnamefont {Oded}\
  \bibnamefont {Zilberberg}},\ }\bibfield  {title} {\enquote {\bibinfo {title}
  {The role of fluctuations in quantum and classical time crystals},}\ }\href
  {\doibase 10.21468/SciPostPhysCore.6.3.053} {\bibfield  {journal} {\bibinfo
  {journal} {SciPost Physics Core}\ }\textbf {\bibinfo {volume} {6}},\ \bibinfo
  {pages} {053} (\bibinfo {year} {2023})}\BibitemShut {NoStop}%
\bibitem [{\citenamefont {Dutt}\ \emph {et~al.}(2020)\citenamefont {Dutt},
  \citenamefont {Lin}, \citenamefont {Yuan}, \citenamefont {Minkov},
  \citenamefont {Xiao},\ and\ \citenamefont {Fan}}]{duttsingle2020}%
  \BibitemOpen
  \bibfield  {author} {\bibinfo {author} {\bibfnamefont {Avik}\ \bibnamefont
  {Dutt}}, \bibinfo {author} {\bibfnamefont {Qian}\ \bibnamefont {Lin}},
  \bibinfo {author} {\bibfnamefont {Luqi}\ \bibnamefont {Yuan}}, \bibinfo
  {author} {\bibfnamefont {Momchil}\ \bibnamefont {Minkov}}, \bibinfo {author}
  {\bibfnamefont {Meng}\ \bibnamefont {Xiao}}, \ and\ \bibinfo {author}
  {\bibfnamefont {Shanhui}\ \bibnamefont {Fan}},\ }\bibfield  {title} {\enquote
  {\bibinfo {title} {A single photonic cavity with two independent physical
  synthetic dimensions},}\ }\href {\doibase 10.1126/science.aaz3071} {\bibfield
   {journal} {\bibinfo  {journal} {Science}\ }\textbf {\bibinfo {volume}
  {367}},\ \bibinfo {pages} {59--64} (\bibinfo {year} {2020})}\BibitemShut
  {NoStop}%
\bibitem [{\citenamefont {Coen}\ \emph {et~al.}(2024)\citenamefont {Coen},
  \citenamefont {Garbin}, \citenamefont {Xu}, \citenamefont {Quinn},
  \citenamefont {Goldman}, \citenamefont {Oppo}, \citenamefont {Erkintalo},
  \citenamefont {Murdoch},\ and\ \citenamefont {Fatome}}]{coenNonlinear2024}%
  \BibitemOpen
  \bibfield  {author} {\bibinfo {author} {\bibfnamefont {St{\'e}phane}\
  \bibnamefont {Coen}}, \bibinfo {author} {\bibfnamefont {Bruno}\ \bibnamefont
  {Garbin}}, \bibinfo {author} {\bibfnamefont {Gang}\ \bibnamefont {Xu}},
  \bibinfo {author} {\bibfnamefont {Liam}\ \bibnamefont {Quinn}}, \bibinfo
  {author} {\bibfnamefont {Nathan}\ \bibnamefont {Goldman}}, \bibinfo {author}
  {\bibfnamefont {Gian-Luca}\ \bibnamefont {Oppo}}, \bibinfo {author}
  {\bibfnamefont {Miro}\ \bibnamefont {Erkintalo}}, \bibinfo {author}
  {\bibfnamefont {Stuart~G.}\ \bibnamefont {Murdoch}}, \ and\ \bibinfo {author}
  {\bibfnamefont {Julien}\ \bibnamefont {Fatome}},\ }\bibfield  {title}
  {\enquote {\bibinfo {title} {Nonlinear topological symmetry protection in a
  dissipative system},}\ }\href {\doibase 10.1038/s41467-023-44640-x}
  {\bibfield  {journal} {\bibinfo  {journal} {Nature Communications}\ }\textbf
  {\bibinfo {volume} {15}},\ \bibinfo {pages} {1398} (\bibinfo {year}
  {2024})}\BibitemShut {NoStop}%
\bibitem [{\citenamefont {Smirnova}\ \emph {et~al.}(2020)\citenamefont
  {Smirnova}, \citenamefont {Leykam}, \citenamefont {Chong},\ and\
  \citenamefont {Kivshar}}]{smirnovaNonlinear2020}%
  \BibitemOpen
  \bibfield  {author} {\bibinfo {author} {\bibfnamefont {Daria}\ \bibnamefont
  {Smirnova}}, \bibinfo {author} {\bibfnamefont {Daniel}\ \bibnamefont
  {Leykam}}, \bibinfo {author} {\bibfnamefont {Yidong}\ \bibnamefont {Chong}},
  \ and\ \bibinfo {author} {\bibfnamefont {Yuri}\ \bibnamefont {Kivshar}},\
  }\bibfield  {title} {\enquote {\bibinfo {title} {Nonlinear topological
  photonics},}\ }\href {\doibase 10.1063/1.5142397} {\bibfield  {journal}
  {\bibinfo  {journal} {Applied Physics Reviews}\ }\textbf {\bibinfo {volume}
  {7}},\ \bibinfo {pages} {021306} (\bibinfo {year} {2020})}\BibitemShut
  {NoStop}%
\bibitem [{\citenamefont {Clark}\ \emph {et~al.}(2019)\citenamefont {Clark},
  \citenamefont {Jia}, \citenamefont {Schine}, \citenamefont {Baum},
  \citenamefont {Georgakopoulos},\ and\ \citenamefont
  {Simon}}]{clarkInteracting2019}%
  \BibitemOpen
  \bibfield  {author} {\bibinfo {author} {\bibfnamefont {Logan~W.}\
  \bibnamefont {Clark}}, \bibinfo {author} {\bibfnamefont {Ningyuan}\
  \bibnamefont {Jia}}, \bibinfo {author} {\bibfnamefont {Nathan}\ \bibnamefont
  {Schine}}, \bibinfo {author} {\bibfnamefont {Claire}\ \bibnamefont {Baum}},
  \bibinfo {author} {\bibfnamefont {Alexandros}\ \bibnamefont
  {Georgakopoulos}}, \ and\ \bibinfo {author} {\bibfnamefont {Jonathan}\
  \bibnamefont {Simon}},\ }\bibfield  {title} {\enquote {\bibinfo {title}
  {Interacting {{Floquet}} polaritons},}\ }\href {\doibase
  10.1038/s41586-019-1354-5} {\bibfield  {journal} {\bibinfo  {journal}
  {Nature}\ }\textbf {\bibinfo {volume} {571}},\ \bibinfo {pages} {532--536}
  (\bibinfo {year} {2019})}\BibitemShut {NoStop}%
\bibitem [{\citenamefont {Lemmer}\ \emph {et~al.}(2018)\citenamefont {Lemmer},
  \citenamefont {Cormick}, \citenamefont {Tamascelli}, \citenamefont {Schaetz},
  \citenamefont {Huelga},\ and\ \citenamefont {Plenio}}]{lemmertrappedion2018}%
  \BibitemOpen
  \bibfield  {author} {\bibinfo {author} {\bibfnamefont {A.}~\bibnamefont
  {Lemmer}}, \bibinfo {author} {\bibfnamefont {C.}~\bibnamefont {Cormick}},
  \bibinfo {author} {\bibfnamefont {D.}~\bibnamefont {Tamascelli}}, \bibinfo
  {author} {\bibfnamefont {T.}~\bibnamefont {Schaetz}}, \bibinfo {author}
  {\bibfnamefont {S.~F.}\ \bibnamefont {Huelga}}, \ and\ \bibinfo {author}
  {\bibfnamefont {M.~B.}\ \bibnamefont {Plenio}},\ }\bibfield  {title}
  {\enquote {\bibinfo {title} {A trapped-ion simulator for spin-boson models
  with structured environments},}\ }\href {\doibase 10.1088/1367-2630/aac87d}
  {\bibfield  {journal} {\bibinfo  {journal} {New Journal of Physics}\ }\textbf
  {\bibinfo {volume} {20}},\ \bibinfo {pages} {073002} (\bibinfo {year}
  {2018})}\BibitemShut {NoStop}%
\bibitem [{\citenamefont {Underwood}\ \emph {et~al.}(2012)\citenamefont
  {Underwood}, \citenamefont {Shanks}, \citenamefont {Koch},\ and\
  \citenamefont {Houck}}]{underwoodLowdisorder2012}%
  \BibitemOpen
  \bibfield  {author} {\bibinfo {author} {\bibfnamefont {D.~L.}\ \bibnamefont
  {Underwood}}, \bibinfo {author} {\bibfnamefont {W.~E.}\ \bibnamefont
  {Shanks}}, \bibinfo {author} {\bibfnamefont {Jens}\ \bibnamefont {Koch}}, \
  and\ \bibinfo {author} {\bibfnamefont {A.~A.}\ \bibnamefont {Houck}},\
  }\bibfield  {title} {\enquote {\bibinfo {title} {Low-disorder microwave
  cavity lattices for quantum simulation with photons},}\ }\href {\doibase
  10.1103/PhysRevA.86.023837} {\bibfield  {journal} {\bibinfo  {journal}
  {Physical Review A}\ }\textbf {\bibinfo {volume} {86}},\ \bibinfo {pages}
  {023837} (\bibinfo {year} {2012})}\BibitemShut {NoStop}%
\bibitem [{\citenamefont {Fedorov}\ \emph {et~al.}(2021)\citenamefont
  {Fedorov}, \citenamefont {Remizov}, \citenamefont {Shapiro}, \citenamefont
  {Pogosov}, \citenamefont {Egorova}, \citenamefont {Tsitsilin}, \citenamefont
  {Andronik}, \citenamefont {Dobronosova}, \citenamefont {Rodionov},
  \citenamefont {Astafiev},\ and\ \citenamefont {Ustinov}}]{fedorovPhoton2021}%
  \BibitemOpen
  \bibfield  {author} {\bibinfo {author} {\bibfnamefont {G.~P.}\ \bibnamefont
  {Fedorov}}, \bibinfo {author} {\bibfnamefont {S.~V.}\ \bibnamefont
  {Remizov}}, \bibinfo {author} {\bibfnamefont {D.~S.}\ \bibnamefont
  {Shapiro}}, \bibinfo {author} {\bibfnamefont {W.~V.}\ \bibnamefont
  {Pogosov}}, \bibinfo {author} {\bibfnamefont {E.}~\bibnamefont {Egorova}},
  \bibinfo {author} {\bibfnamefont {I.}~\bibnamefont {Tsitsilin}}, \bibinfo
  {author} {\bibfnamefont {M.}~\bibnamefont {Andronik}}, \bibinfo {author}
  {\bibfnamefont {A.~A.}\ \bibnamefont {Dobronosova}}, \bibinfo {author}
  {\bibfnamefont {I.~A.}\ \bibnamefont {Rodionov}}, \bibinfo {author}
  {\bibfnamefont {O.~V.}\ \bibnamefont {Astafiev}}, \ and\ \bibinfo {author}
  {\bibfnamefont {A.~V.}\ \bibnamefont {Ustinov}},\ }\bibfield  {title}
  {\enquote {\bibinfo {title} {Photon {{Transport}} in a {{Bose-Hubbard Chain}}
  of {{Superconducting Artificial Atoms}}},}\ }\href {\doibase
  10.1103/PhysRevLett.126.180503} {\bibfield  {journal} {\bibinfo  {journal}
  {Physical Review Letters}\ }\textbf {\bibinfo {volume} {126}},\ \bibinfo
  {pages} {180503} (\bibinfo {year} {2021})}\BibitemShut {NoStop}%
\bibitem [{\citenamefont {Jouanny}\ \emph {et~al.}(2024)\citenamefont
  {Jouanny}, \citenamefont {Frasca}, \citenamefont {Weibel}, \citenamefont
  {Peyruchat}, \citenamefont {Scigliuzzo}, \citenamefont {Oppliger},
  \citenamefont {De~Palma}, \citenamefont {Sbroggio}, \citenamefont {Beaulieu},
  \citenamefont {Zilberberg},\ and\ \citenamefont
  {Scarlino}}]{jouannyBand2024}%
  \BibitemOpen
  \bibfield  {author} {\bibinfo {author} {\bibfnamefont {Vincent}\ \bibnamefont
  {Jouanny}}, \bibinfo {author} {\bibfnamefont {Simone}\ \bibnamefont
  {Frasca}}, \bibinfo {author} {\bibfnamefont {Vera~Jo}\ \bibnamefont
  {Weibel}}, \bibinfo {author} {\bibfnamefont {Leo}\ \bibnamefont {Peyruchat}},
  \bibinfo {author} {\bibfnamefont {Marco}\ \bibnamefont {Scigliuzzo}},
  \bibinfo {author} {\bibfnamefont {Fabian}\ \bibnamefont {Oppliger}}, \bibinfo
  {author} {\bibfnamefont {Franco}\ \bibnamefont {De~Palma}}, \bibinfo {author}
  {\bibfnamefont {Davide}\ \bibnamefont {Sbroggio}}, \bibinfo {author}
  {\bibfnamefont {Guillaume}\ \bibnamefont {Beaulieu}}, \bibinfo {author}
  {\bibfnamefont {Oded}\ \bibnamefont {Zilberberg}}, \ and\ \bibinfo {author}
  {\bibfnamefont {Pasquale}\ \bibnamefont {Scarlino}},\ }\bibfield  {title}
  {\enquote {\bibinfo {title} {Band engineering and study of disorder using
  topology in compact high kinetic inductance cavity arrays},}\ }\href
  {\doibase 10.48550/arXiv.2403.18150} {\  (\bibinfo {year} {2024}),\
  10.48550/arXiv.2403.18150},\ \Eprint {http://arxiv.org/abs/2403.18150}
  {2403.18150 [cond-mat, physics:quant-ph]} \BibitemShut {NoStop}%
\bibitem [{\citenamefont {Frisk~Kockum}\ \emph {et~al.}(2019)\citenamefont
  {Frisk~Kockum}, \citenamefont {Miranowicz}, \citenamefont {De~Liberato},
  \citenamefont {Savasta},\ and\ \citenamefont
  {Nori}}]{friskkockumUltrastrong2019}%
  \BibitemOpen
  \bibfield  {author} {\bibinfo {author} {\bibfnamefont {Anton}\ \bibnamefont
  {Frisk~Kockum}}, \bibinfo {author} {\bibfnamefont {Adam}\ \bibnamefont
  {Miranowicz}}, \bibinfo {author} {\bibfnamefont {Simone}\ \bibnamefont
  {De~Liberato}}, \bibinfo {author} {\bibfnamefont {Salvatore}\ \bibnamefont
  {Savasta}}, \ and\ \bibinfo {author} {\bibfnamefont {Franco}\ \bibnamefont
  {Nori}},\ }\bibfield  {title} {\enquote {\bibinfo {title} {Ultrastrong
  coupling between light and matter},}\ }\href {\doibase
  10.1038/s42254-018-0006-2} {\bibfield  {journal} {\bibinfo  {journal} {Nature
  Reviews Physics}\ }\textbf {\bibinfo {volume} {1}},\ \bibinfo {pages}
  {19--40} (\bibinfo {year} {2019})}\BibitemShut {NoStop}%
\bibitem [{\citenamefont {Bonifacio}\ \emph {et~al.}(2020)\citenamefont
  {Bonifacio}, \citenamefont {Dom{\'i}nguez},\ and\ \citenamefont
  {S{\'a}nchez}}]{bonifacioLandauZenerSt2020}%
  \BibitemOpen
  \bibfield  {author} {\bibinfo {author} {\bibfnamefont {Mariano}\ \bibnamefont
  {Bonifacio}}, \bibinfo {author} {\bibfnamefont {Daniel}\ \bibnamefont
  {Dom{\'i}nguez}}, \ and\ \bibinfo {author} {\bibfnamefont
  {Mar{\'i}a~Jos{\'e}}\ \bibnamefont {S{\'a}nchez}},\ }\bibfield  {title}
  {\enquote {\bibinfo {title} {Landau-{{Zener-St}}{\"u}ckelberg interferometry
  in dissipative circuit quantum electrodynamics},}\ }\href {\doibase
  10.1103/PhysRevB.101.245415} {\bibfield  {journal} {\bibinfo  {journal}
  {Physical Review B}\ }\textbf {\bibinfo {volume} {101}},\ \bibinfo {pages}
  {245415} (\bibinfo {year} {2020})}\BibitemShut {NoStop}%
\bibitem [{\citenamefont {Lescanne}\ \emph {et~al.}(2020)\citenamefont
  {Lescanne}, \citenamefont {Villiers}, \citenamefont {Peronnin}, \citenamefont
  {Sarlette}, \citenamefont {Delbecq}, \citenamefont {Huard}, \citenamefont
  {Kontos}, \citenamefont {Mirrahimi},\ and\ \citenamefont
  {Leghtas}}]{Lescanne2020}%
  \BibitemOpen
  \bibfield  {author} {\bibinfo {author} {\bibfnamefont {Raphaël}\
  \bibnamefont {Lescanne}}, \bibinfo {author} {\bibfnamefont {Marius}\
  \bibnamefont {Villiers}}, \bibinfo {author} {\bibfnamefont {Théau}\
  \bibnamefont {Peronnin}}, \bibinfo {author} {\bibfnamefont {Alain}\
  \bibnamefont {Sarlette}}, \bibinfo {author} {\bibfnamefont {Matthieu}\
  \bibnamefont {Delbecq}}, \bibinfo {author} {\bibfnamefont {Benjamin}\
  \bibnamefont {Huard}}, \bibinfo {author} {\bibfnamefont {Takis}\ \bibnamefont
  {Kontos}}, \bibinfo {author} {\bibfnamefont {Mazyar}\ \bibnamefont
  {Mirrahimi}}, \ and\ \bibinfo {author} {\bibfnamefont {Zaki}\ \bibnamefont
  {Leghtas}},\ }\bibfield  {title} {\enquote {\bibinfo {title} {Exponential
  suppression of bit-flips in a qubit encoded in an oscillator},}\ }\href
  {\doibase 10.1038/s41567-020-0824-x} {\bibfield  {journal} {\bibinfo
  {journal} {Nature Physics}\ }\textbf {\bibinfo {volume} {16}},\ \bibinfo
  {pages} {509–513} (\bibinfo {year} {2020})}\BibitemShut {NoStop}%
\bibitem [{\citenamefont {Damski}(2005)}]{BogdanPRL05}%
  \BibitemOpen
  \bibfield  {author} {\bibinfo {author} {\bibfnamefont {Bogdan}\ \bibnamefont
  {Damski}},\ }\bibfield  {title} {\enquote {\bibinfo {title} {The simplest
  quantum model supporting the kibble-zurek mechanism of topological defect
  production: Landau-zener transitions from a new perspective},}\ }\href
  {\doibase 10.1103/PhysRevLett.95.035701} {\bibfield  {journal} {\bibinfo
  {journal} {Phys. Rev. Lett.}\ }\textbf {\bibinfo {volume} {95}},\ \bibinfo
  {pages} {035701} (\bibinfo {year} {2005})}\BibitemShut {NoStop}%
\bibitem [{\citenamefont {{Higuera-Quintero}}\ \emph
  {et~al.}(2022)\citenamefont {{Higuera-Quintero}}, \citenamefont
  {Rodr{\'i}guez}, \citenamefont {Quiroga},\ and\ \citenamefont
  {{G{\'o}mez-Ruiz}}}]{higuera-quinteroExperimental2022b}%
  \BibitemOpen
  \bibfield  {author} {\bibinfo {author} {\bibfnamefont {Santiago}\
  \bibnamefont {{Higuera-Quintero}}}, \bibinfo {author} {\bibfnamefont
  {Ferney~J.}\ \bibnamefont {Rodr{\'i}guez}}, \bibinfo {author} {\bibfnamefont
  {Luis}\ \bibnamefont {Quiroga}}, \ and\ \bibinfo {author} {\bibfnamefont
  {Fernando~J.}\ \bibnamefont {{G{\'o}mez-Ruiz}}},\ }\bibfield  {title}
  {\enquote {\bibinfo {title} {Experimental validation of the {{Kibble-Zurek}}
  mechanism on a digital quantum computer},}\ }\href {\doibase
  10.3389/frqst.2022.1026025} {\bibfield  {journal} {\bibinfo  {journal}
  {Frontiers in Quantum Science and Technology}\ }\textbf {\bibinfo {volume}
  {1}} (\bibinfo {year} {2022}),\ 10.3389/frqst.2022.1026025}\BibitemShut
  {NoStop}%
\bibitem [{\citenamefont {Kreikebaum}\ \emph {et~al.}(2020)\citenamefont
  {Kreikebaum}, \citenamefont {O'Brien}, \citenamefont {Morvan},\ and\
  \citenamefont {Siddiqi}}]{kreikebaumImproving2020}%
  \BibitemOpen
  \bibfield  {author} {\bibinfo {author} {\bibfnamefont {J~M}\ \bibnamefont
  {Kreikebaum}}, \bibinfo {author} {\bibfnamefont {K~P}\ \bibnamefont
  {O'Brien}}, \bibinfo {author} {\bibfnamefont {A}~\bibnamefont {Morvan}}, \
  and\ \bibinfo {author} {\bibfnamefont {I}~\bibnamefont {Siddiqi}},\
  }\bibfield  {title} {\enquote {\bibinfo {title} {Improving wafer-scale
  {{Josephson}} junction resistance variation in superconducting quantum
  coherent circuits},}\ }\href {\doibase 10.1088/1361-6668/ab8617} {\bibfield
  {journal} {\bibinfo  {journal} {Superconductor Science and Technology}\
  }\textbf {\bibinfo {volume} {33}},\ \bibinfo {pages} {06LT02} (\bibinfo
  {year} {2020})}\BibitemShut {NoStop}%
\bibitem [{\citenamefont {Chen}\ \emph {et~al.}(2022)\citenamefont {Chen},
  \citenamefont {Partanen}, \citenamefont {Fesquet}, \citenamefont {Honasoge},
  \citenamefont {Kronowetter}, \citenamefont {Nojiri}, \citenamefont {Renger},
  \citenamefont {Fedorov}, \citenamefont {Marx}, \citenamefont {Deppe},\ and\
  \citenamefont {Gross}}]{chenScattering2022a}%
  \BibitemOpen
  \bibfield  {author} {\bibinfo {author} {\bibfnamefont {Qi-Ming}\ \bibnamefont
  {Chen}}, \bibinfo {author} {\bibfnamefont {Matti}\ \bibnamefont {Partanen}},
  \bibinfo {author} {\bibfnamefont {Florian}\ \bibnamefont {Fesquet}}, \bibinfo
  {author} {\bibfnamefont {Kedar~E.}\ \bibnamefont {Honasoge}}, \bibinfo
  {author} {\bibfnamefont {Fabian}\ \bibnamefont {Kronowetter}}, \bibinfo
  {author} {\bibfnamefont {Yuki}\ \bibnamefont {Nojiri}}, \bibinfo {author}
  {\bibfnamefont {Michael}\ \bibnamefont {Renger}}, \bibinfo {author}
  {\bibfnamefont {Kirill~G.}\ \bibnamefont {Fedorov}}, \bibinfo {author}
  {\bibfnamefont {Achim}\ \bibnamefont {Marx}}, \bibinfo {author}
  {\bibfnamefont {Frank}\ \bibnamefont {Deppe}}, \ and\ \bibinfo {author}
  {\bibfnamefont {Rudolf}\ \bibnamefont {Gross}},\ }\bibfield  {title}
  {\enquote {\bibinfo {title} {Scattering coefficients of superconducting
  microwave resonators. {{II}}. {{System-bath}} approach},}\ }\href {\doibase
  10.1103/PhysRevB.106.214506} {\bibfield  {journal} {\bibinfo  {journal}
  {Physical Review B}\ }\textbf {\bibinfo {volume} {106}},\ \bibinfo {pages}
  {214506} (\bibinfo {year} {2022})}\BibitemShut {NoStop}%
\bibitem [{\citenamefont {Khalil}\ \emph {et~al.}(2012)\citenamefont {Khalil},
  \citenamefont {Stoutimore}, \citenamefont {Wellstood},\ and\ \citenamefont
  {Osborn}}]{khalilAnalysisMethodAsymmetric2012}%
  \BibitemOpen
  \bibfield  {author} {\bibinfo {author} {\bibfnamefont {M.~S.}\ \bibnamefont
  {Khalil}}, \bibinfo {author} {\bibfnamefont {M.~J.~A.}\ \bibnamefont
  {Stoutimore}}, \bibinfo {author} {\bibfnamefont {F.~C.}\ \bibnamefont
  {Wellstood}}, \ and\ \bibinfo {author} {\bibfnamefont {K.~D.}\ \bibnamefont
  {Osborn}},\ }\bibfield  {title} {\enquote {\bibinfo {title} {An analysis
  method for asymmetric resonator transmission applied to superconducting
  devices},}\ }\href {\doibase 10.1063/1.3692073} {\bibfield  {journal}
  {\bibinfo  {journal} {Journal of Applied Physics}\ }\textbf {\bibinfo
  {volume} {111}},\ \bibinfo {pages} {054510} (\bibinfo {year}
  {2012})}\BibitemShut {NoStop}%
\bibitem [{\citenamefont {Eichler}\ and\ \citenamefont
  {Wallraff}(2014)}]{eichlerControlling2014}%
  \BibitemOpen
  \bibfield  {author} {\bibinfo {author} {\bibfnamefont {Christopher}\
  \bibnamefont {Eichler}}\ and\ \bibinfo {author} {\bibfnamefont {Andreas}\
  \bibnamefont {Wallraff}},\ }\bibfield  {title} {\enquote {\bibinfo {title}
  {Controlling the dynamic range of a {{Josephson}} parametric amplifier},}\
  }\href {\doibase 10.1140/epjqt2} {\bibfield  {journal} {\bibinfo  {journal}
  {EPJ Quantum Technology}\ }\textbf {\bibinfo {volume} {1}},\ \bibinfo {pages}
  {1--19} (\bibinfo {year} {2014})}\BibitemShut {NoStop}%
\bibitem [{\citenamefont {Kos}\ \emph {et~al.}(2018)\citenamefont {Kos},
  \citenamefont {Ljubotina},\ and\ \citenamefont
  {Prosen}}]{kos_many-body_2018}%
  \BibitemOpen
  \bibfield  {author} {\bibinfo {author} {\bibfnamefont {Pavel}\ \bibnamefont
  {Kos}}, \bibinfo {author} {\bibfnamefont {Marko}\ \bibnamefont {Ljubotina}},
  \ and\ \bibinfo {author} {\bibfnamefont {Tomaž}\ \bibnamefont {Prosen}},\
  }\bibfield  {title} {\enquote {\bibinfo {title} {Many-{Body} {Quantum}
  {Chaos}: {Analytic} {Connection} to {Random} {Matrix} {Theory}},}\ }\href
  {\doibase 10.1103/PhysRevX.8.021062} {\bibfield  {journal} {\bibinfo
  {journal} {Physical Review X}\ }\textbf {\bibinfo {volume} {8}},\ \bibinfo
  {pages} {021062} (\bibinfo {year} {2018})}\BibitemShut {NoStop}%
\bibitem [{\citenamefont {Akemann}\ \emph {et~al.}(2019)\citenamefont
  {Akemann}, \citenamefont {Kieburg}, \citenamefont {Mielke},\ and\
  \citenamefont {Prosen}}]{AkemannPRL19}%
  \BibitemOpen
  \bibfield  {author} {\bibinfo {author} {\bibfnamefont {Gernot}\ \bibnamefont
  {Akemann}}, \bibinfo {author} {\bibfnamefont {Mario}\ \bibnamefont
  {Kieburg}}, \bibinfo {author} {\bibfnamefont {Adam}\ \bibnamefont {Mielke}},
  \ and\ \bibinfo {author} {\bibfnamefont {Tomaz}\ \bibnamefont {Prosen}},\
  }\bibfield  {title} {\enquote {\bibinfo {title} {Universal {Signature} from
  {Integrability} to {Chaos} in {Dissipative} {Open} {Quantum} {Systems}},}\
  }\href {\doibase 10.1103/PhysRevLett.123.254101} {\bibfield  {journal}
  {\bibinfo  {journal} {Physical Review Letters}\ }\textbf {\bibinfo {volume}
  {123}},\ \bibinfo {pages} {254101} (\bibinfo {year} {2019})}\BibitemShut
  {NoStop}%
\bibitem [{\citenamefont {Serbyn}\ and\ \citenamefont
  {Moore}(2016)}]{serbyn_spectral_2016}%
  \BibitemOpen
  \bibfield  {author} {\bibinfo {author} {\bibfnamefont {Maksym}\ \bibnamefont
  {Serbyn}}\ and\ \bibinfo {author} {\bibfnamefont {Joel~E.}\ \bibnamefont
  {Moore}},\ }\bibfield  {title} {\enquote {\bibinfo {title} {Spectral
  statistics across the many-body localization transition},}\ }\href {\doibase
  10.1103/PhysRevB.93.041424} {\bibfield  {journal} {\bibinfo  {journal}
  {Physical Review B}\ }\textbf {\bibinfo {volume} {93}},\ \bibinfo {pages}
  {041424} (\bibinfo {year} {2016})}\BibitemShut {NoStop}%
\bibitem [{\citenamefont {Bordia}\ \emph {et~al.}(2017)\citenamefont {Bordia},
  \citenamefont {Lüschen}, \citenamefont {Schneider}, \citenamefont {Knap},\
  and\ \citenamefont {Bloch}}]{bordia_periodically_2017}%
  \BibitemOpen
  \bibfield  {author} {\bibinfo {author} {\bibfnamefont {Pranjal}\ \bibnamefont
  {Bordia}}, \bibinfo {author} {\bibfnamefont {Henrik}\ \bibnamefont
  {Lüschen}}, \bibinfo {author} {\bibfnamefont {Ulrich}\ \bibnamefont
  {Schneider}}, \bibinfo {author} {\bibfnamefont {Michael}\ \bibnamefont
  {Knap}}, \ and\ \bibinfo {author} {\bibfnamefont {Immanuel}\ \bibnamefont
  {Bloch}},\ }\bibfield  {title} {\enquote {\bibinfo {title} {Periodically
  driving a many-body localized quantum system},}\ }\href {\doibase
  10.1038/nphys4020} {\bibfield  {journal} {\bibinfo  {journal} {Nature
  Physics}\ }\textbf {\bibinfo {volume} {13}},\ \bibinfo {pages} {460--464}
  (\bibinfo {year} {2017})}\BibitemShut {NoStop}%
\bibitem [{\citenamefont {Abanin}\ \emph {et~al.}(2019)\citenamefont {Abanin},
  \citenamefont {Altman}, \citenamefont {Bloch},\ and\ \citenamefont
  {Serbyn}}]{abanin_colloquium_2019}%
  \BibitemOpen
  \bibfield  {author} {\bibinfo {author} {\bibfnamefont {Dmitry~A.}\
  \bibnamefont {Abanin}}, \bibinfo {author} {\bibfnamefont {Ehud}\ \bibnamefont
  {Altman}}, \bibinfo {author} {\bibfnamefont {Immanuel}\ \bibnamefont
  {Bloch}}, \ and\ \bibinfo {author} {\bibfnamefont {Maksym}\ \bibnamefont
  {Serbyn}},\ }\bibfield  {title} {\enquote {\bibinfo {title}
  {\textit{{Colloquium}} : {Many}-body localization, thermalization, and
  entanglement},}\ }\href {\doibase 10.1103/RevModPhys.91.021001} {\bibfield
  {journal} {\bibinfo  {journal} {Reviews of Modern Physics}\ }\textbf
  {\bibinfo {volume} {91}},\ \bibinfo {pages} {021001} (\bibinfo {year}
  {2019})}\BibitemShut {NoStop}%
\bibitem [{\citenamefont {Breuer}\ and\ \citenamefont
  {Petruccione}(2007)}]{breuer_theory_2007}%
  \BibitemOpen
  \bibfield  {author} {\bibinfo {author} {\bibfnamefont {Heinz-Peter}\
  \bibnamefont {Breuer}}\ and\ \bibinfo {author} {\bibfnamefont {Francesco}\
  \bibnamefont {Petruccione}},\ }\href {\doibase
  10.1093/acprof:oso/9780199213900.001.0001} {\emph {\bibinfo {title} {The
  {Theory} of {Open} {Quantum} {Systems}}}},\ \bibinfo {edition} {1st}\ ed.\
  (\bibinfo  {publisher} {Oxford University PressOxford},\ \bibinfo {year}
  {2007})\BibitemShut {NoStop}%
\bibitem [{\citenamefont {Markum}\ \emph {et~al.}(1999)\citenamefont {Markum},
  \citenamefont {Pullirsch},\ and\ \citenamefont {Wettig}}]{MarkumPRL99}%
  \BibitemOpen
  \bibfield  {author} {\bibinfo {author} {\bibfnamefont {H.}~\bibnamefont
  {Markum}}, \bibinfo {author} {\bibfnamefont {R.}~\bibnamefont {Pullirsch}}, \
  and\ \bibinfo {author} {\bibfnamefont {T.}~\bibnamefont {Wettig}},\
  }\bibfield  {title} {\enquote {\bibinfo {title} {Non-{Hermitian} {Random}
  {Matrix} {Theory} and {Lattice} {QCD} with {Chemical} {Potential}},}\ }\href
  {\doibase 10.1103/PhysRevLett.83.484} {\bibfield  {journal} {\bibinfo
  {journal} {Physical Review Letters}\ }\textbf {\bibinfo {volume} {83}},\
  \bibinfo {pages} {484--487} (\bibinfo {year} {1999})}\BibitemShut {NoStop}%
\bibitem [{Note5()}]{Note5}%
  \BibitemOpen
  \bibinfo {note} {We set the cutoff $c_{\protect \textrm {min}} = \protect
  \bar {C}/1000$ where $\protect \bar {C}$ is the average of the spectral
  weights in \protect \mbox {Eq.~(\ref {eq:spectralrho})}, as detailed in
  Ref.~\cite {FerrariPRR25}. For the Floquet Liouvillian, we found that some
  of the $|c_j|$ were very large (order of magnitudes bigger than one). As the
  average procedure of the spectral weights would have been affected by those
  outliers, we restrict the mean to the ones such that $|c_j|\le 1$. Such a
  choice is justified as a spectral coefficient $|c_j| > 1$ will be for sure
  chosen with the SSQT protocol, and we get a meaningful $c_{\protect \rm min}$
  as in Ref.~\cite {FerrariPRR25}.}\BibitemShut {Stop}%
\bibitem [{\citenamefont {Walls}\ and\ \citenamefont
  {Milburn}(2008)}]{WallsBOOK08}%
  \BibitemOpen
  \bibinfo {editor} {\bibfnamefont {D.F.}\ \bibnamefont {Walls}}\ and\ \bibinfo
  {editor} {\bibfnamefont {Gerard~J.}\ \bibnamefont {Milburn}},\ eds.,\ \href
  {\doibase 10.1007/978-3-540-28574-8} {\emph {\bibinfo {title} {Quantum
  {Optics}}}}\ (\bibinfo  {publisher} {Springer Berlin Heidelberg},\ \bibinfo
  {address} {Berlin, Heidelberg},\ \bibinfo {year} {2008})\BibitemShut
  {NoStop}%
\bibitem [{\citenamefont {Wiseman}\ and\ \citenamefont
  {Milburn}(2009)}]{WisemanBOOK09}%
  \BibitemOpen
  \bibfield  {author} {\bibinfo {author} {\bibfnamefont {Howard~M.}\
  \bibnamefont {Wiseman}}\ and\ \bibinfo {author} {\bibfnamefont {Gerard~J.}\
  \bibnamefont {Milburn}},\ }\href {\doibase 10.1017/CBO9780511813948} {\emph
  {\bibinfo {title} {Quantum {Measurement} and {Control}}}},\ \bibinfo
  {edition} {1st}\ ed.\ (\bibinfo  {publisher} {Cambridge University Press},\
  \bibinfo {year} {2009})\BibitemShut {NoStop}%
\bibitem [{\citenamefont {Maragkou}\ \emph {et~al.}(2013)\citenamefont
  {Maragkou}, \citenamefont {S\'anchez-Mu\~noz}, \citenamefont
  {Lazi\ifmmode~\acute{c}\else \'{c}\fi{}}, \citenamefont {Chernysheva},
  \citenamefont {van~der Meulen}, \citenamefont {Gonz\'alez-Tudela},
  \citenamefont {Tejedor}, \citenamefont {Mart\'{\i}nez}, \citenamefont
  {Prieto}, \citenamefont {Postigo},\ and\ \citenamefont
  {Calleja}}]{MaragkouPRB13}%
  \BibitemOpen
  \bibfield  {author} {\bibinfo {author} {\bibfnamefont {M.}~\bibnamefont
  {Maragkou}}, \bibinfo {author} {\bibfnamefont {C.}~\bibnamefont
  {S\'anchez-Mu\~noz}}, \bibinfo {author} {\bibfnamefont {S.}~\bibnamefont
  {Lazi\ifmmode~\acute{c}\else \'{c}\fi{}}}, \bibinfo {author} {\bibfnamefont
  {E.}~\bibnamefont {Chernysheva}}, \bibinfo {author} {\bibfnamefont {H.~P.}\
  \bibnamefont {van~der Meulen}}, \bibinfo {author} {\bibfnamefont
  {A.}~\bibnamefont {Gonz\'alez-Tudela}}, \bibinfo {author} {\bibfnamefont
  {C.}~\bibnamefont {Tejedor}}, \bibinfo {author} {\bibfnamefont {L.~J.}\
  \bibnamefont {Mart\'{\i}nez}}, \bibinfo {author} {\bibfnamefont
  {I.}~\bibnamefont {Prieto}}, \bibinfo {author} {\bibfnamefont {P.~A.}\
  \bibnamefont {Postigo}}, \ and\ \bibinfo {author} {\bibfnamefont {J.~M.}\
  \bibnamefont {Calleja}},\ }\bibfield  {title} {\enquote {\bibinfo {title}
  {Bichromatic dressing of a quantum dot detected by a remote second quantum
  dot},}\ }\href {\doibase 10.1103/PhysRevB.88.075309} {\bibfield  {journal}
  {\bibinfo  {journal} {Phys. Rev. B}\ }\textbf {\bibinfo {volume} {88}},\
  \bibinfo {pages} {075309} (\bibinfo {year} {2013})}\BibitemShut {NoStop}%
\bibitem [{\citenamefont {Macr\`{\i}}\ \emph {et~al.}(2022)\citenamefont
  {Macr\`{\i}}, \citenamefont {Mercurio}, \citenamefont {Nori}, \citenamefont
  {Savasta},\ and\ \citenamefont {S\'anchez Mu\~noz}}]{MacriPRL22}%
  \BibitemOpen
  \bibfield  {author} {\bibinfo {author} {\bibfnamefont {Vincenzo}\
  \bibnamefont {Macr\`{\i}}}, \bibinfo {author} {\bibfnamefont {Alberto}\
  \bibnamefont {Mercurio}}, \bibinfo {author} {\bibfnamefont {Franco}\
  \bibnamefont {Nori}}, \bibinfo {author} {\bibfnamefont {Salvatore}\
  \bibnamefont {Savasta}}, \ and\ \bibinfo {author} {\bibfnamefont {Carlos}\
  \bibnamefont {S\'anchez Mu\~noz}},\ }\bibfield  {title} {\enquote {\bibinfo
  {title} {Spontaneous scattering of raman photons from cavity-qed systems in
  the ultrastrong coupling regime},}\ }\href {\doibase
  10.1103/PhysRevLett.129.273602} {\bibfield  {journal} {\bibinfo  {journal}
  {Phys. Rev. Lett.}\ }\textbf {\bibinfo {volume} {129}},\ \bibinfo {pages}
  {273602} (\bibinfo {year} {2022})}\BibitemShut {NoStop}%
\bibitem [{\citenamefont {Minganti}\ and\ \citenamefont
  {Huybrechts}(2022)}]{MingantiQuantum22}%
  \BibitemOpen
  \bibfield  {author} {\bibinfo {author} {\bibfnamefont {Fabrizio}\
  \bibnamefont {Minganti}}\ and\ \bibinfo {author} {\bibfnamefont {Dolf}\
  \bibnamefont {Huybrechts}},\ }\bibfield  {title} {\enquote {\bibinfo {title}
  {Arnoldi-lindblad time evolution: Faster-than-the-clock algorithm for the
  spectrum of time-independent and floquet open quantum systems},}\ }\href
  {\doibase 10.22331/q-2022-02-10-649} {\bibfield  {journal} {\bibinfo
  {journal} {Quantum}\ }\textbf {\bibinfo {volume} {6}},\ \bibinfo {pages}
  {649} (\bibinfo {year} {2022})}\BibitemShut {NoStop}%
\bibitem [{Note6()}]{Note6}%
  \BibitemOpen
  \bibinfo {note} {The resonator exhibits deviation from the pure Kerr
  nonlinearity prediction due to non-negligible effects of higher
  nonlinearities, e.g., terms of the form $\chi ^{(5)} (\protect \hat
  {a}^\dagger )^3 \protect \hat {a}^3$. We estimate $\chi ^{(5)}/2\pi \approx
  -1.1\protect \,$MHz$\simeq 5\% \chi /2\pi $. As such, although the $\chi
  ^{(5)}$ term produces small changes compared to the model in Eq.~(1) of the
  main text, all the relevant physical features of the LZSM interference are
  captured by the Kerr model. We also note additional nonlinear effects due to
  the large values of flux modulation used to obtain the wanted $\zeta
  $.}\BibitemShut {Stop}%
\bibitem [{\citenamefont {Wu}\ \emph {et~al.}(2019)\citenamefont {Wu},
  \citenamefont {Zhou}, \citenamefont {Xu}, \citenamefont {Liu},\ and\
  \citenamefont {Li}}]{wuLandau2019}%
  \BibitemOpen
  \bibfield  {author} {\bibinfo {author} {\bibfnamefont {Tong}\ \bibnamefont
  {Wu}}, \bibinfo {author} {\bibfnamefont {Yuxuan}\ \bibnamefont {Zhou}},
  \bibinfo {author} {\bibfnamefont {Yuan}\ \bibnamefont {Xu}}, \bibinfo
  {author} {\bibfnamefont {Song}\ \bibnamefont {Liu}}, \ and\ \bibinfo {author}
  {\bibfnamefont {Jian}\ \bibnamefont {Li}},\ }\bibfield  {title} {\enquote
  {\bibinfo {title} {Landau--{{Zener}}--{{St{\"u}ckelberg Interference}} in
  {{Nonlinear Regime}}},}\ }\href {\doibase 10.1088/0256-307X/36/12/124204}
  {\bibfield  {journal} {\bibinfo  {journal} {Chinese Physics Letters}\
  }\textbf {\bibinfo {volume} {36}},\ \bibinfo {pages} {124204} (\bibinfo
  {year} {2019})}\BibitemShut {NoStop}%
\end{thebibliography}
\end{document}